\documentclass[10pt,rmp,preprintnumbers,amsmath,amssymb,floatfix,twocolumn]{revtex4}
\pdfoutput=1
\usepackage{graphicx}% Include figure files
\usepackage{epstopdf}
\usepackage{dcolumn}% Align table columns on decimal point
\usepackage{bm}% bold maths
\usepackage{longtable}
\usepackage{graphics}
\usepackage{natbib}
\usepackage{amssymb}
\usepackage{amsmath}
\usepackage{xspace}
\usepackage{epsfig}
\newcommand {\ibid}{{\it ibid}. }
\newcommand {\etal}{{\it et al}. }

\begin{document}
\title{An introduction to effective low-energy Hamiltonians in condensed matter physics and chemistry}
\author{B. J. Powell\hspace*{5pt}}\affiliation{Centre for Organic Photonics and Electronics, School of Mathematics and Physics,
The University of Queensland, Queensland 4072,
Australia}\email{bjpowell@gmail.com}

\maketitle

\section{Aims and scope}

These lecture notes introduce some simple effective Hamiltonians (also known as semi-empirical models) that have widespread applications to solid state and molecular systems. They are aimed as an introduction to a beginning graduate student. I also hope that it may help to break down the divide between the physics and chemistry literatures. 

After a brief introduction to second quantisation notation (section \ref{sect:2}), which is used extensively, I focus of the ``four H's": the H\"uckel (or tight binding; section \ref{sect:Huckel}), Hubbard  (section \ref{sect:Hubbard}), Heisenberg  (section \ref{sect:Heisenberg}) and Holstein (section \ref{sect:Holstein}) models. These models play central roles in our understanding of condensed matter physics, particularly for materials where electronic correlations are important, but are less well known to the chemistry community. Some other related models, such as the Pariser-Parr-Pople model, the extended Hubbard model, multi-orbital models and the ionic Hubbard model, are also discussed in section \ref{sect:other}. As well as their practical applications these models allow us to systematically investigate electronic correlations by `turning on' various interactions in the Hamiltonian one at a time. Finally, in section \ref{eh_v_se}, I discuss the epistemological basis of effective Hamiltonians and compare and contrast this approach with \emph{ab initio} methods before discussing the problem of the parameterisation of effective Hamiltonians.

As these notes are intended to be introductory, I will not attempt to make frequent comparisons to the latest research problems, rather I compare the predictions of model Hamiltonians with simple systems chosen for pedagogical reasons. Likewise, references have been chosen for their pedagogical and historical value rather than on the basis of scientific priority. 

Given the similarity in the problems addressed by theoretical chemistry and theoretical condensed matter physics, there are surprisingly few advanced texts discussing the interface of two subjects. This, unfortunately, leads to many cultural differences between the fields. Nevertheless, some textbooks do try to bridge the gap, and the reader in search of more than the introductory material presented here is referred to Refs. \onlinecite{Fulde} and \onlinecite{Jeff}.

\section{A brief introduction to second quantisation notation}\label{sect:2}

The models discussed in these notes are easiest to understand if
one employs the so-called second quantisation formalism. In this
section we briefly and informally introduce this formalism. More
details can be found in many textbooks, (e.g. Refs. \onlinecite{SchatzRatner} and \onlinecite{Mahan}). Readers
already familiar with this notation may wish to skip this section, however, the last two paragraphs do define some nomenclature that is used throughout these notes. %, as
%it does not contain any material that is not well known to those who
%know it well.

\subsection{The simple harmonic oscillator}\label{sho}

Let us begin by considering a particle of mass $m$ moving in a
one-dimensional harmonic potential:
\begin{eqnarray}
V(x)=\frac12kx^2.\label{eqn:Vsho}
\end{eqnarray}
This may be familiar  as the potential of an ideal spring displaced
from its equilibrium position by a distance $x$, in
which context $k$ is known as the spring constant \cite{Goldstein}. Eq. \ref{eqn:Vsho} is also the
potential felt by an atom as it is displaced (by a small amount) from its equilibrium
position in molecule \cite{Atkins}. Classically this problem is
straightforward to solve \cite{Goldstein} and, as well as the
trivial solution, one finds that the particle may oscillate with a 
resonant frequency $\omega=\sqrt{k/m}$. The time-independent
Schr\"odinger equation for a simple harmonic oscillator is therefore,
\begin{eqnarray}
\hat{\cal H}_\textrm{sho}\psi_n\equiv\left( \frac{\hat p^2}{2m}+\frac12m\omega^2\hat x^2 \right)\psi_n=E_n\psi_n, \label{hamsho}
\end{eqnarray}
where $\hat p=\frac{\hbar}{i}\frac{\partial}{\partial x}$ is the particle's momentum and
$\psi_n$ is the n$^\textrm{th}$ wavefunction or eigenfunction, which has energy, or eigenvalue, $E_n$.

This problem is solved in many introductory texts on quantum
mechanics \cite{Rae} using the standard methods of `first quantised'
quantum mechanics. However, a more elegant way to solve this problem
is to introduce the `ladder operator',
\begin{subequations}
\begin{eqnarray}
\hat a\equiv \sqrt{\frac{m\omega}{2\hbar}}\hat x+i\frac{\hat p}{\sqrt{2m\hbar\omega}} \label{def-ann}
\end{eqnarray}
and its hermitian conjugate
\begin{eqnarray}
\hat a^\dagger\equiv \sqrt{\frac{m\omega}{2\hbar}} \hat x-i\frac{\hat p}{\sqrt{2m\hbar\omega}}. \label{def-cre}
\end{eqnarray}
\end{subequations}

One of the most important features of quantum mechanics is that
momentum and position do not commute \cite{Rae}, i.e., $[\hat p,\hat x]\equiv
\hat p\hat x-\hat x\hat p=-i\hbar$. From this commutation relation it is straightforward
to show that
\begin{eqnarray}
\hat{\cal H}_\textrm{sho}=\hbar\omega\left(\hat a^\dagger\hat a +
\frac12\right)
\end{eqnarray}
and
\begin{eqnarray}
[\hat a,\hat a^\dagger]\equiv \hat a \hat a^\dagger - \hat
a^\dagger\hat a=1.\label{comm}
\end{eqnarray}
One can also show that $[\hat{\cal H}_\textrm{sho},\hat
a]=[\hbar\omega(\hat a^\dagger\hat a+\frac12),\hat
a]=\hbar\omega[\hat a^\dagger,\hat a]\hat a=-\hbar\omega\hat a$ in a
similar manner. Therefore $[{\cal H}_\textrm{sho},\hat a]
\psi_n=-\hbar\omega\hat a  \psi_n$ and hence
\begin{eqnarray}
\hat{\cal H}_\textrm{sho} \hat a \psi_n = (E_n - \hbar\omega) \hat a \psi_n. \label{newstate}
\end{eqnarray}
Eq. \ref{newstate} tells us that $\hat a \psi_n$ is an eigenstate of $\hat {\cal H}_\textrm{sho}$ with energy $E_n-\hbar\omega$, provided $\hat a \psi_n\ne0$. That is, the operator $\hat a$ moves the system from one eigenstate to another whose energy is lower by $\hbar\omega$, thus $\hat a$ is known as the lowering or destruction operator.

Note that for any wavefunction, $\phi$,
$\langle\phi|\hat p^2|\phi\rangle\geq0$ and
$\langle\phi|\hat x^2|\phi\rangle\geq0$. Therefore, it follows from Eq.
\ref{hamsho} that $E_n\geq0$ for all $n$. Hence, there is a lowest
energy state, or ground state, which we will denote as $\psi_0$.
Therefore there is a limit to how often we can keep lowering the
energy of the state, i.e., $\hat a\psi_0=0$. We can now calculate
the ground state energy of the harmonic oscillator,
\begin{eqnarray}
\hat{\cal H}_\textrm{sho} \psi_0 = \hbar\omega\left(\hat a^\dagger\hat a + \frac12\right) \psi_0=\frac12\hbar\omega.
\end{eqnarray}

In the same way as we derived Eq. \ref{newstate}, one can easily show
that ${\cal H}_\textrm{sho} \hat a^\dagger \psi_n = (E_n +
\hbar\omega) \hat a^\dagger \psi_n$. Therefore $\hat a^\dagger$
moves us up the ladder of states that $\hat a$ moved us down. Hence
$\hat a^\dagger$ is known as a raising or creation operator. Thus we
have
\begin{subequations}
\begin{eqnarray}
\hat a^\dagger \psi_n &=& \sqrt{n+1} \, \psi_{n+1},\\
\hat a \psi_n &=& \sqrt{n} \, \psi_{n-1}
\end{eqnarray}
\end{subequations}
where the terms inside the radicals are required for the correct normalisation of the wavefunctions \cite{Gasiorowicz}. 
Therefore $\psi_n = \frac{1}{\sqrt{n!}} \, (\hat a^\dagger)^n \psi_0$ and
\begin{eqnarray}
E_n=\hbar\omega\left(n+\frac12\right).
\end{eqnarray}

Notice that above we solved the simple harmonic oscillator, i.e.,
calculated the energies of all of the eigenstates, without needing to find
explicit expressions for any of the first quantised eigenfunctions,
$\psi_n$. This general feature of the second quantised approach is extremely advantageous when we are dealing with
the complex many-body wavefunctions typical in condensed matter
physics and chemistry.

\subsection{Second quantisation for light and matter}\label{2lm}

We can extend the second quantisation formalism to light and matter. Let us first consider bosons, which are not subject to the Pauli exclusion principle, e.g., phonons, photons, deuterium nuclei, $^4$He atoms, etc. We define the bosonic `field operator' $\hat b^\dagger(\mathbf r)$ as creating a boson at position $\mathbf r$, similarly, $\hat b(\mathbf r)$ annihilates a boson at position $\mathbf r$. The bosonic field operators obey the commutation relations $[\hat b(\mathbf r), \hat b({\mathbf r}')]=0$,
$[\hat b ^\dagger(\mathbf r), \hat b^\dagger({\mathbf r}')]=0$, and
\begin{eqnarray}
[\hat b(\mathbf r), \hat b^\dagger({\mathbf r}')]=\delta({\mathbf r}-{\mathbf r}').
\end{eqnarray}
This is just the generalisation of Eq. \ref{comm} for the field operators.
We can create any state by acting products, or sums of products, of the $\hat b ^\dagger(\mathbf r)$ on the vacuum state, i.e., the state that does not contain any bosons, which is usually denoted as $|0\rangle$.

Many body wavefunctions for fermions, e.g. electrons, protons,
neutrons, $^3$He atoms, etc., are complicated by the need for the
antisymmetrisation of the wavefunction, i.e., the wavefunction must
change sign under the exchange of any two fermions. Therefore, if we
introduce the fermionic field operators $\hat \psi^\dagger(\mathbf
r)$ and $\hat \psi(\mathbf r)$, which, respectively, create and
annihilate fermions at position $\mathbf r$, we must make sure that any
wavefunction that we can make by acting some set of these operators
on the vacuum state is properly antisymmetrised. This is ensured \cite{JordanWigner} if one insists that the field operators
anti-commute, i.e., if
\begin{subequations}
\begin{eqnarray}
\{\hat \psi(\mathbf r), \hat \psi^\dagger({\mathbf r}')\} &\equiv& \hat \psi(\mathbf r) \hat \psi^\dagger({\mathbf r}') + \hat \psi^\dagger({\mathbf r}') \hat \psi(\mathbf r) =\delta({\mathbf r}-{\mathbf r}'), \notag\\ \\
\{\hat \psi(\mathbf r), \hat \psi({\mathbf r}')\} &=& 0, \\
\{\hat \psi^\dagger(\mathbf r), \hat \psi^\dagger({\mathbf r}')\} &=& 0.
\end{eqnarray}
\end{subequations}
This guarantee of an antisymmetrised wavefunction is one of the most obvious advantages of the second quantisation formalism as it is much easier than having to deal with the Slater determinants that are typically used to ensure the antisymmetrisation of the many-body wavefunction in the first quantised formalism \cite{SchatzRatner}.

For any practical calculation one needs to work with a particular basis set, $\{\phi_i({\mathbf r})\}$. The field operators can be expanded in an arbitrary basis set as
\begin{subequations}
\begin{eqnarray}
\hat\psi(\mathbf r) &=& \sum_i\hat c_i\phi_i(\mathbf r),\\
\hat\psi^\dagger(\mathbf r) &=& \sum_i\hat
c_i^\dagger\phi_i^*(\mathbf r).
\end{eqnarray}
\end{subequations}
Thus $\hat c_i^{(\dagger)}$ annihilates (creates) a fermion in the state $\phi_i(\mathbf r)$. These operators also obey fermionic anticommutation relations,
\begin{subequations}
\begin{eqnarray}
\{\hat c_i, \hat c_j^\dagger\} &=& \delta_{ij}, \\
\{\hat c_i, \hat c_j\} &=& 0, \\
\{\hat c_i^\dagger, \hat c_j^\dagger\} &=& 0.
\end{eqnarray}
\end{subequations}

As fermions obey the Pauli exclusion principle there can be at most one fermion in a given state. We will denote a state in which the $i^\textrm{th}$ basis function contains zero (one) particles by $|0_i\rangle$ ($|1_i\rangle$). Therefore
\begin{eqnarray}
\begin{array}{cc}
\hat c_i|0_i\rangle=0  & \hspace*{1cm}   \hat c_i|1_i\rangle=|0_i\rangle   \\
\hat c_i^\dagger|0_i\rangle=|1_i\rangle  & \hspace*{1cm}   \hat c_i^\dagger|1_i\rangle=0.
\end{array}\label{0s1s}
\end{eqnarray}
It is important to realise that the number 0 is very different from the state $|0_i\rangle$.

Any operator acting on a system of fermions can be expressed in
terms of the $\hat c$ operators. A particularly important example is
the `number operator', $\hat n_{i}\equiv\hat c_i^\dagger\hat c_i$, which simply
counts the number of particles in the state $i$ - as can be
confirmed by explicit calculation from Eqs. \ref{0s1s}.  The total number of particles in the system is therefore simply the
expectation value of the operator $\hat N=\sum_i\hat n_{i}=\sum_i\hat c_i^\dagger\hat
c_i$. Importantly, because we can write any operator in terms of the $\hat c$ operators, we can calculate any observable from the expectation value of some set of $\hat c$ operators. Thus we have access to a complete description of the system from the second quantisation formalism. Further, we can always write the wave function in terms the $\hat c$ operators if an explicit description of the wavefunction is required. For example the sum of Slater determinants, 
\begin{eqnarray}
\Psi({\bf r}_1,{\bf r}_2)=\alpha
\left|
\begin{array}{cc}
\phi_1({\bf r}_1)  & \phi_2({\bf r}_1)   \\
\phi_1({\bf r}_2)  & \phi_2({\bf r}_2)
\end{array}
\right|
+\beta\left|
\begin{array}{cc}
\phi_3({\bf r}_1)  & \phi_4({\bf r}_1)   \\
\phi_3({\bf r}_2)  & \phi_4({\bf r}_2)
\end{array}
\right|,\notag\\
\end{eqnarray}
describes the same state as 
\begin{eqnarray}
|\Psi\rangle=\left(\alpha \hat c_1 \hat c_2
+\beta \hat c_3 \hat c_4\right)|0\rangle,
\end{eqnarray}
where $|0\rangle=|0_1,0_2,0_3,0_4,\dots\rangle$ is the vacuum state, as $\Psi({\bf r}_1,{\bf r}_2)=\langle{\bf r}_1,{\bf r}_2|\Psi\rangle$ (cf. Ref. \onlinecite{Gasiorowicz}).

Often, in order to describe solid state and chemical systems, one needs to describe a set of $N$ electrons whose behaviour is governed by a Hamiltonian of the form
\begin{eqnarray}
{\cal H}=\sum_{n=1}^N\left[ -\frac{\hbar^2 {\mathbf \nabla_n}^2}{2m} +
U({\mathbf r}_n) + \frac12 \sum_{m\ne n}V({\mathbf r}_n-{\mathbf
r}_m) \right], \label{eqn:latticemodelH}
\end{eqnarray}
where $V({\mathbf r}_n-{\mathbf r}_m)$ is the potential describing the interactions between electrons and $U({\mathbf r}_i)$ is an external potential (including interactions with ions or nuclei, which may often be considered to be stationary on the time scales relevant to electronic processes - although we will discuss effects due to the displacement of the nuclei in section \ref{sect:Holstein}). In terms of our second quantisation operators this Hamiltonian may be written as
\begin{eqnarray}
\hat{\cal H}=-\sum_{ij}t_{ij}\hat c_i^\dagger \hat c_j + \frac12\sum_{ijkl}V_{ijkl}\hat c_i^\dagger\hat c_k^\dagger\hat c_l\hat c_j, \label{basicHam}
\end{eqnarray}
where
\begin{eqnarray}
t_{ij}&=& -\int d^3{\mathbf r} \, \phi_i^*({\mathbf r}) \left[ -\frac{\hbar^2 {\mathbf \nabla}^2}{2m} + U({\mathbf r})  \right]  \phi_j({\mathbf r})\\
V_{ijkl}&=& \int d^3{\mathbf r}_1  \int d^3{\mathbf r}_2 \, \phi_i^*({\mathbf r}_1) \phi_j({\mathbf r}_1) \\&&\notag\hspace{2.5cm}\times  V({\mathbf r}_1-{\mathbf r}_2) \phi_k^*({\mathbf r}_2) \phi_l({\mathbf r}_2), \label{eqn:V}
\end{eqnarray}
and the labels $i$, $j$, $k$ and $l$ are taken to define the spin as
well as the basis function. This is exact provided we have an
infinite complete basis. But practical calculations require the use of
finite basis sets and often use incomplete basis sets. The
simplest approach is to just ignore this problem and calculate
$t_{ij}$ and $V_{ijkl}$ directly from the finite basis set. However,
this is often not the best approach. We will delay a detailed
discussion of why this is and of the deep philosophical issues raised
by this until section \ref{eh_v_se}. We also delay discussion of
how to calculate these parameters until section \ref{eh_v_se}. Until
then  we will simply
assume that $t_{ij}$, $V_{ijkl}$ and other similar parameters
required are known and instead focus on how to perform practical
calculations using models of the form of Eq.
\ref{basicHam} and closely related Hamiltonians.

In what follows we will assume that the states created by the $\hat c^\dagger_{i}$ operators form an orthonormal basis. This greatly simplifies the mathematics, but differs from the approach usually taken in introductory chemistry textbooks as most quantum chemical calculations are performed in non-orthogonal bases for reasons of computational expedience.

\section{The H\"uckel or tight-binding model}\label{sect:Huckel}

The simplest model with the form of Eq. \ref{basicHam} is  usual
called the H\"uckel model in the context of molecular systems \cite{Lowe} and
the tight-binding model in the context of crystals \cite{A&M}. In these models
one makes the approximation that $V_{ijkl}=0$ for all $i$, $j$, $k$,
and $l$. Therefore these models explicitly neglect the interactions between electrons. Both models are identical, but slightly different notation
is standard in the different contexts.   We assume that our basis set consists of orbitals
centred on particular sites, as we will in all of the models considered in
these notes. These sites might be, for example, atoms in a molecule or solid, chemical groups in a molecule, p-d hybrid states in a transition metal oxide, entire molecules in a molecular crystal, or even larger strucutures. Clearly the simplest problem has only one orbital per spin state on each site,  in which case,
\begin{eqnarray}
\hat{\cal H}_\textrm{tb} =-\sum_{ij\sigma} t_{ij} \hat c_{i\sigma}^\dagger  \hat c_{j\sigma} ,
\end{eqnarray}
where  $\hat c^{(\dagger)}_{i\sigma}$ annihilates (creates) an electron with spin $\sigma$ in an orbital centred on site $i$.

\subsection{Molecules (the H\"uckel model)}

%In molecular systems one usually speaks of the H\"uckel model rather than the tight binding model, and a slightly different notation is usually used.
The standard notation in this context is $t_{ii}=-\alpha_i$, $t_{ij}=-\beta_{ij}$ if site $i$ and site $j$ are connected by a chemical bond; one assumes that $t_{ij}=0$ otherwise. Note that the subscripts on $\alpha$ and $\beta$ are also often dropped, but they are usually implicit; if the molecule contains more than one species of atom the $\alpha$s will clearly be different on the different species and the $\beta$s will depend on the species of each of the atoms that the electron is hopping between. Therefore,
\begin{eqnarray}
\hat{\cal H}_\textrm{H\"uckel} = \sum_{i\sigma} \alpha_i \hat c_{i\sigma}^\dagger  \hat c_{i\sigma} + \sum_{\langle ij\rangle\sigma} \beta_{ij} \hat c_{i\sigma}^\dagger  \hat c_{j\sigma},
\end{eqnarray}
where $\langle ij\rangle$ serves to remind us that the sum is only over those pairs of atoms joined by a chemical bond.
Note that $\beta_{ij}$ is typically negative.

\subsubsection{Molecular hydrogen}\label{section:H2Huck}

\begin{figure}
\includegraphics[width=8cm]{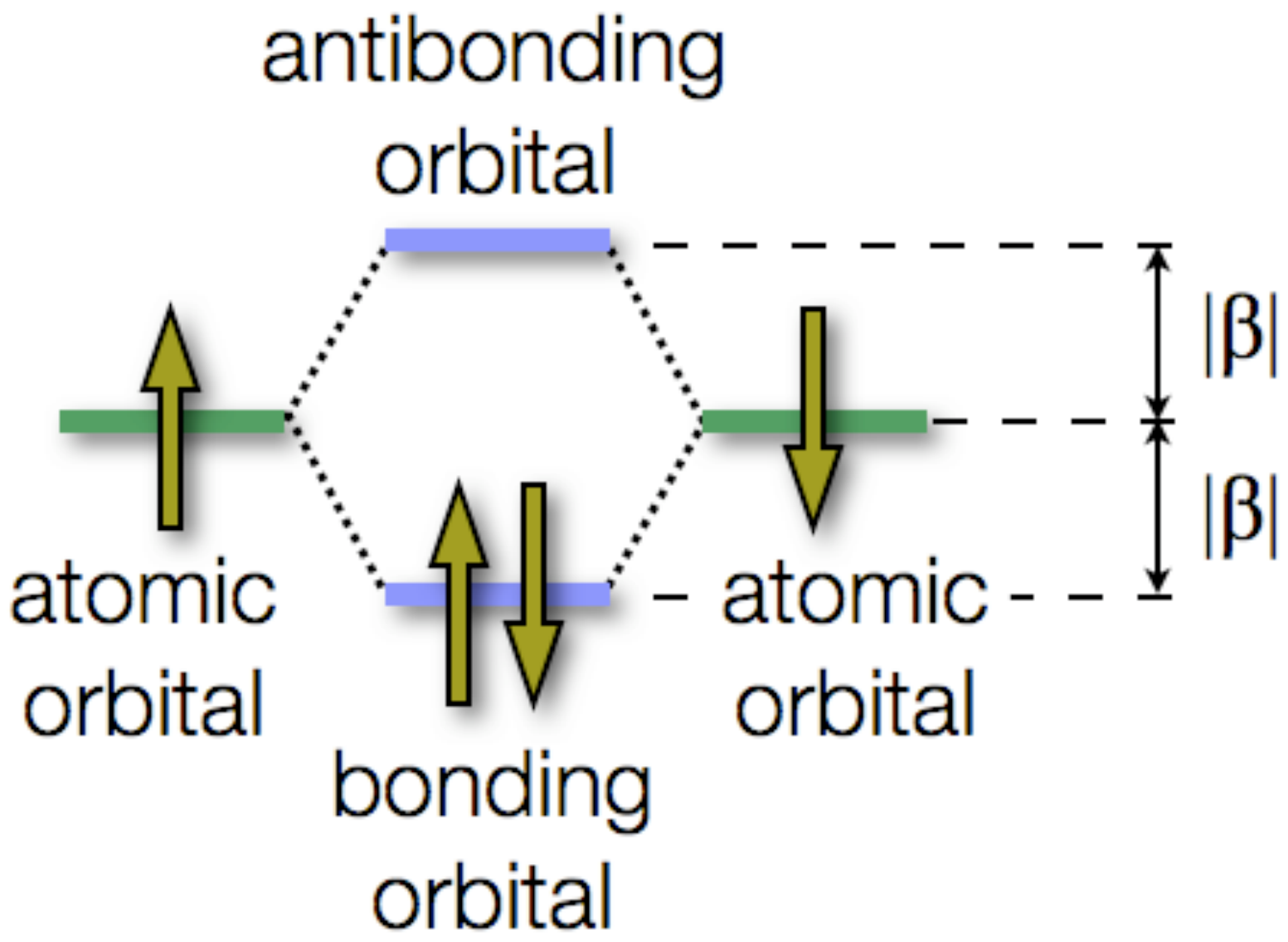}
\caption{The energy levels of the atomic and molecular orbitals in the H\"uckel description of H$_2$. The bonding orbital is $|\beta|$ lower in energy than the atomic orbital, whereas the antibonding orbital  is $|\beta|$ higher in energy than the atomic orbital. Therefore, neutral H$_2$ is stabilised by 2$|\beta|$ relative to 2H.} \label{H2Huckel}
\end{figure}

Clearly, in H$_2$ there is only a single atomic species. In this
case one can set $\alpha_i=\alpha$ for all $i$ without loss of generality. Further, as there is also only a single bond, we may also choose
$\beta_{ij}=\beta$ giving
\begin{eqnarray}
\hat{\cal H}_\textrm{H\"uckel} = \alpha\sum_\sigma(\hat n_{1\sigma}+\hat n_{2\sigma})+\beta\sum_\sigma \Big( \hat c_{1\sigma}^\dagger  \hat c_{2\sigma} + \hat c_{2\sigma}^\dagger  \hat c_{1\sigma}\Big),\notag\\
\end{eqnarray}
where we have labelled the two atomic sites 1 and 2. This Hamiltonian has two eigenstates: one is known as the bonding state,
\begin{eqnarray}
|\psi_{b\sigma}\rangle=\frac1{\sqrt 2}(\hat c_{1\sigma}^\dagger+\hat c_{2\sigma}^\dagger)|0\rangle,
\end{eqnarray}
and the other is known as antibonding state,
\begin{eqnarray}
|\psi_{a\sigma}\rangle=\frac1{\sqrt 2}(\hat c_{1\sigma}^\dagger-\hat c_{2\sigma}^\dagger)|0\rangle.
\end{eqnarray}
The bonding state has energy $\alpha+\beta$, whereas the antibonding state has energy $\alpha-\beta$, recall that $\beta<0$. Therefore every electron in the bonding state stabilises the  molecule by an amount $|\beta|$, whereas electrons in the antibonding state destabilise the  molecule by an amount $|\beta|$, hence the nomenclature.\footnote{Note that in a non-orthogonal basis the antibonding orbital may be destabilised by a greater amount than the bonding orbital is stabilised.} This is sketched in Fig. \ref{H2Huckel}.

Because $V_{ijkl}=0$ the  electrons are non-interacting and so the
molecular orbitals are not dependent on the occupation of other orbitals.
Therefore to calculate the total energy of the ground state of the
molecule one simply fills up the states starting with the lowest
energy states and respecting the Pauli exclusion principle. If the two protons are infinitely separated $\beta=0$ and the system has total energy $N\alpha$, where $N$ is the total number of electrons. H$_2^+$
has only one electron, which, in the ground state, will occupy the
bonding orbital, and so H$_2^+$ has a binding energy of $\beta$.
H$_2$ has two electrons; in the ground state these electrons have
opposite spin and therefore can both occupy the bonding orbital.
Thus H$_2$ has a binding energy of $2\beta$. H$_2^-$ has three
electrons, so while two can occupy the bonding state one must be in
the antibonding state, therefore the binding energy is only $\beta$.
Finally, H$_2^{2-}$ has four electrons so one finds two in the each
molecular orbital. Therefore the bonding energy is zero: the
molecule is predicted to be unstable.

Thus the H\"uckel model makes several predictions: neutral H$_2$ is predicted to be significantly more stable than any of the ionic states; the two singly ionic species are predicted to be equally stable; the doubly cationic species is predicted to be unstable. Further, the lowest optical absorption is expected to correspond to the transition between the bonding orbital and the antibonding orbital. The energy gap for this transition is $2|\beta|$. Therefore, the lowest optical absorption is predicted to be the same in the neutral species and the singly cationic species. Further, this absorption is predicted to occur at a frequency with the same energy as the heat of formation for the neutral species.  While these predictions do capture qualitatively what is observed experimentally, they are certainly not within chemical accuracy (i.e. within $k_BT\sim1$ kcal mol$^{-1}\sim0.03$ eV for $T=300$ K). For example the experimentally determined binding energies \cite{Lowe} are 2.27 eV for H$_2^+$, 4.74 eV for H$_2$, 1.7 eV for H$_2^-$, while $H_2^{2-}$ is indeed unstable.

\subsubsection{$\pi$-H\"uckel theory of benzene}

\begin{figure}
\includegraphics[width=8cm]{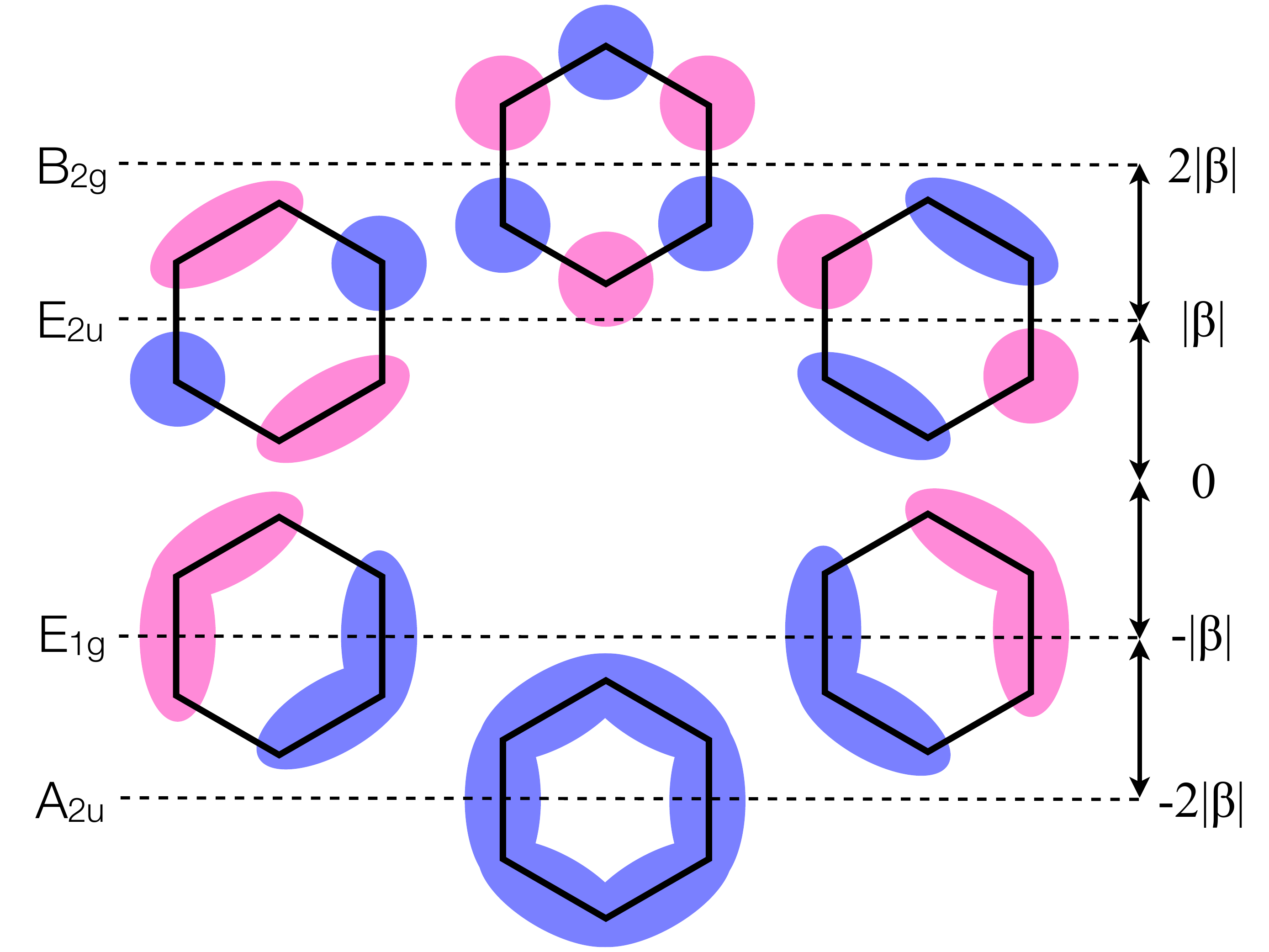}
\caption{The molecular orbitals for benzene from $\pi$-H\"uckel theory. Different colours indicate a change in sign of the wavefunction. In the neutral molecule the A$_{2u}$ and both E$_{1g}$ states are occupied, while the B$_{2g}$ and E$_{2u}$ states are virtual. Note that we have taken real superpositions \cite{Lowe} of the two-fold degenerate states in order to facilitate these plots.} \label{benzene}
\end{figure}

For many organic molecules a model known as $\pi$-H\"uckel theory is very useful. In $\pi$-H\"uckel theory one considers only the $\pi$ electrons. A simple example is a benzene molecule. The hydrogen atoms have no $\pi$ electrons and are therefore not represented in the model. This leaves only the carbon atoms, so again we can set $\alpha_i=\alpha$ and $\beta_{ij}=\beta$. Because of the ring geometry of benzene (and assuming that the molecule is planar) the Hamiltonian becomes
\begin{eqnarray}
\hat{\cal H}_\textrm{H\"uckel} = \alpha\sum_{i\sigma}\hat n_{i\sigma}+\beta\sum_{i\sigma} \Big( \hat c_{i\sigma}^\dagger  \hat c_{i+1\sigma} + \hat c_{i+1\sigma}^\dagger  \hat c_{i\sigma}\Big),
\end{eqnarray}
where the addition in the site index is defined modulo six, i.e., site number seven \emph{is} site number one. For benzene we have six solutions per spin state, which are 
\begin{subequations}
\begin{eqnarray}
&&|\psi_{A_{2u}}\rangle=\frac1{\sqrt{6}}\left(\hat c^\dagger_{1\sigma} + \hat c^\dagger_{2 \sigma} + \hat c^\dagger_{3 \sigma} + \hat c^\dagger_{4 \sigma} + \hat c^\dagger_{5 \sigma} + \hat c^\dagger_{6 \sigma} \right) |0\rangle,\notag\\
&&|\psi_{E_{1g}} \rangle
\notag\\&&\hspace{10pt}=\frac1{\sqrt{6}}\left(\hat c^\dagger_{1 \sigma} + \epsilon\hat c^\dagger_{2 \sigma} + \epsilon^2\hat c^\dagger_{3 \sigma} - \hat c^\dagger_{4 \sigma} - \epsilon\hat c^\dagger_{5 \sigma} - \epsilon^2\hat c^\dagger_{6 \sigma} \right) |0\rangle, \notag\\
&&|\psi_{E_{1g}}^\prime \rangle
\notag\\&&\hspace{10pt}=\frac1{\sqrt{6}}\left(\hat c^\dagger_{1 \sigma} -\epsilon^2 \hat c^\dagger_{2 \sigma} -\epsilon \hat c^\dagger_{3 \sigma} -\epsilon^2 \hat c^\dagger_{4 \sigma} +\epsilon \hat c^\dagger_{5 \sigma} + \hat c^\dagger_{6 \sigma} \right) |0\rangle, \notag\\
&&|\psi_{E_{2u}} \rangle
\notag\\&&\hspace{10pt}=\frac1{\sqrt{6}}\left(\hat c^\dagger_{1 \sigma} + \epsilon^2\hat c^\dagger_{2 \sigma} -\epsilon \hat c^\dagger_{3 \sigma} + \hat c^\dagger_{4 \sigma} +\epsilon^2 \hat c^\dagger_{5 \sigma} - \epsilon\hat c^\dagger_{6 \sigma} \right) |0\rangle, \notag\\
&&|\psi_{E_{2u}}^\prime \rangle
\notag\\&&\hspace{10pt}=\frac1{\sqrt{6}}\left(\hat c^\dagger_{1 \sigma} -\epsilon \hat c^\dagger_{2 \sigma} + \epsilon^2\hat c^\dagger_{3 \sigma} + \hat c^\dagger_{4 \sigma} - \epsilon\hat c^\dagger_{5 \sigma} + \epsilon^2\hat c^\dagger_{6 \sigma} \right) |0\rangle \notag
\end{eqnarray}
and
\begin{eqnarray}
|\psi_{B_{2g}} \rangle=\frac1{\sqrt{6}}\left(\hat c^\dagger_{1 \sigma} - \hat c^\dagger_{2 \sigma} + \hat c^\dagger_{3 \sigma} - \hat c^\dagger_{4 \sigma} + \hat c^\dagger_{5 \sigma} - \hat c^\dagger_{6 \sigma} \right) |0\rangle, \notag
\end{eqnarray}
\end{subequations}
where $\epsilon=e^{i\pi/3}$. These wavefunctions are sketched in Fig. \ref{benzene}. The energies of these states are $E_{A_{2u}}=\alpha-2|\beta|$, $E_{E_{1g}}=E_{E_{1g}}^\prime= \alpha-|\beta|$, $E_{E_{2u}}=E_{E_{2u}}^\prime= \alpha +|\beta|$ and $E_{B_{2g}}= \alpha +2|\beta|$. The subscripts are symmetry labels \cite{Tinkham,Lax} for the group $D_{6h}$ and  one should recall that, because we are dealing with $\pi$-orbitals, all of the orbitals sketched here are antisymmetric under reflection through the plane of the page. 
 The degenerate (E$_\textrm{1g}$ and E$_\textrm{2u}$) orbitals are typically written/drawn rather differently (cf. Ref. \onlinecite{Lowe}). However, any linear combination of degenerate eigenstates is also an eigenstate; this representation was chosen as it highlights the symmetry of the problem. For a more detailed discussion of this problem see Ref. \onlinecite{Coulson}.

%predictions vs expt

\subsubsection{Electronic interactions and the parameterisation of the H\"uckel model}\label{Huckel-params}

As noted above the H\"uckel model does not explicitly include interactions between electrons. This leads to serious qualitative and quantitative failures of the model, some of which we have seen above and which will discuss further below. However, given the (mathematical and conceptual) simplicity and the computational economy of the method one would like to improve the method as far as possible. So far we have treated the theory as parameter free. However, if we treat the model as a semi-empirical method instead one can include some of the effects due to electron-electron interactions without greatly increasing the computational cost of the method. For example, one can make $\alpha$ dependent on the charge on the atom. This is reasonable, as the more electrons we put on an atom the harder it is to add another due to the additional Coulomb repulsion from the extra electrons. The simplest way to account for this is the `$\omega$ technique' \cite{Lowe} where one replaces 
\begin{eqnarray}
\alpha_i\rightarrow\alpha_i'=\alpha_i+\omega(q_0-q_i)\beta, \label{omega-method}
\end{eqnarray}
where $q_i$ is the charge on atom $i$, $q_0$ is a (fixed) reference charge and $\omega$ is a parameter. The $\omega$ technique  suppresses the unphysical fluctuations of the electron density, which are often predicted by the H\"uckel model (cf. the discussion of H$_2$ above). Similar techniques can also be applied to $\beta$. These parameterisations only slightly complicate the model and do not lead to a major inflation of the computational cost, but can significantly improve the accuracy  of the predictions of the H\"uckel  model \cite{Brogli}.

\subsection{Crystals (the tight binding model)}\label{sect:tb}

For infinite systems it is necessary
to work with a fixed chemical potential rather than a fixed particle
number. Therefore before we discuss the tight binding model we will briefly review the chemical potential (also see Ref.  \onlinecite{Atkins} for a discussion of the chemical potential in a chemical context).

\subsubsection{The chemical potential}

When one is dealing with a large system keeping track of the number of particles can become difficult. This is particularly true in the thermodynamic limit where the number of electrons $N_e\equiv\langle\hat N\rangle\rightarrow\infty$ and the volume of the system $V\rightarrow\infty$ in such a way so as to ensure that the electronic density, $n_e=N_e/V$, remains constant. Lagrange multipliers \cite{Arfkin} are a powerful and general method for imposing constraints on differential equations (such as the Schr\"odinger equation) without requiring the solution of integro-differential equations. Briefly, consider a function, $f(x,y,z,\dots)$ that we wish to extremise (minimise or maximise) subject to a constraint that means that $x$, $y$, $z\dots$ are no longer independent. In general we may write the constraint in the form $\phi(x,y,z,\dots)=0$. This allows us to define the function $g(x,y,z\dots,\lambda)\equiv f(x,y,z\dots)+\lambda\phi(x,y,z\dots)$, where $\lambda$ is known as a Lagrange multiplier. One may show \cite{Arfkin} that the extremum of $g(x,y,z\dots,\lambda)$ with respect to $x$, $y$, $z\dots$ and $\lambda$ is the extremum of $f(x,y,z\dots)$ with respect to $x$, $y$, $z\dots$ subject to the constraint that $\phi(x,y,z,\dots)=0$.

Typically the problem we wish to solve in chemistry and condensed matter physics is to minimise the free energy, $F$, (which reduces to the energy, $E$, at $T=0$) subject to the constraint of having a fixed number of electrons (determined by the chemistry of the material in question).
This suggests that one should simply introduce a Lagrange multiplier to resolve the difficulty of constraining the number of electrons in the thermodynamic limit. A suitable constraint could be introduced by adding the term $\lambda(N_0-\hat N)$ to the Hamiltonian, where $N_0$ is the chemically required number of electrons, and requiring the the free energy is an extremum with respect to $\lambda$. However, one can also impose the same constraint and achieve additional physical insight  by subtracting the term $\mu\hat N$ from the Hamiltonian and requiring that 
\begin{eqnarray}
N_0=-\frac{\partial F}{\partial \mu}.
\end{eqnarray}
The chemical potential (for electrons), $\mu$, is then given by 
\begin{eqnarray}
\mu=-\frac{\partial F}{\partial N_e}.\label{chem}
\end{eqnarray}
Therefore, specifying a system's chemical potential is equivalent to specifying the number of electrons, but provides a far more powerful approach for bulk systems. 

Physically this approach is equivalent to thinking of the system as being attached to to an infinite bath of electrons, i.e., one is working in the grand canonical ensemble \cite{Mandl}. Thus, the Fermi distribution for the system is given by
\begin{eqnarray}
f(E,T)=\frac1{1+e^{\frac{(E-\mu)}{k_BT}}} \label{Fermi}
\end{eqnarray}
Therefore at $T=0$ all of the states with energies lower than the chemical potential are occupied, and all of the states with energies greater than the chemical potential are unoccupied. Therefore, the Fermi energy, $E_F=\mu(T=0)$. Note that as $F$ is temperature dependent Eq. \ref{chem} shows that, in general, $\mu$ will also be temperature dependent.\footnote{In contrast, as $E_F$ is only defined at $T=0$ it is not temperature dependent.} Nevertheless Eq. \ref{Fermi} gives a clear interpretation of the chemical potential at any temperature: $\mu(T)$ is the energy of a state with a 50\% probability of occupation at temperature $T$.

\subsubsection{The tight binding model}

For periodic systems (crystals) one usually refers to the H\"uckel model as
the tight binding model. Often one considers models with only
`nearest neighbour' terms, that is one takes $t_{ii}=-\epsilon_i$,
$t_{ij}=t$ if $i$ and $j$ are at nearest neighbour sites, and
$t_{ij}=0$ otherwise.  Thus, for nearest neighbour hopping only,
\begin{eqnarray}
\hat{\cal H}_\textrm{tb} - \mu\hat N=-t\sum_{\langle ij\rangle\sigma}  \hat c_{i\sigma}^\dagger  \hat c_{j\sigma} +\sum_{i\sigma} (\epsilon_i - \mu) \hat c_{i\sigma}^\dagger  \hat c_{i\sigma},
\end{eqnarray}
where $\mu$ is the chemical potential and $\langle ij\rangle$ indicates that the sum is over nearest neighbours only.  Further, if we consider materials with only a single atomic species we can set $\epsilon_i=0$ yielding
\begin{eqnarray}
\hat{\cal H}_\textrm{tb} - \mu\hat N=-t\sum_{\langle ij\rangle\sigma}  \hat c_{i\sigma}^\dagger  \hat c_{j\sigma} -\mu\sum_{i\sigma}  \hat c_{i\sigma}^\dagger  \hat c_{i\sigma}.
\end{eqnarray}

\subsubsection{The one dimensional chain}

\begin{figure*}
\includegraphics[width=7.75cm]{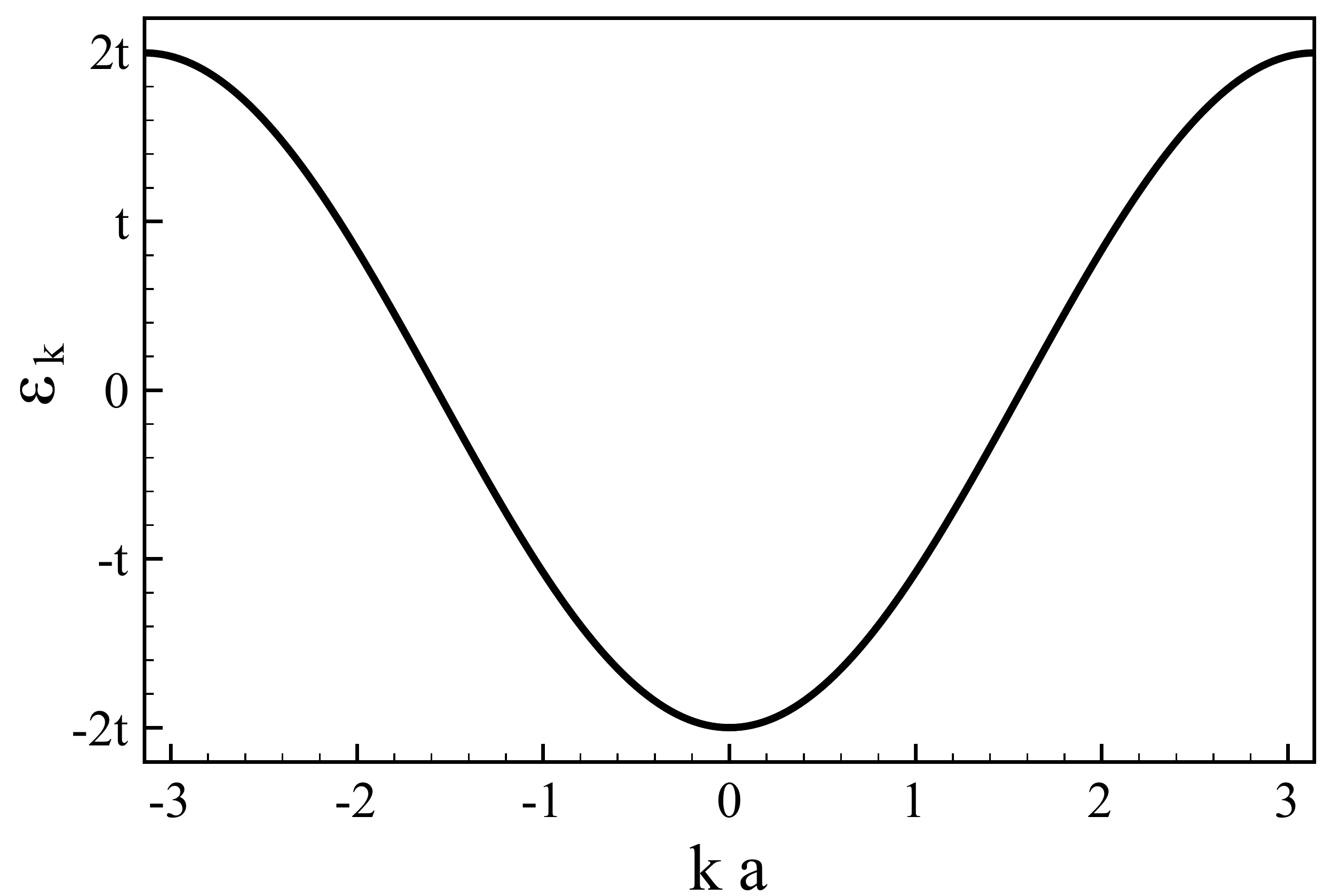}\hspace{0.5cm}
\includegraphics[width=7.75cm]{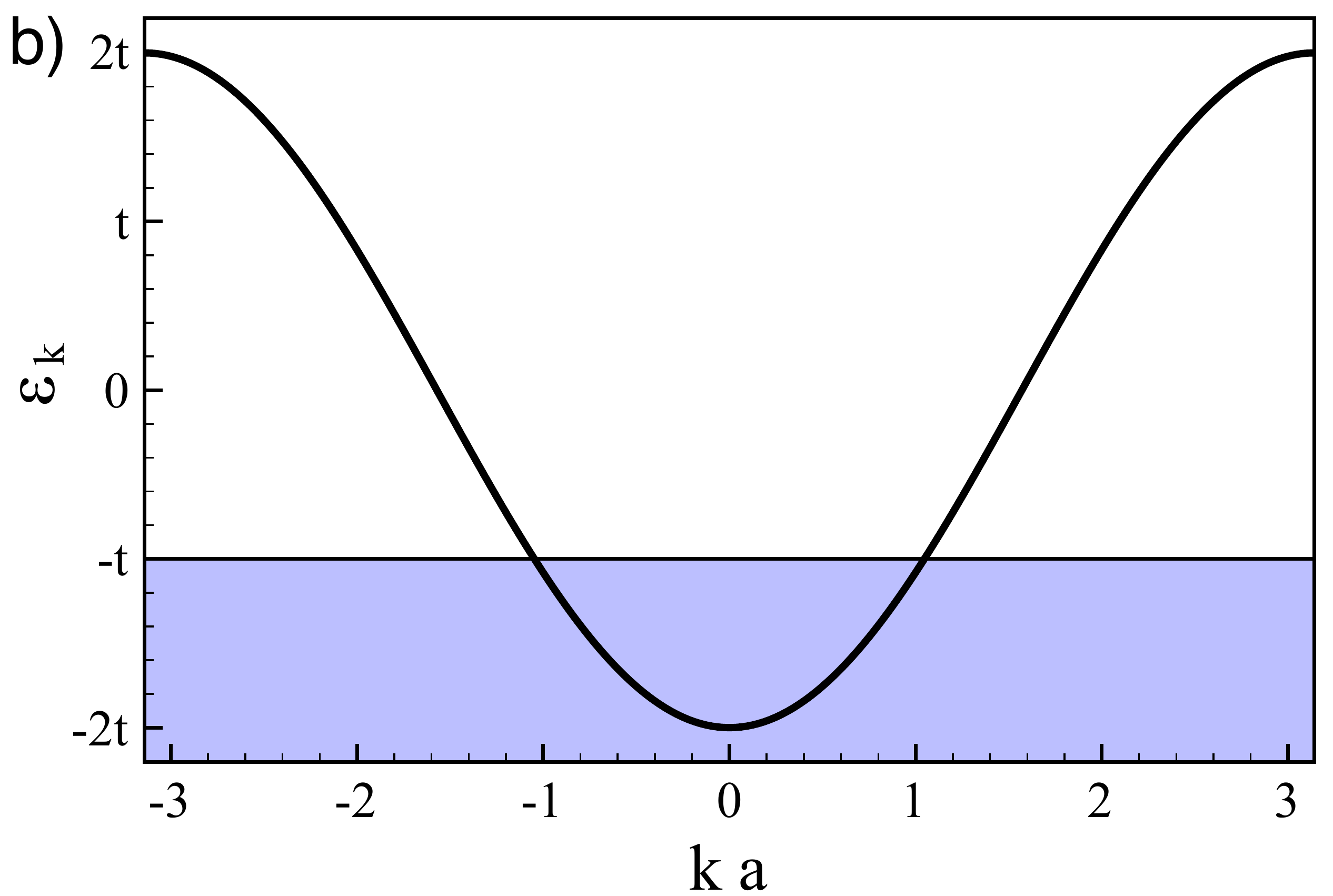}
\includegraphics[width=7.75cm]{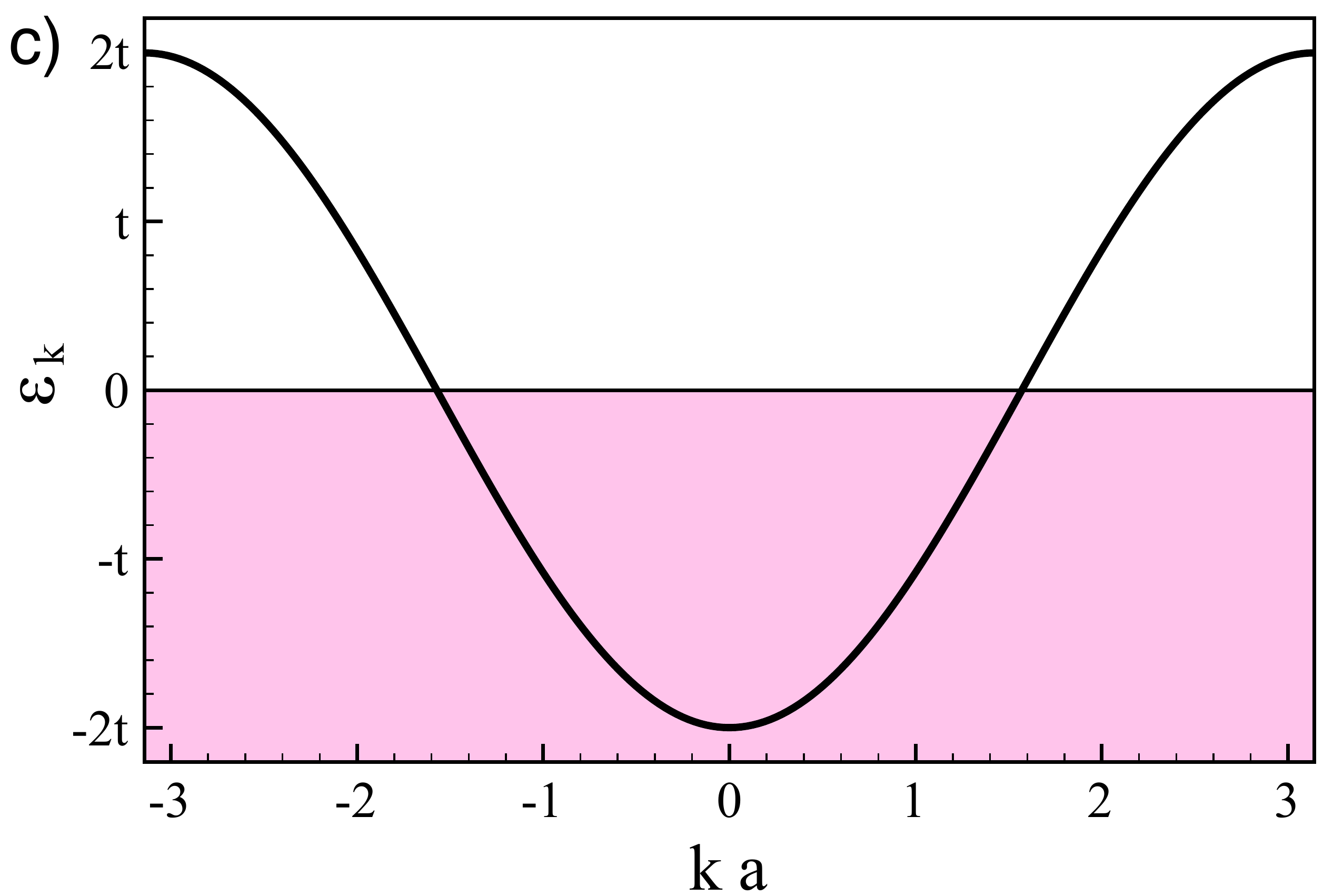}\hspace{0.5cm}
\includegraphics[width=7.75cm]{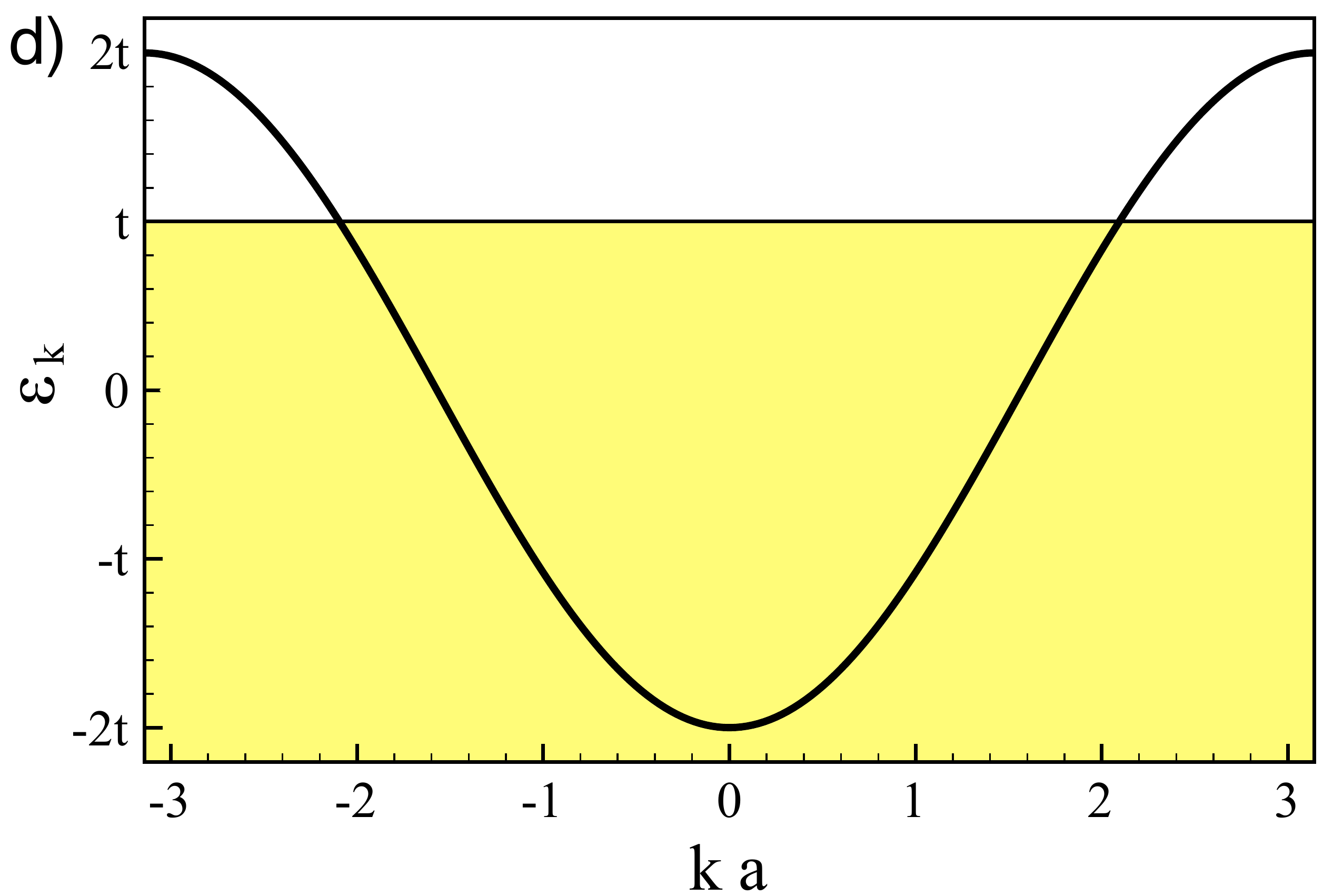}
\caption{(a) The dispersion relation, $\varepsilon_k=-2t\cos(ka)$, of the one dimensional tight binding chain with nearest neighbour hopping only. (b) Shaded area shows the filled states for $\mu=-t$. (c) Shaded area shows the filled states for $\mu=0$. (d) Shaded area shows the filled states for $\mu=t$.} \label{1d-dispersion}
\end{figure*}

The simplest infinite system is a  chain with nearest neighbour
hopping only. As we are on a chain the sites have a natural ordering
and the Hamiltonian may be written as
\begin{equation}
\hat{\cal H}_\textrm{tb} - \mu\hat N=-t\sum_{i\sigma} \Big( \hat c_{i\sigma}^\dagger  \hat c_{i+1\sigma} +\hat c_{i+1\sigma}^\dagger  \hat c_{i\sigma} \Big)-\mu\sum_{i\sigma}  \hat c_{i\sigma}^\dagger  \hat c_{i\sigma},
\end{equation}

We can solve this model exactly by performing a lattice Fourier transform. We begin by introducing the reciprocal space creation and annihilation operators:
\begin{subequations}
\begin{eqnarray}
\hat c_{i\sigma} = \frac{1}{\sqrt N} \sum_k \hat c_{k\sigma} e^{ikR_i},\\
\hat c_{i\sigma}^\dagger = \frac{1}{\sqrt N} \sum_k \hat c_{k\sigma}^\dagger e^{-ikR_i},
\end{eqnarray}
\end{subequations}
where $k$ is the lattice wavenumber or crystal momentum and $R_i$ is the position of the $i^\textrm{th}$ lattice site. Therefore,
\begin{eqnarray}
\hat{\cal H}_\textrm{tb} - \mu\hat N&=&
\frac{1}{N}\sum_{ikk'\sigma}  \hat c_{k\sigma}^\dagger \hat c_{k'\sigma}  e^{i(k'-k)R_i} \notag\\&&\hspace{1.2cm}\times\Big[   -t ( e^{ik'a}
 +     e^{-ika} ) - \mu \Big],
\end{eqnarray}
where $a$ is the lattice constant, i.e., the distance between neighbouring sites $R_i$
and $R_{i+1}$. $\frac1N\sum_ie^{i(k'-k)R_i}=\delta(k'-k)$ \cite{Arfkin799};
therefore
\begin{eqnarray}
\hat{\cal H}_\textrm{tb} - \mu\hat N&=&
\sum_{k\sigma}  \Big[   -2t\cos(ka) \,  \hat c_{k\sigma}^\dagger \hat c_{k\sigma}   - \mu   \hat c_{k\sigma}^\dagger \hat c_{k\sigma} \Big]\notag\\
&=&
\sum_{k\sigma} (  \varepsilon_k -\mu) \hat c_{k\sigma}^\dagger \hat c_{k\sigma}, \label{chain-soln}
\end{eqnarray}
where $\varepsilon_k=-2t\cos(ka)$ is known as the dispersion relation. Notice that Eq. \ref{chain-soln} is diagonal, i.e., it only depends on the number operator terms, $n_{k\sigma}=\hat c_{k\sigma}^\dagger \hat c_{k\sigma}$. Therefore the energy is just the sum of $\varepsilon_k$ for the states $k\sigma$ that are occupied, and we have solved the problem. We plot the dispersion relation in Fig. \ref{1d-dispersion}a. For a tight binding model calculating the dispersion relation is equivalent to solving the problem.

The chemical potential, $\mu$, must be chosen to ensure that there are the physically required number of electrons. Changing the chemical potential has the effect of moving the Fermi energy up or down the band and hence changing the number of electrons in the system. For example (see Fig. \ref{1d-dispersion}b-d), in the above problem the half filled band corresponds to $\mu=0$; the quarter filled band corresponds to $\mu=-t$; and the three quarters filled band corresponds to $\mu=t$.

\subsubsection{The square, cubic and hypercubic lattices}

In more than one dimension the notation becomes slightly more
complicated, but the mathematics does not, necessarily, become any
more difficult. The simplest generalisation of the chain we have
solved above is the two dimensional square lattice where
\begin{eqnarray}
\hat{\cal H}_\textrm{tb} - \mu\hat N=-t\sum_{\langle ij\rangle\sigma} \hat c_{i\sigma}^\dagger  \hat c_{j\sigma} -\mu\sum_{i\sigma}  \hat c_{i\sigma}^\dagger  \hat c_{i\sigma}. \label{hamtb}
\end{eqnarray}
Recall that  $\langle ij\rangle$ indicates that the sum is over nearest neighbours only.
To solve this problem we simply generalise our reciprocal lattice operators to
\begin{subequations}
\begin{eqnarray}
\hat c_{i\sigma} = \frac{1}{\sqrt N} \sum_\mathbf{k} \hat c_{\mathbf{k}\sigma} e^{i\mathbf{k}\cdot\mathbf{R}_i},\\
\hat c_{i\sigma}^\dagger = \frac{1}{\sqrt N} \sum_\mathbf{k} \hat c_{\mathbf{k}\sigma}^\dagger e^{-i\mathbf{k}\cdot\mathbf{R}_i},
\end{eqnarray}
\end{subequations}
where $\mathbf{k}=(k_x,k_y)$ is the lattice wavevector or crystal momentum and $\mathbf{R}_i=(x_i,y_i)$ is the position of the $i^\textrm{th}$ lattice site. We then simply repeat the process we used to solve the one dimensional chain. As the lattice only contains bonds in perpendicular directions the calculations for the $x$ and $y$ directions go through independently and one finds that
\begin{eqnarray}
\hat{\cal H}_\textrm{tb} - \mu\hat N=\sum_{\mathbf{k}\sigma} (  \varepsilon_\mathbf{k} -\mu) \hat c_{\mathbf{k}\sigma}^\dagger \hat c_{\mathbf{k}\sigma},  \label{2dsoln}
\end{eqnarray}
where the dispersion relation is now $\varepsilon_ \mathbf{k} = -2t[\cos(k_xa_x)+\cos(k_ya_y)]$ and $a_\nu$ is the lattice constants in the $\nu$ direction.

A three dimensional cubic lattice is not any more difficult. In this
case $\mathbf{k}=(k_x,k_y,k_z)$ and the solution is of the form of
Eq. \ref{2dsoln} but with $\varepsilon_ \mathbf{k} =
-2t[\cos(k_xa_x)+\cos(k_ya_y)+\cos(k_za_z)]$. Indeed so long as we
keep all the bonds mutually perpendicular one can keep generalising
this solution to higher dimensions. This may sound somewhat academic
as no materials live in more than three dimensions, but the infinite
dimensional hypercubic lattice has become important in recent years
because many models that include interactions can be solved exactly
in infinite dimensions as we will discuss in section \ref{DMFT}.

\subsubsection{The hexagonal and honeycomb lattices}

\begin{figure*}
\includegraphics[width=12cm]{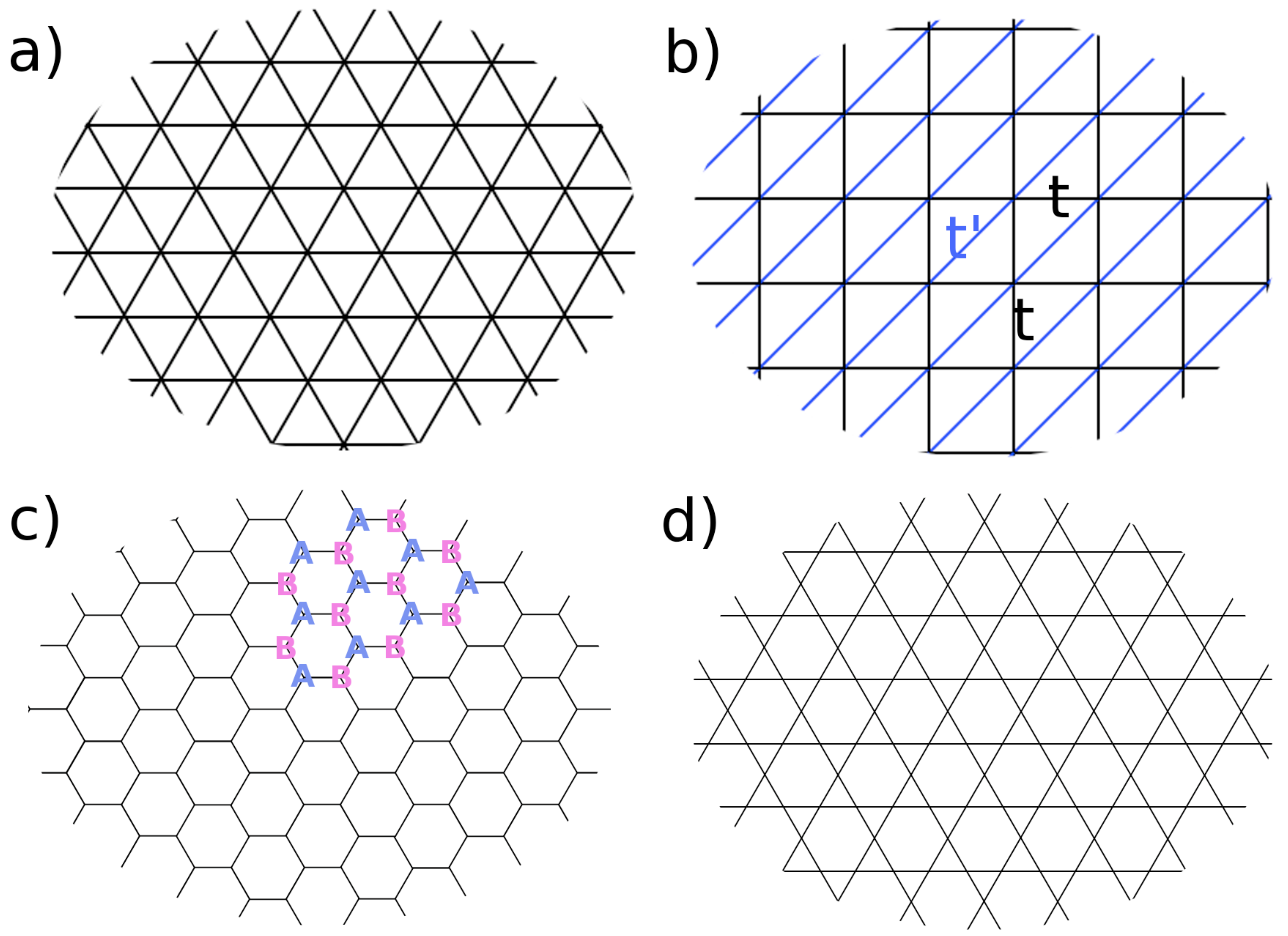}
\caption{The (a) hexagonal (triangular), (b) anisotropic triangular, (c) honeycomb and (d) kagome lattices. The hexagonal lattice contains two inequivalent types of lattice site, some of which are labelled A and B. The sets of equivalent sites are referred to as sublattices.} \label{lattices}
\end{figure*}

Even if the bonds are not all mutually perpendicular the solution to the tight-binding model can still be found by Fourier transforming the Hamiltonian. Three important examples of such lattices are the hexagonal lattice (which is often referred to as the triangular lattice, although this is formally incorrect), the anisotropic triangular lattice, and the honeycomb lattice, which are sketched in Fig. \ref{lattices}. For each lattice the solution is of the form of  Eq. \ref{2dsoln}. For the hexagonal lattice
\begin{equation}
 \varepsilon_\mathbf{k} = -2t\cos(k_xa_x)-4t\cos\left(\frac{\sqrt 3}2k_ya_y\right)\cos\left(\frac{k_xa_x}2\right).
 \end{equation}
 For the anisotropic triangular lattice
\begin{equation}
 \varepsilon_\mathbf{k} = -2t\Big[\cos(k_xa_x)+\cos(k_ya_y)\Big]-2t'\cos(k_xa_x+k_ya_y).
 \end{equation}

\begin{figure}
\includegraphics[width=8cm]{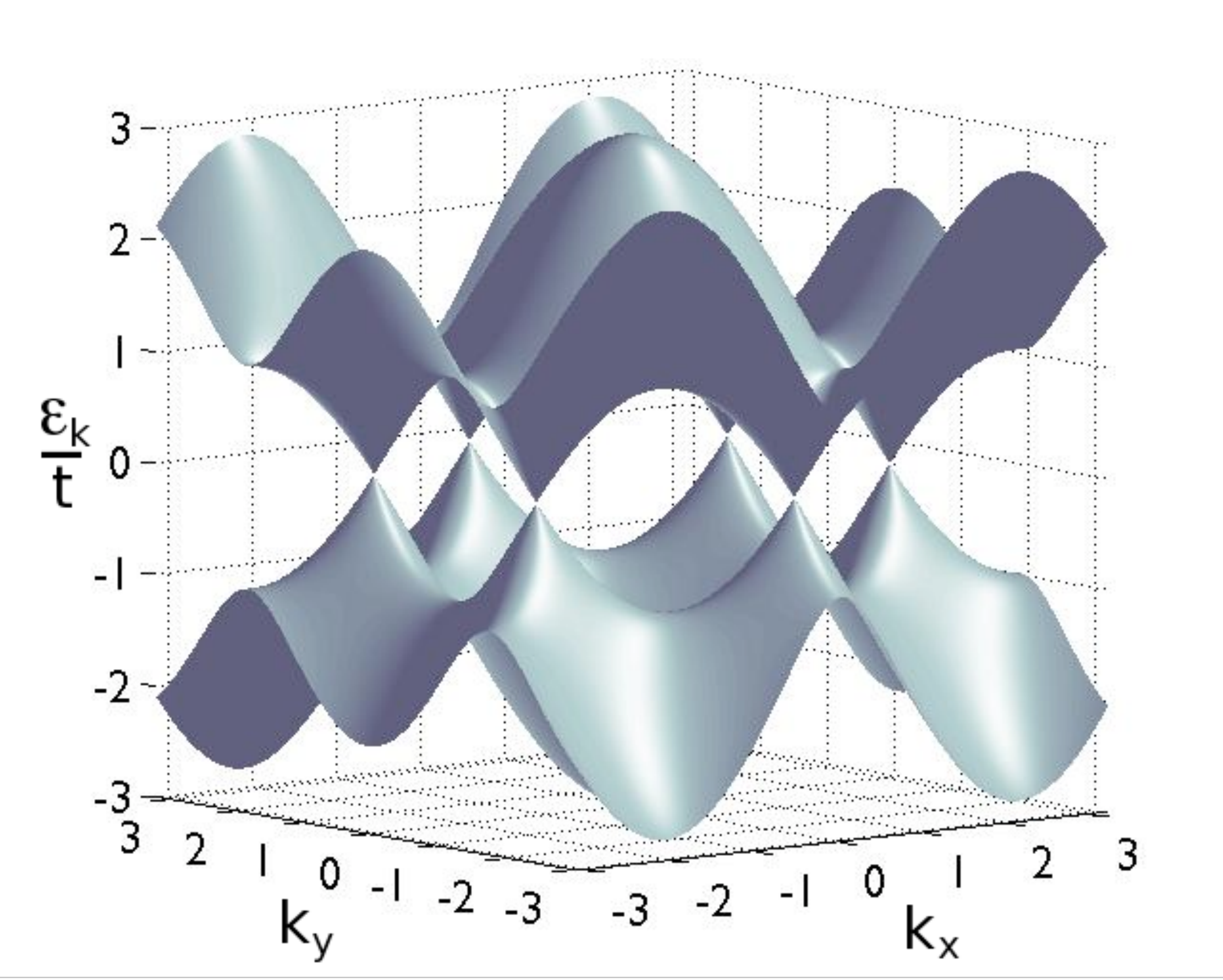}
\caption{The Dirac dispersion of the honeycomb lattice} \label{fig:Dirac}
\end{figure}

The honeycomb lattice has an important additional subtlety, that
there are two inequivalent types of lattice site (cf. Fig. \ref{lattices}c), which
it is worthwhile to work through. We begin by introducing new
operators, $\hat c_{i\nu\sigma}$, which annihilate an electron with
spin $\sigma$ on the $\nu^\textrm{th}$ sublattice in the
$i^\textrm{th}$ unit cell, where $\nu=A$ or $B$. Therefore we can rewrite
Eq. \ref{hamtb} as
\begin{widetext}
\begin{eqnarray}
\hat{\cal H}_\textrm{tb}
&=&-t\sum_{\langle ij\rangle\sigma} \hat c_{iA\sigma}^\dagger  \hat c_{jB\sigma} +\hat c_{jB\sigma}^\dagger  \hat c_{iA\sigma} \notag\\
&=&-t\sum_{\langle ij\rangle\sigma}
\left(
\begin{array}{c}
 \hat c_{iA\sigma} \\
 \hat c_{iB\sigma}
\end{array}
\right)^\dagger
\left(
\begin{array}{cc}
 0 & 1\\
 1 & 0
\end{array}
\right)
\left(
\begin{array}{c}
 \hat c_{iA\sigma} \\
 \hat c_{iB\sigma}
\end{array}
\right)
\\ \notag
&=&-t\sum_{{\bf k}\sigma}
\left(
\begin{array}{c}
 \hat c_{{\bf k}A\sigma} \\
 \hat c_{{\bf k}B\sigma}
\end{array}
\right)^\dagger
\left(
\begin{array}{cc}
 0 & h_{\bf k} \\
 h_{\bf k}^* & 0
\end{array}
\right)
\left(
\begin{array}{c}
 \hat c_{{\bf k}A\sigma} \\
 \hat c_{{\bf k}B\sigma}
\end{array}
\right),
\end{eqnarray}
where $h_{\bf k}=e^{ik_xa} + e^{-i(k_x+\sqrt{3}k_y)a/2}+ e^{-i(k_x-\sqrt{3}k_y)a/2}$. Therefore
\begin{eqnarray}
 \varepsilon_\mathbf{k} &=& \pm t\left|h_{\bf k}\right|=\pm t\sqrt{3+2\cos(\sqrt3k_ya)+4\cos\left(\frac{\sqrt3k_ya}2\right)\cos\left(\frac{3k_xa}2\right)}
 \end{eqnarray}
 \end{widetext}
 We plot this dispersion relation in Fig. \ref{fig:Dirac}.

The most interesting features of this band structure are called the
`Dirac points'. The Dirac points
are located at ${\bf k}=n{\bf K}+m{\bf K}'$ where $n$ and $m$ are
integers, ${\bf K}=(2\pi/3a, 2\pi/3\sqrt{3}a)$, and ${\bf
K}'=(2\pi/3a,-2\pi/3\sqrt{3}a)$. In order to see why these points
are interesting consider a point ${\bf K}+{\bf q}$  in the
neighbourhood of ${\bf K}$. Recalling that $\cos(K+q)=\cos K-q\sin
K+\frac12q^2\cos K+\dots$ one finds that, for small $|{\bf q}|$, 
\begin{eqnarray}
 \varepsilon_{\mathbf{K}+\mathbf{q}} = \hbar v_F|\mathbf{q}|+\dots
 \end{eqnarray}
where $v_F=3ta/2\hbar$ is known as the Fermi velocity.

This result should be compared with the relativistic result
\begin{eqnarray}
 E_\mathbf{k}^2=m^2c^4 + \hbar^2 c^2 |\mathbf{k}|^2,
\end{eqnarray}
where $m$ is a particles rest mass and $c$ is the speed of light. This reduces to the famous $E=mc^2$ for $k=0$, but for massless particles, such as photons, one finds that
%\begin{eqnarray}
 $E_\mathbf{k}=\hbar c |\mathbf{k}|$.
%\end{eqnarray}
Thus the low-energy electronic excitations on a honeycomb lattice behave as if they are massless relativistic particles, with the Fermi velocity playing the role of the speed of light in the theory. Therefore much excitement \cite{Castro} has been caused by the recent synthesis of atomically thick sheets of graphene \cite{Geim}, in which carbon atoms form a honeycomb lattice. In graphene $v_F\simeq1\times10^6$ ms$^{-1}$, two orders smaller than the speed of light in the vacuum. This has opened the possibility of exploring and controlling `relativistic' effects in a solid state system \cite{Castro}.

%\subsection{Larger basis sets}

\section{The Hubbard model}\label{sect:Hubbard}

So far we have neglected electron-electron interactions. In real materials the electrons repel each other due to the Coulomb interaction between them. The most obvious extension to the tight binding model that describes some of the electron-electron interactions is to allow only on-site interactions, i.e., if $V_{ijkl}\ne0$ if and only if $i$, $j$, $k$ and $l$ all refer to the same orbital. For one orbital per site we then have the Hubbard model, 
\begin{eqnarray}
\hat{\cal H}_\textrm{Hubbard} =-t\sum_{\langle ij\rangle\sigma} \hat c_{i\sigma}^\dagger  \hat c_{j\sigma} +U \sum_{i} \hat c_{i\uparrow}^\dagger  \hat c_{i \uparrow} \hat c_{i \downarrow}^\dagger  \hat c_{i \downarrow}, \label{ham:hubbard}
\end{eqnarray}
where we have assumed nearest neighbour hopping only. It follows from Eq. \ref{eqn:V} that $U>0$, i.e., electrons repel one another.

\subsection{The two site Hubbard model: molecular hydrogen}

The two site Hubbard model is a nice context in which to consider some of the
basic properties of the chemical bond. The two body term in the Hubbard
model greatly complicates the problem relative to tight binding
model. Therefore the Hubbard model also presents a nice context in
which to introduce one of the most important tools in theoretical
physics and chemistry: mean-field theory.

\subsubsection{Mean-field theory, the Hartree-Fock approximation \& molecular orbital theory}\label{sect:HF2s}

To construct  a mean-field theory of any two, as yet unspecified, physical
quantities, $m=\overline m + \delta m$ and $n=\overline n + \delta n$, where $\overline n$ ($\overline m$) is the mean value of $n$ ($m$) and $\delta n$ ($\delta m$) are the fluctuations about the mean, which are assumed to be small, one notes that
\begin{eqnarray}
m  n &=& (\overline m + \delta m) (\overline n + \delta n) \notag\\&=& \overline m \, \overline n + \overline m \, \delta n + \delta m \, \overline n + \delta m \, \delta n \notag\\
&\approx& \overline m \, \overline n + \overline m \, \delta n + \delta m \, \overline n.
\end{eqnarray}
Thus mean-field approximations neglect terms that are quadratic in the fluctuations.

Hartree theory is a mean-field in the electron density, i.e.,
\begin{widetext}
\begin{eqnarray}
\hat c_{\alpha}^\dagger  \hat c_{\beta} \hat c_{\gamma}^\dagger  \hat c_{\delta}
 &=& \Big[ \langle \hat c_{\alpha}^\dagger  \hat c_{\beta} \rangle + \big( \hat c_{\alpha}^\dagger  \hat c_{\beta} - \langle \hat c_{\alpha}^\dagger  \hat c_{\beta} \rangle \big)  \Big] \notag
\Big[  \langle \hat c_{\gamma}^\dagger  \hat c_{\delta} \rangle + \big( \hat c_{\gamma}^\dagger  \hat c_{\delta} -   \langle \hat c_{\gamma}^\dagger  \hat c_{\delta} \rangle \big) \Big]\\
&\approx& \langle \hat c_{\alpha}^\dagger  \hat c_{\beta} \rangle  \hat c_{\gamma}^\dagger  \hat c_{\delta}
+ \hat c_{\alpha}^\dagger  \hat c_{\beta} \langle \hat c_{\gamma}^\dagger  \hat c_{\delta} \rangle
- \langle \hat c_{\alpha}^\dagger  \hat c_{\beta} \rangle \langle\hat c_{\gamma}^\dagger  \hat c_{\delta} \rangle.
\end{eqnarray}
However, it was quickly realised that this does not allow for electron exchange, i.e., one should also include averages such as $\langle \hat c_\alpha^\dagger\hat c_\delta\rangle$, therefore a better mean-field theory is Hartree-Fock theory, which includes these terms. However, because of the limited interactions included in the Hubbard model the Hartree theory is identical to the Hartree-Fock theory if one assumes that spin-flip terms are negligible, i.e., that $\langle \hat c_{i\uparrow}^\dagger\hat c_{i\downarrow}\rangle=0$, which we will.

The Hartree-Fock approximation to the Hubbard Hamiltonian is therefore
\begin{eqnarray}
\hat{\cal H}_\textrm{HF}
&=&-t\sum_{\langle ij\rangle\sigma} \hat c_{i\sigma}^\dagger  \hat c_{j\sigma}
+U \sum_{i} \Big[ \langle \hat c_{i\uparrow}^\dagger  \hat c_{i \uparrow} \rangle \hat c_{i \downarrow}^\dagger  \hat c_{i \downarrow}
+ \hat c_{i\uparrow}^\dagger  \hat c_{i \uparrow} \langle \hat c_{i \downarrow}^\dagger  \hat c_{i \downarrow} \rangle
- \langle \hat c_{i\uparrow}^\dagger  \hat c_{i \uparrow} \rangle \langle \hat c_{i \downarrow}^\dagger  \hat c_{i \downarrow} \rangle
\Big] \notag\\
&=&-t\sum_{\langle ij\rangle\sigma} \hat c_{i\sigma}^\dagger  \hat c_{j\sigma}
+U \sum_{i} \Big[ n_{i\uparrow} \hat c_{i \downarrow}^\dagger  \hat c_{i \downarrow}
+  n_{i \downarrow} \hat c_{i\uparrow}^\dagger  \hat c_{i \uparrow}
-  n_{i \uparrow}  n_{i \downarrow}\Big],%\\
%&=&\sum_{\langle ij\rangle\sigma} \Big[ -t + Un_{i \overline\sigma} \Big] \hat c_{i\sigma}^\dagger  \hat c_{j\sigma}
%-U \sum_{i}  n_{i \uparrow}  n_{i \downarrow},
\end{eqnarray}
where $ n_{i \sigma}=\langle \hat c_{i \sigma}^\dagger  \hat c_{i \sigma} \rangle$.
Thus we have a Hamiltonian for a single electron moving in the mean-field of the other electrons. Note that this Hamiltonian is equivalent to the $\omega$ method parameterisation of the H\"uckel model (cf. section \ref{Huckel-params} and particularly Eq. \ref{omega-method}) if we set $\omega=U/\beta$. Thus the $\omega$ method is just a parameterisation of the Hubbard model solved in the Hartree-Fock approximation.

The Hubbard model with two sites and two electrons can be taken as a model for molecular hydrogen.  In the Hartree-Fock  ground state, $|\Psi_\textrm{HF}^0\rangle$, the two electrons have opposite spin and each occupy the bonding state, which we found to be the ground state of the H\"uckel model in section \ref{section:H2Huck}:
\begin{subequations}
\begin{eqnarray}
|\Psi_\textrm{HF}^0\rangle
&=& |\psi_{b\downarrow}\rangle \otimes |\psi_{b\uparrow}\rangle = \frac12(\hat c^\dagger_{1\uparrow}+\hat c^\dagger_{2\uparrow})(\hat c^\dagger_{1\downarrow}+\hat c^\dagger_{2\downarrow})|0\rangle \label{eqn:HFfactor}\\
&=& \frac12(\hat c^\dagger_{1\uparrow}\hat c^\dagger_{1\downarrow}+\hat c^\dagger_{1\uparrow}\hat c^\dagger_{2\downarrow}-\hat c^\dagger_{1\downarrow}\hat c^\dagger_{2\uparrow}+\hat c^\dagger_{2\uparrow}\hat c^\dagger_{2\downarrow})|0\rangle. \label{eq:HF2}
\end{eqnarray}
\end{subequations}
\end{widetext}
Notice that $|\Psi_\textrm{HF}^0\rangle$ is just a product of two single particle wavefunctions (one for the spin up electron and another for the spin down electron; cf. Eq. \ref{eqn:HFfactor}). Thus we say that the wavefunction is \emph{uncorrelated} and that the two electrons are unentangled.

An important prediction of the Hartree-Fock theory is that if we pull the protons apart we are equally likely to get two hydrogen atoms (H+H) or two hydrogen ions (H$^+$+H$^-$). This is not what is observed experimentally. In reality the former is far more likely.

\subsubsection{The Heitler-London wavefunction \& valence bond theory}

Just a year after Schr\"odinger wrote down his wave equation \cite{Schrodinger}, Heitler and London \cite{HeitlerLondon} proposed a theory of the chemical
bond based on the new quantum mechanics. Explaining the nature of
the chemical bond remains one of the greatest achievement of quantum
mechanics. Heitler and London's theory led to the valence bond
theory of the chemical bond \cite{Pauling}. The two site
Hubbard model of H$_2$ is the simplest context in which to study this
theory.

The Heitler-London wavefunction is
\begin{eqnarray}
|\Psi_\textrm{HL}^0\rangle
= \frac1{\sqrt{2}}(\hat c^\dagger_{1\uparrow}\hat c^\dagger_{2\downarrow}-\hat c^\dagger_{1\downarrow}\hat c^\dagger_{2\uparrow})|0\rangle.\label{eq:HL2}
\end{eqnarray}
Notice that the wavefunction is \emph{correlated} as it cannot be written as a product of a wavefunction for each of the particles. Equivalently one can say that the two electrons are entangled. The Heitler London wavefunction overcorrects the physical errors in the Hartree-Fock molecular orbital wavefunction as it predicts zero probability of H$_2$ dissociating to an ionic state, but is, nevertheless, a significant improvement on molecular orbital theory.

\subsubsection{Exact solution of the two site Hubbard model}

The Hilbert space of the two site, two electron Hubbard model is sufficiently small that we can solve it analytically; nevertheless this problem can be greatly simplified by using the symmetry properties of the Hamiltonian. Firstly, note that the total spin operator commutes with the Hamiltonian \ref{ham:hubbard}, as none of the terms in the Hamiltonian cause spin flips. Therefore the energy eigenstates must also be spin eigenstates. For two electrons this means that all of the eigenstates will either be singlets ($S=0$) or triplets ($S=1$).

Let us begin with the triplet states, $|\Psi_1^m\rangle$. Consider a
state with two spin up electrons, $|\Psi_1^1\rangle$. Because there
is only one orbital per site the Pauli exclusion principal ensures
that there will be exactly one electron per site, i.e.,
$|\Psi_1^1\rangle=\hat c^\dagger_{1\uparrow}\hat c^\dagger_{2\uparrow}|0\rangle$.
The electrons cannot hop  between sites as the presence of the other electron and
the Pauli principle forbid it. Therefore, $\langle \Psi_1^1 | (-t\,
\hat c_{1\sigma}^\dagger  \hat c_{2\sigma}) |\Psi_1^1\rangle=\langle
\Psi_1^1 | (-t\, \hat c_{2\sigma}^\dagger  \hat c_{1\sigma})
|\Psi_1^1\rangle=0$ for $\sigma=\uparrow$ or $\downarrow$. There is
exactly one electron on each site so $\langle \Psi_1^1 | U \sum_{i}
\hat c_{i\uparrow}^\dagger  \hat c_{i \uparrow} \hat c_{i
\downarrow}^\dagger  \hat c_{i \downarrow}  |\Psi_1^1\rangle=0$.
Thus the total energy of this state is $E_1^1=0$.

The same chain of reasoning shows that $|\Psi_1^{-1}\rangle=\hat c^\dagger_{1\downarrow}\hat c^\dagger_{2\downarrow}|0\rangle$ and $E_1^{-1}=0$. It then follows from spin rotation symmetry that $|\Psi_1^{0}\rangle=\frac1{\sqrt 2}(\hat c^\dagger_{1\uparrow}\hat c^\dagger_{2\downarrow} +\hat c^\dagger_{1\downarrow}\hat c^\dagger_{2\uparrow})|0 \rangle$ and $E_1^{0}=0$.

As the Hilbert space contains six states, this leaves three singlet
states. A convenient basis of these is formed by the Heitler-London state and the two charge transfer states:
$|\Psi_\textrm{HL}\rangle=\frac1{\sqrt 2}(\hat c^\dagger_{1\uparrow}\hat c^\dagger_{2\downarrow}
-\hat c^\dagger_{1\downarrow}\hat c^\dagger_{2\uparrow})|0 \rangle$,
$|\Psi_\textrm{ct+}\rangle=\frac1{\sqrt 2}(\hat c^\dagger_{1\uparrow}\hat c^\dagger_{1\downarrow}
+ \hat c^\dagger_{2\uparrow}\hat c^\dagger_{2\downarrow})|0 \rangle$ and
$|\Psi_\textrm{ct-}\rangle=\frac1{\sqrt 2}(\hat c^\dagger_{1\uparrow}\hat c^\dagger_{1\downarrow}
- \hat c^\dagger_{2\uparrow}\hat c^\dagger_{2\downarrow})|0 \rangle$. Note that
$|\Psi_\textrm{HL}\rangle$ and $|\Psi_\textrm{ct+}\rangle$ are even
under `inversion' symmetry,\footnote{It may not be immediately obvious that $|\Psi_\textrm{HL}\rangle$ is even under inversion symmetry, but this is easily confirmed as 
%\begin{eqnarray}
${\cal I}|\Psi_\textrm{HL}\rangle =  \frac1{\sqrt 2}(\hat c^\dagger_{2\uparrow}\hat c^\dagger_{1\downarrow}
-\hat c^\dagger_{2\downarrow}\hat c^\dagger_{1\uparrow})|0 \rangle
=\frac1{\sqrt 2}(-\hat c^\dagger_{1\downarrow}\hat c^\dagger_{2\uparrow}
+\hat c^\dagger_{1\uparrow}\hat c^\dagger_{2\downarrow})|0 \rangle = |\Psi_\textrm{HL}\rangle$,
%\notag
%\end{eqnarray}
where ${\cal I}$ is the inversion operator, which swaps the labels 1 and 2.} which swaps the site labels
$1\leftrightarrow2$, whereas $|\Psi_\textrm{ct-}\rangle$ is odd
under inversion symmetry. As the Hamiltonian is symmetric under
inversion the eigenstates will have a definite parity so
$|\Psi_\textrm{ct-}\rangle$ is an eigenstate, with energy
$E_\textrm{ct-}=U$. The other two singlet states are not
distinguished by any symmetry of the Hamiltonian and so they do
couple, yielding the Hamiltonian matrix
\begin{eqnarray}
{\cal H}&=&
\left(
\begin{array}{cc}
\langle \Psi_\textrm{HL}| \hat{\cal H}_\textrm{Hubbard} |\Psi_\textrm{HL}\rangle
&  \langle \Psi_\textrm{HL}| \hat{\cal H}_\textrm{Hubbard} |\Psi_\textrm{ct+}\rangle  \\
   \langle \Psi_\textrm{ct+}| \hat{\cal H}_\textrm{Hubbard} |\Psi_\textrm{HL}\rangle
 &   \langle \Psi_\textrm{ct+}| \hat{\cal H}_\textrm{Hubbard} |\Psi_\textrm{ct+}\rangle
\end{array}
\right)\notag\\
&=&
\left(
\begin{array}{cc}
0
&  -2t  \\
   -2t
 &   U
\end{array}
\right).
\end{eqnarray}
This has eigenvalues, $E_\textrm{CF}=\frac12(U-\sqrt{U^2+16t^2})$ and  $E_\textrm{S$^2$}=\frac12(U+\sqrt{U^2+16t^2})$. The corresponding eigenstates are 
\begin{subequations}
\begin{eqnarray}
|\Psi_\textrm{CF}\rangle &=& \cos\theta |\Psi_\textrm{HL}\rangle + \sin\theta |\Psi_\textrm{ct+}\rangle \notag\\
&=&\left[\frac{\cos\theta}{\sqrt2}\left(\hat c^\dagger_{1\uparrow}\hat c^\dagger_{2\downarrow}-\hat c^\dagger_{1\downarrow}\hat c^\dagger_{2\uparrow}\right) \right. \notag\\&& \hspace{1.0cm}  +\left. \frac{\sin\theta}{\sqrt2}\left(\hat c^\dagger_{1\downarrow}\hat c^\dagger_{1\uparrow}+\hat c^\dagger_{2\downarrow}\hat c^\dagger_{2\uparrow}\right)\right]|0\rangle \label{exactgs}\\
|\Psi_\textrm{S$^2$}\rangle &=& \sin\theta |\Psi_\textrm{HL}\rangle + \cos\theta |\Psi_\textrm{ct+}\rangle\notag\\
&=&\left[\frac{\sin\theta}{\sqrt2}\left(\hat c^\dagger_{1\uparrow}\hat c^\dagger_{2\downarrow}-\hat c^\dagger_{1\downarrow}\hat c^\dagger_{2\uparrow}\right) \right. \notag\\&& \hspace{1.0cm}  +\left. \frac{\cos\theta}{\sqrt2}\left(\hat c^\dagger_{1\downarrow}\hat c^\dagger_{1\uparrow}+\hat c^\dagger_{2\downarrow}\hat c^\dagger_{2\uparrow}\right)\right]|0\rangle,
\end{eqnarray}
\end{subequations}
where $\tan\theta=(U-\sqrt{U^2+16t^2})/4t$. For $U>0$, as is
required physically, the  state
$|\Psi_\textrm{CF}\rangle$ is the ground state for all values of
$U/t$. $|\Psi_\textrm{CF}\rangle$ is often called the
Coulson-Fischer wavefunction.

Inspection of Eq. \ref{exactgs} reveals that for $U/t\rightarrow\infty$ the Coulson-Fischer state tends to the Heitler-London wavefunction, while for $U/t\rightarrow0$ we regain the molecular orbital picture (Hartree-Fock wavefunction).

\subsection{Mott insulators \& the Mott-Hubbard metal-insulator transition}

In 1949  Mott \cite{Mott} asked an apparently simple question with a profound and surprising answer. As we have seen above, for the two site Hubbard model both the
molecular orbital (Hartree-Fock) and valence bond (Heitler-London)
wavefunctions  are just approximations to the exact
(Coulson-Fischer) wavefunction. Mott asked whether the equivalent
statement is true in an infinite solid, and, surprisingly, found that
the answer is no. Further, Mott showed that the Hartree-Fock and
Heitler-London wavefunctions predict very different properties for
crystals.

One of the most important properties of a crystal is its
conductivity. In a metal the conductivity is high and increases as
the temperature is lowered. Whereas in a semiconductor or an
insulator the conductivity is low and decreases as the temperature
is lowered. These behaviours arise because of fundamental
differences between the electronic structures of metals and
semiconductors/insulators \cite{A&M}. In metals there are excited states at
arbitrarily low energies above the Fermi energy. This means that,
even at the lowest temperatures, electrons can move in response to
an applied electric field. In semiconductors and insulators there is
an energy gap between the highest occupied electronic state and the
lowest unoccupied electronic state at zero temperature. This means that a thermal
activation energy must be provided if electrons are to move in response
to an applied field. The difference between semiconductors and
insulators is simply the size of the gap; therefore we will
not distinguish between the two below and will refer to any material with a gap
as an insulator.

Consider a Hubbard model at `half-filling', i.e., with the same
number of electrons as lattice sites. In
order for a macroscopic current to flow, an electron must move from
one lattice site (leaving an empty site with a net positive charge)
to a \emph{distant} site (creating a doubly occupied site with a net
negative charge). The net charges may move through collective
motions of the electrons. One could keep track of this by describing
the movement of all of the electrons, but it is easier to introduce
an equivalent description where we treat the net charges as
particles moving in a neutral background. Therefore, we will refer
to the positive charge as a holon and the negative charge as a
doublon. In the ground state of the valence bond theory all of the
sites are neutral and there are no holons or doublons (cf. Eq.
\ref{eq:HL2}). However, it is reasonable to postulate that there are low lying
excited states and hence thermal states that contain a few doublons and holons.
These doublons and holons would interact via the Coulomb potential,
$V(r)=-e^{2}/\kappa r$, where $\kappa$ is the dielectric constant of
the crystal. We know from the theory of the hydrogen atom (or, better,
positronium, cf. Ref. \onlinecite{Gasiorowicz}) that this potential gives rise
to bound states. Therefore one expects that, in the valence bond
theory, holons and doublons are bound and separating  holon-doublon
pairs costs a significant amount of energy. Thus one expects the number of distant holon-doublon
pairs to decrease as the temperature is lowered. Therefore, the valence bond theory predicts
that the half-filled Hubbard model is an insulator.

In contrast the molecular orbital theory has large numbers of holons
and doublons (cf. Eq. \ref{eq:HF2}, which suggests that for an
$N$-site model there will be $N/2$ neutral sites, $N/4$ empty sites,
and $N/4$ doubly occupied sites). Mott reasoned that if there are many
holon-doublon pairs ``it no longer follows that work must
necessarily be done to form some more''. This is because the holon
and doublon now interact via a screened potential,
$V(r)=-(e^{2}/\kappa r) \exp(-qr)$, where $q$ is the Thomas-Fermi wavevector (cf. Ref. \onlinecite{A&M}). For
sufficiently large $q$ there will be no bound states and the
molecular orbital theory predicts that the half-filled Hubbard model
is metallic.

\begin{figure}
\includegraphics[width=8cm]{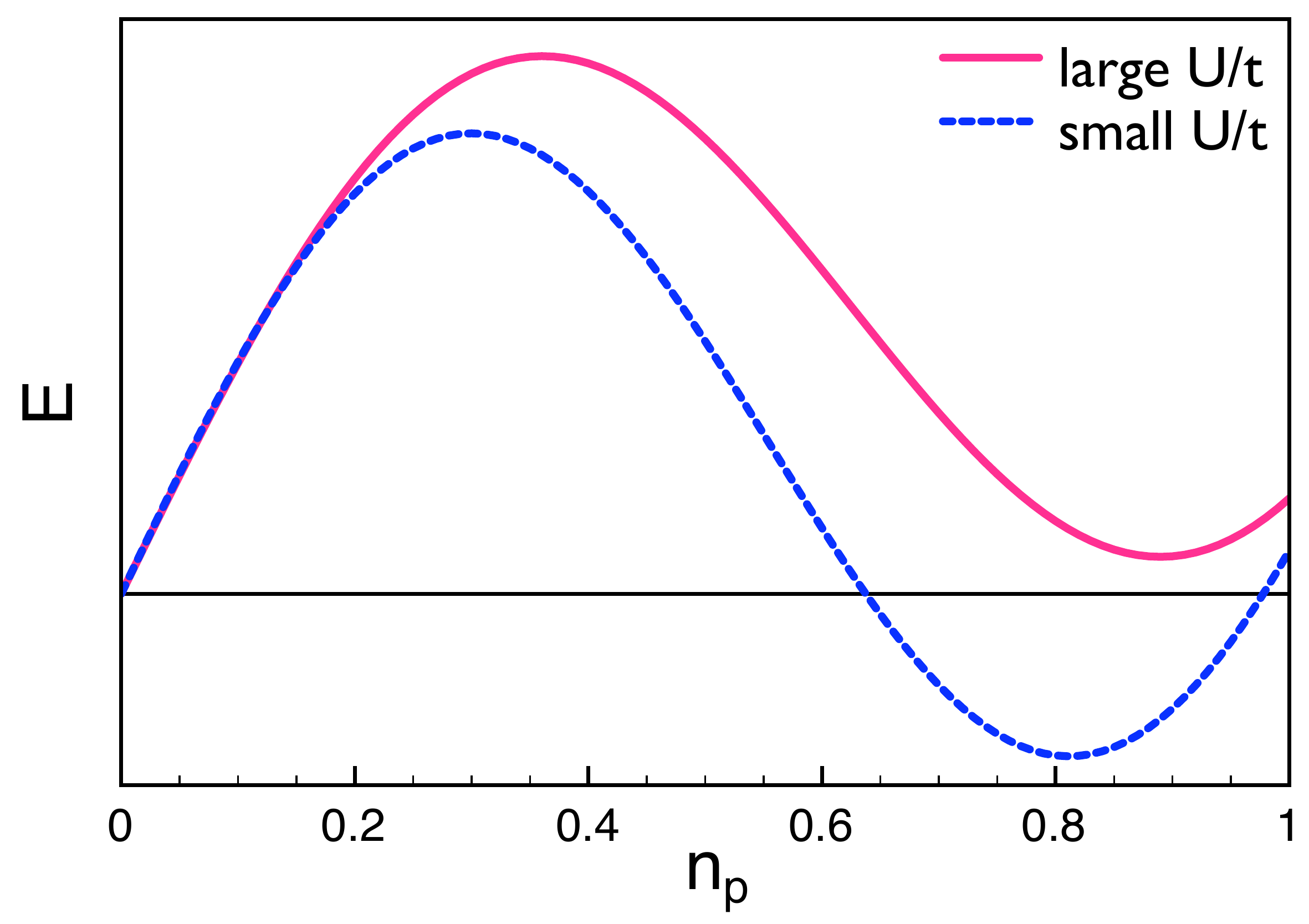}
\caption{Sketch of Mott's proposal for the energy of the Hubbard model as a function of the number of holon-doublon pairs, $n_p$, at low (zero) temperature(s) for large and small $U/t$.} \label{fig:Mott}
\end{figure}

Thus, Mott argued that their are two (local) minima of the free
energy in a crystal (cf. Fig. \ref{fig:Mott}). One of the minima corresponds to a
state with no holon-doublon pairs that is well approximated by a valence
bond wavefunction and is now known as the Mott insulating state. The
second minimum corresponds to a state with many doublon-holon pairs
that is well approximated by a molecular orbital wavefunction
and is metallic. As we saw
above, valence bond theory works well for $U\gg t$ and molecular
orbital theory works well for $U\ll t$. Therefore, in the
half-filled Hubbard model we expect a Mott insulator for large $U/t$
and a metal for small $U/t$. Further the `double well' structure of
the energy predicted by Mott's argument (Fig. \ref{fig:Mott}) suggests
that there is a first order metal-insulator phase transition, known
as the Mott transition. Mott predicted that this metal-insulator transition can
be driven by applying pressure to a Mott insulator. This has now
been observed in a number of systems; perhaps the purest examples
are the organic charge transfer salts (BEDT-TTF)$_2X$ \cite{JPCMrev}.

It is interesting to note that this infusion of chemical ideas into
condensed matter physics has remained important in studies of the
Mott transition. Of particular note is Anderson's resonating valence
bond theory of superconductivity in the high temperature
superconductors \cite{Anderson87,Anderson-who-or-what}, which
describes superconductivity in a doped Mott insulator in terms of a
generalisation of the valence bond theory discussed above. This
theory can also be modified to describe superconductivity on the
metallic side of the Mott transition for a half-filled lattice. This
theory then provides a good description of the superconductivity
observed in the (BEDT-TTF)$_2X$ salts \cite{RVBorganics}.

Note that theories, such as Hartree-Fock theory or density functional theory \cite{Yang}, that do not include the strong electronic correlations present in the Hubbard model do not predict a Mott insulating state. Thus weakly correlated theories make the qualitatively incorrect prediction that materials such as NiO, V$_2$O$_3$, La$_{2}$CuO$_4$  and $\kappa$-(BEDT-TTF)$_2$Cu[N(CN)$_2$]Cl are metals, whereas experimentally all are insulators.

We will discuss a quantitative theory of the Mott transition is
section \ref{slavebosons}.

\subsection{Mean-field theories for crystals}

\subsubsection{Hartree-Fock theory of the Hubbard model: Stoner ferromagnetism}\label{sect:Stoner}

\begin{figure}
\includegraphics[width=8cm]{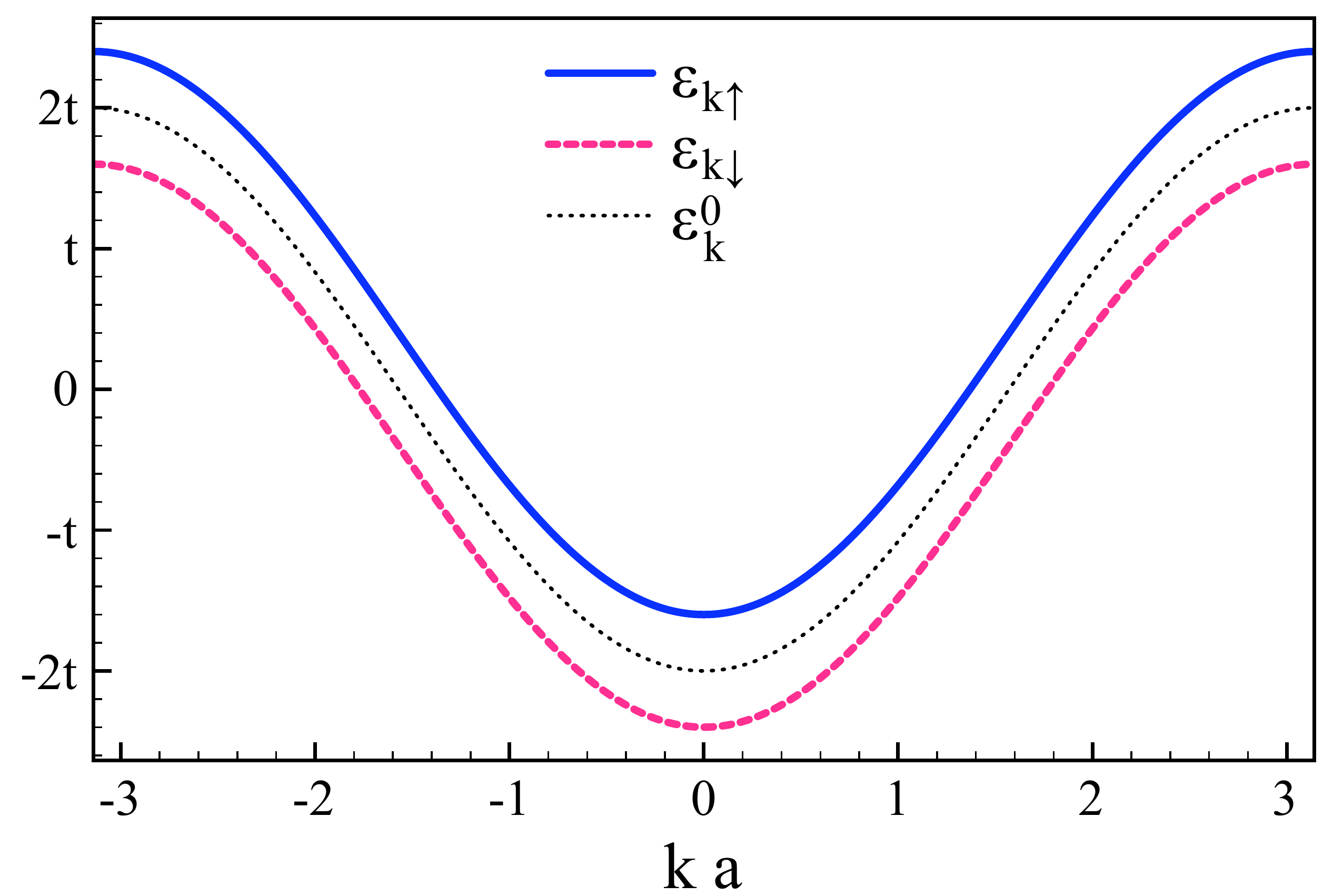}
\caption{The dispersion relations for spin-up and spin-down electrons in the Hartree-Fock theory of the Hubbard chain (Stoner model of ferromagnetism) with $m=0.8t/U$.} \label{fig:Stoner-bands}
\end{figure}

In a similar manner to that in which  we constructed the
Hartree-Fock  mean-field theory for the
two site Hubbard model in section \ref{sect:HF2s} we can also
construct a Hartree-Fock theory of the infinite lattice Hubbard model.
Again, we simply replace the number operators in the two body term
by their mean values, $n_{i\sigma}\equiv\langle\hat c_{i\sigma}^\dagger\hat
c_{i\sigma}\rangle$, plus the fluctuations about the mean,
$\big(\hat c_{i\sigma}^\dagger\hat c_{i\sigma}-n_{i\sigma}\big)$,
and neglect terms that are quadratic in the fluctuations, {\it
viz}.,
\begin{widetext}
\begin{eqnarray}
U \sum_{i} \hat c_{i\uparrow}^\dagger  \hat c_{i \uparrow} \hat c_{i \downarrow}^\dagger  \hat c_{i \downarrow}
&=&U \sum_{i} \Big[n_{i \uparrow} + \big(\hat c_{i\uparrow}^\dagger  \hat c_{i \uparrow} - n_{i \uparrow} \big) \Big]
 \Big[n_{i \downarrow} + \big(\hat c_{i\downarrow}^\dagger  \hat c_{i \downarrow} - n_{i \downarrow} \big) \Big] \notag\\
&\simeq&U \sum_{i} \Big[  n_{i \downarrow} \hat c_{i\uparrow}^\dagger  \hat c_{i \uparrow} + n_{i \uparrow} \hat c_{i\downarrow}^\dagger  \hat c_{i \downarrow} - n_{i \uparrow} n_{i \downarrow}  \Big].
\end{eqnarray}
If we make the additional  approximation that $n_{i\sigma}=n_\sigma$
for all $i$, i.e., that the system is homogeneous and does not spontaneously break translational symmetry, we find that the
Hartree-Fock Hamiltonian for the Hubbard model is
\begin{eqnarray}
\hat{\cal H}_\textrm{HF} - \mu\hat N=-t\sum_{\langle ij\rangle\sigma} \hat c_{i\sigma}^\dagger  \hat c_{j\sigma}  + \sum_{i\sigma} (Un_{\overline\sigma} - \mu)  \hat c_{i\sigma}^\dagger  \hat c_{i\sigma} -U N n_{\uparrow} n_{\downarrow},
\end{eqnarray}
where $N$ is the number of lattice sites and $\overline\sigma$ is the opposite spin to $\sigma$. It is convenient to write this Hamiltonian in terms of the total electron density, $n=n_\uparrow+n_\downarrow$ and the magnetisation density, $m=n_\uparrow-n_\downarrow$, which gives,
\begin{eqnarray}
\hat{\cal H}_\textrm{HF} - \mu\hat N &=&-t\sum_{\langle
ij\rangle\sigma} \hat c_{i\sigma}^\dagger  \hat c_{j\sigma}
 -\mu\sum_{i\sigma}  \hat c_{i\sigma}^\dagger  \hat c_{i\sigma}
 \notag\\&& \notag
 +U \sum_{i} \Big[\frac12  (n-m) \hat c_{i\uparrow}^\dagger  \hat c_{i \uparrow} + \frac12  (n+m)  \hat c_{i\downarrow}^\dagger  \hat c_{i \downarrow} - \frac14  (n+m)  (n-m)   \Big], \\
%&=&-t\sum_{\langle ij\rangle\sigma} \hat c_{i\sigma}^\dagger  \hat c_{j\sigma} +\frac{U}{2} \sum_{i} \Big[n (\hat c_{i\uparrow}^\dagger \hat c_{i \uparrow} + \hat c_{i\downarrow}^\dagger  \hat c_{i \downarrow} ) - m (\hat c_{i\uparrow}^\dagger \hat c_{i \uparrow} - \hat c_{i\downarrow}^\dagger  \hat c_{i \downarrow} ) \Big] \notag\\&&
%- \frac{NU}4  (n^2-m^2)
% -\mu\sum_{i\sigma}  \hat c_{i\sigma}^\dagger  \hat c_{i\sigma}\\
%&=&-t\sum_{\langle ij\rangle\sigma} \hat c_{i\sigma}^\dagger  \hat c_{j\sigma} -\frac{Um}{2} \sum_{i} (\hat c_{i\uparrow}^\dagger \hat c_{i \uparrow} - \hat c_{i\downarrow}^\dagger  \hat c_{i \downarrow} ) %
%- \frac{NU}4  (n^2-m^2)   \notag\\&&
%+\left(\frac{Un}2 -\mu\right)\sum_{i\sigma}  \hat c_{i\sigma}^\dagger  \hat c_{i\sigma} \notag\\
&=&\sum_{{\mathbf k}\sigma} \left(\varepsilon_{\mathbf k}^0 +\overline\sigma\frac{Um}{2}\right)  \hat n_{{\mathbf k}\sigma} -\left(\mu- \frac{Un}2  \right) \sum_{{\mathbf k}\sigma} \hat n_{{\mathbf k} \sigma}
- \frac{NU}4  (n^2-m^2)
\end{eqnarray}
\end{widetext}
where $\varepsilon_{\mathbf k}^0$ is the dispersion relation for $U=0$ and $\sigma=\pm1=\uparrow\downarrow$. The last term is just a constant and will not concern us greatly. The penultimate term is the `renormalised' chemical potential, i.e., the chemical potential, $\mu$, of the system with $U=0$ is decreased  by $Un/2$ due to the interactions. The first term is just the renormalised dispersion relation, in particular we find that if the magnetisation density is non-zero the dispersion relation for spin-up electrons is different from that for spin-down electrons (cf. Fig. \ref{fig:Stoner-bands}). It is important to note that the Hartree-Fock approximation has reduced the problem to a single particle (single determinant) theory. Thus we can write
\begin{equation}
\hat{\cal H}_\textrm{HF} - \mu\hat N
=\sum_{{\mathbf k}\sigma} \left(\varepsilon_{{\mathbf k}\sigma}^*   -\mu^*  \right) \hat n_{{\mathbf k} \sigma}
- \frac{NU}4  (n^2-m^2),
\end{equation}
where $\varepsilon_{{\mathbf k}\sigma}^*=\varepsilon_{\mathbf k}^0 -\frac12\sigma{Um}$ and $\mu^*=\mu-\frac12Un$.

\begin{figure}
\includegraphics[width=8cm]{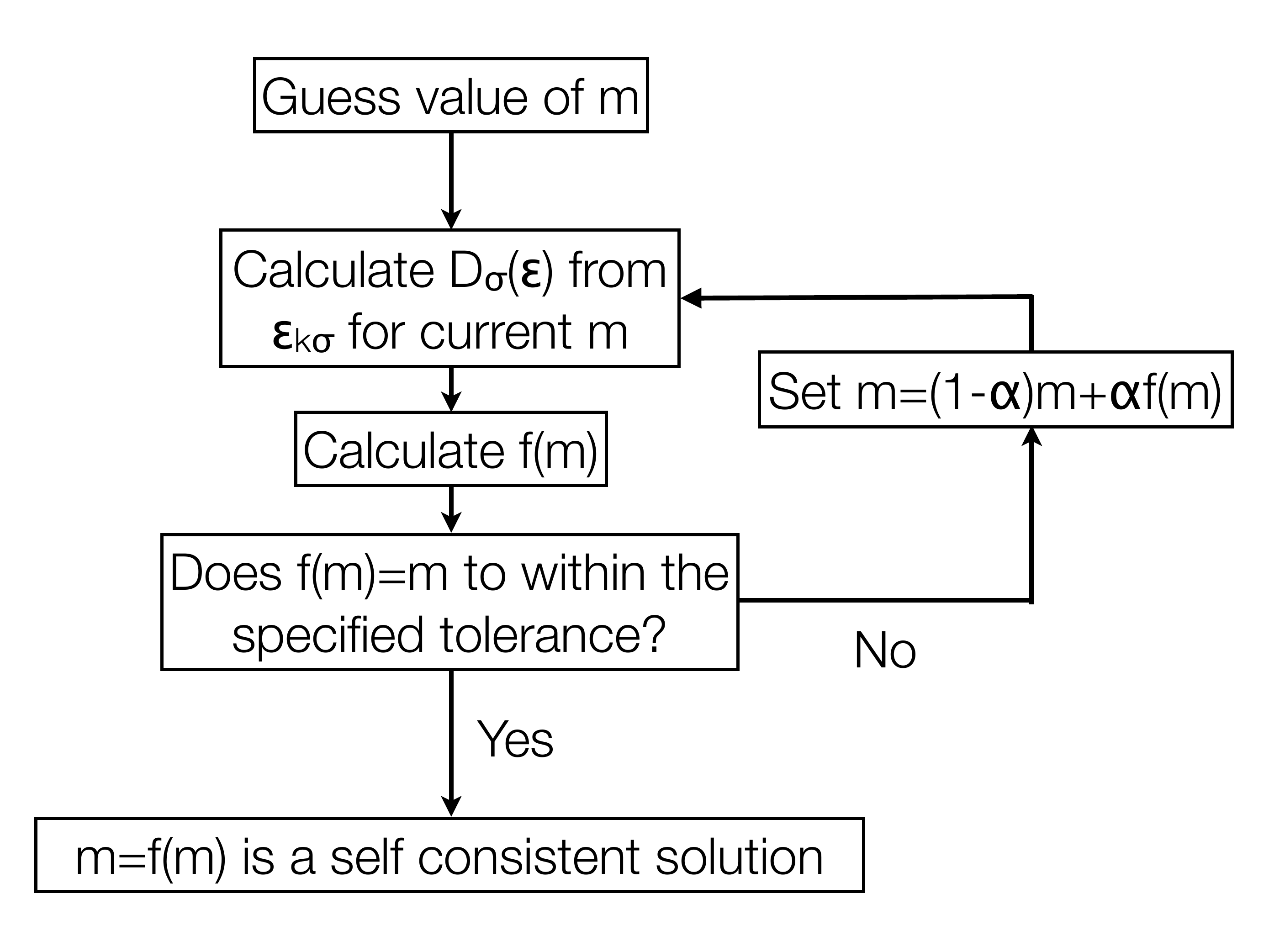}
\caption{How to find the self consistent solution of Eq. \ref{magnetisation}. If the convergence works well one can take $\alpha=1$, but for some problems convergence can be reached more reliably with a small value of $\alpha$ (often a value as small as $\sim0.05$ is used).} \label{selfconstsoln}
\end{figure}

We can now calculate the magnetisation density:
\begin{eqnarray}
m&=&n_\uparrow-n_\downarrow \notag\\
&=& \int_{-\infty}^0 d\epsilon \left[ D_\uparrow\left(\epsilon-\mu^*\right) - D_\downarrow \left(\epsilon-\mu^* \right) \right] \notag\\
&=& \int_{-\infty}^0 d\epsilon \left[ D_0 \left(\epsilon-\frac12Um+\frac12Un-\mu \right)  \right.\notag\\&& \hspace{2cm} \left. - D_0 \left(\epsilon+\frac12Um+\frac12Un-\mu \right) \right] \notag\\
& \equiv & f(m)  = D_0(0)Um+{\cal O}(m^2), \label{magnetisation}
\end{eqnarray}
where $D_0(\epsilon)=\partial N_{0}(\epsilon)/\partial\epsilon|_\epsilon$ is the density of states (DOS; cf. Ref. \onlinecite{A&M}) per spin for $U=0$, $N_{0}(\epsilon)$ is the number of electrons (per spin species) for which $\varepsilon_{{\mathbf k}}^0\leq\epsilon$ for $U=0$, $D_\sigma(\epsilon)=\partial N_{\sigma}(\epsilon)/\partial\epsilon|_\epsilon$ is the full interacting DOS for spin $\sigma$ electrons, and $N_{\sigma}(\epsilon)$ is the number of electrons with spin $\sigma$ for which $\varepsilon_{{\mathbf k}\sigma}\leq\epsilon$. The standard way to solve mean-field theories, known as the method of self consistent solution, is illustrated  in Fig. \ref{selfconstsoln}. The major difficulty with self consistent solutions is that it is not possible to establish whether or not one has found all of the self consistent solutions and therefore it is not possible to establish whether or not one has found the global minimum. Therefore it is prudent to try a wide range of initial guesses for $m$ (or whatever variable the initial guess is made in).

\begin{figure}
\includegraphics[width=8cm]{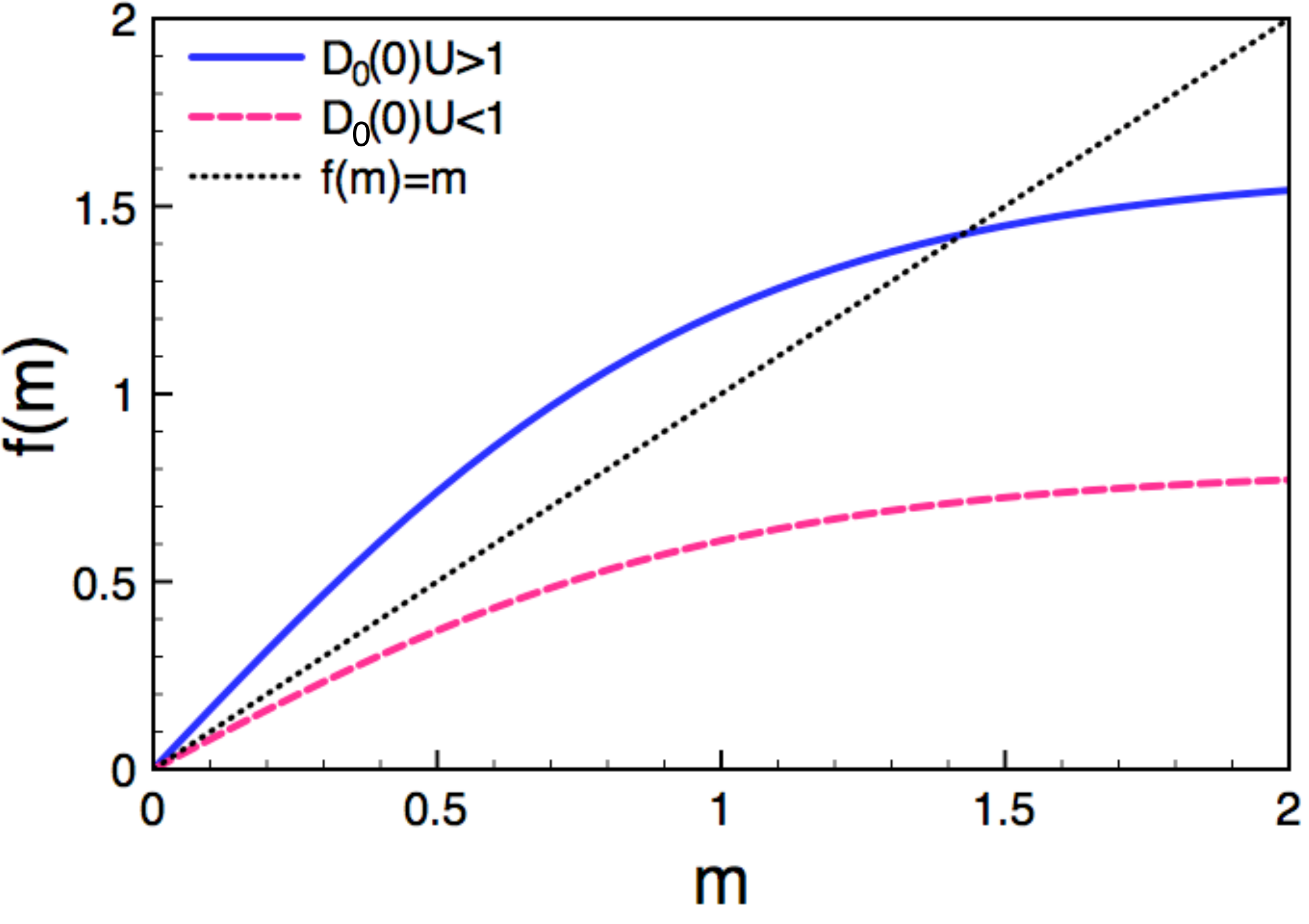}
\caption{Graphical solution of the self-consistency equation (Eq. \ref{magnetisation}) for the Stoner model of ferromagnetism.} \label{Stoner-graphical-soln}
\end{figure}

Clearly $m=0$ is always a solution of Eq. \ref{magnetisation}, and for $UD_0(0)<1$ this turns out to be the only solution. But, for $UD_0(0)>1$ there are additional solutions with $m\ne0$. This is easily understood from the sketch in Fig. \ref{Stoner-graphical-soln}. Furthermore, the $m\ne0$ solutions typically have lower energy than the $m=0$ solution and therefore for $UD_0(0)>1$ the ground state is ferromagnetic. $UD_0(0)\geq1$ is known as the Stoner condition for ferromagnetism. In order for the Stoner condition to be satisfied a system must have narrow bands [small $t$, and hence large $D(0)$] and strong interactions (large $U$).

There are three elemental ferromagnets, Fe, Co and Ni, each of which is also metallic. As the Hartree-Fock theory of the Hubbard model predicts metallic magnetism if the Stoner criterion is satisfied and these materials have narrow bands of strongly interacting electrons it is natural to ask whether this is a good description of these materials. However, if one extends the above treatment to finite temperatures \cite{Rossler} one finds that the the Hartree-Fock theory of the Hubbard model does not provide a good theory of the three elemental magnets. The Curie temperatures, $T_C$, (i.e., the temperature at which the material becomes ferromagnetic) of Fe, Co and Ni are $\sim$1000 K (see, e.g., table 33.1 of Ref. \onlinecite{A&M}). The Hartree-Fock theory predicts that $T_c\sim Um_0$, where $m_0$ is the magnetisation at $T=0$. If the parameters in the Hubbard model are chosen so that Hartree-Fock theory reproduces the observed $m_0$ then the predicted critical temperature is $\sim$10,000 K. This, order-of-magnitude, disagreement with experiment results from the failure of the mean-field Hartree-Fock approximation to properly account for the fluctuations in the local magnetisation. This is closely related to the (incorrect) prediction of the Hartree-Fock approximation that there are no local moments above $T_c$. (Experimentally local moments \emph{are} observed above $T_c$.) However, for weak ferromagnets, such as ZrZn$_{2}$ ($T_{C}\sim30$ K) the Hartree-Fock theory of the Hubbard model provides an excellent description of the observed behaviour \cite{Mohn}.

The effects missed by Hartree-Fock theory are referred to as \emph{electronic correlations}. The dramatic failure of Hartree-Fock theory in Fe, Co and Ni shows that electron correlations are very important in these materials, as do other comparisons of theory and experiment \cite{Jacko}. However, it is important to note that mean-field theory is not limited to Hartree-Fock theory (although the terms are often, but incorrectly, used synonymously). Rather Hartree-Fock theory is the mean-field theory of the electronic density. By constructing mean-field theories of other properties it is possible to construct mean-field theories that capture (some) electronic correlations. We will now consider an example of a rather different mean-field theory.

\subsubsection{The Gutzwiller approximation, slave bosons \& the Brinkman-Rice metal-insulator
transition}\label{slavebosons}

In 1963 Gutzwiller \cite{Gutzwiller} proposed a variational wavefunction for the Hubbard model:
\begin{eqnarray}
|\Psi_G\rangle &=& \prod_i(1-\alpha \hat n_{i\uparrow} \hat n_{i\downarrow})|\Psi_0\rangle \notag\\&=& \exp\left( -g\sum_i \hat n_{i\uparrow} \hat n_{i\downarrow}\right)|\Psi_0\rangle,\label{Gutz}
\end{eqnarray}
where $g=-\ln(1-\alpha)$ is a variational parameter and $|\Psi_0\rangle$ is the ground state for uncorrelated electrons. One should note that the Gutzwiller wavefunction is closely related to the coupled cluster ansatz \cite{Fulde}, which is widely used in both physics and chemistry. Gutzwiller used this ansatz to study the problem of itinerant ferromagnetism. This leads to an improvement over the Hartree-Fock theory discussed above. However, in 1970 Brinkman and Rice \cite{BR} showed that this wavefunction also describes a metal-insulator transition, now referred to as a Brinkman-Rice transition. Rather than studying this wavefunction in detail we will instead use an equivalent technique known as `slave bosons'. This has the advantage of making it clear that the Brinkman-Rice transition is just a mean-field description of the Mott transition. 

The $i^\textrm{th}$ site in a Hubbard model has four possible states: the site can be empty, $|e_i\rangle$; contain a single spin $\sigma$ ($=\uparrow$ or $\downarrow$) electron $|\sigma_i\rangle$; or two electrons, $|d_i\rangle$. The Kotliar-Ruckenstein slave boson technique introduces an over-complete description of these states:
\begin{subequations}
\begin{eqnarray}
|e_i\rangle&=&\hat e_i^\dagger|0_i\rangle, \\
|\sigma_i\rangle&=&\hat p_{i\sigma}^\dagger\hat c_{i\sigma}^\dagger|0_i\rangle,
\end{eqnarray}
 and
\begin{eqnarray}
|d_i\rangle&=&\hat d_i^\dagger\hat c_{i\uparrow}^\dagger\hat c_{i\downarrow}^\dagger|0_i\rangle,
\end{eqnarray}
\end{subequations}
where $\hat e_i^\dagger$, $\hat p_{i\sigma}^\dagger$, and $\hat d_i^\dagger$ are bosonic creation operators which correspond to empty, partially filled, and doubly occupied sites. $|0_i\rangle$ is a state with no fermions and no bosons on site $i$; note that this is not a physically realisable state. This transformation is not only kosher, but also exact, so long as we also introduce the constraints
\begin{subequations}
\begin{eqnarray}
\hat e_i^\dagger\hat e_i + \sum_\sigma \hat p_{i\sigma}^\dagger\hat p_{i\sigma} + \hat d_i^\dagger\hat d_i &=&1,
\end{eqnarray}
which ensures that there is exactly one boson per site and therefore that each site is either empty, partially occupied or doubly occupied,  and
\begin{eqnarray}
\hat c_{i\sigma}^\dagger\hat c_{i\sigma} - \hat p_{i\sigma}^\dagger\hat p_{i\sigma} - \hat d_i^\dagger\hat d_i  &=&0,
\end{eqnarray}
\end{subequations}
which ensures that if a site contains a spin $\sigma$ electron then it is either singly occupied (with spin $\sigma$) or doubly occupied.

Writing the Hubbard Hamiltonian in terms of the slave bosons yields
\begin{eqnarray}
\hat{\cal H}_\textrm{Hubbard} =-t\sum_{\langle ij\rangle\sigma} \hat z_{i\sigma}^\dagger\hat c_{i\sigma}^\dagger  \hat c_{j\sigma} \hat z_{j\sigma} +U \sum_{i} \hat d_{i}^\dagger  \hat d_{i}  ,
\end{eqnarray}
where
%\begin{eqnarray}
 $\hat z_{j\sigma} = \hat e_j^\dagger\hat p_{j\sigma} + \hat p_{j\overline\sigma}^\dagger\hat d_j$.
%\end{eqnarray}

We now make a mean-field approximation and replace the bosonic operators by the expectation values: $\langle e_i\rangle=e$, $\langle p_{i\uparrow}\rangle=\langle p_{i\downarrow}\rangle=p$, $\langle d_i\rangle=d$. Note that we have additionally assumed that the system is homogeneous (the expectation values do not depend on $i$) and paramagnetic ($\langle p_{i\uparrow}\rangle=\langle p_{i\downarrow}\rangle$). Therefore
%where $N_0$ is the number of lattice sites and $|z|^2=|p|^2|e+d|^2$
%
%Further
the constraints reduce to
\begin{subequations}
\begin{eqnarray}
|e|^2 + 2  |p|^2 +  |d|^2 &=&1
\end{eqnarray}
 and
\begin{eqnarray}
|p|^2 + |d|^2 &=&  \langle \hat c_{i\sigma}^\dagger\hat c_{i\sigma} \rangle = \frac{n}2,
\end{eqnarray}
\end{subequations}
where $n$ is the average number of electrons per site. This amounts to only enforcing the constraints on average. 

This theory does not reproduce the correct result for $U=0$. However, this deficiency can be fixed if $\hat z_{j\sigma}$ is replaced by the `renormalised' quantity, $\tilde z_{j\sigma}$, defined such that
\begin{eqnarray}
\langle \tilde z_{j\sigma}^\dagger \tilde z_{j\sigma} \rangle = \frac{\frac{n}2-|d|^2}{(1-\frac{n}2)\frac{n}2} \left( d + \sqrt{1-n+|d|^2} \right).
\end{eqnarray}

Let us specialise to a `half filled' band, $n=1$.
The constraints now allow us to eliminate $|p|^2=\frac12-|d|^2$ and $|e|^2=|d|^2$.
Thus we find that
\begin{eqnarray}
\hat{\cal H}_\textrm{Hubbard} &\simeq& -t\sum_{\langle ij\rangle\sigma} \frac18(|d|^2-2|d|^4) \hat c_{i\sigma}^\dagger  \hat c_{j\sigma}  +U N_0 |d|^2\notag\\
&=& \frac18(|d|^2-2|d|^4) \sum_{{\bf k}\sigma} \varepsilon_{\bf k}^0   \hat n_{{\bf k}\sigma}  +U N_0 |d|^2,
\end{eqnarray}
where $\varepsilon_{\bf k}^0$ is the dispersion for $U=0$ and $N$ is the number of lattice sites. Recall that $|d|^2=\langle d_i^\dagger d_i\rangle$, i.e., $|d|^2$ is the probability of site being doubly occupied. We construct a variational theory by ensuring that the energy is minimised with respect to $|d|$, which yields
\begin{eqnarray}
\frac{\partial E}{\partial |d|}= \frac14(|d|-4|d|^3) \sum_{{\bf k}\sigma} \varepsilon_{\bf k}^0 \langle \hat n_{{\bf k}\sigma} \rangle +2U N_0 |d| =0 \label{sbsc}.
\end{eqnarray}
Eq. \ref{sbsc} allows one to solve the problem self consistently, cf. Fig. \ref{selfconstsoln}. For small $U$ this equation has more than one minimum and the lowest energy state has $|d|^2>0$, which corresponds to a correlated metallic state (the details of this minimum depend on $\varepsilon_{\bf k}^0$). But, above some critical $U$ the ground state has $|d|^2=0$, which corresponds to no doubly occupied states, i.e. the Mott insulator. Thus the dependence of the energy on the number holon-doublon pairs ($n_p=|d|^2$) calculated from the mean-field slave boson theory is exactly as Mott predicted on rather general grounds (shown in Fig. \ref{fig:Mott}).

\subsection{Exact solutions of the Hubbard model}

\subsubsection{One dimension}

Lieb and Wu \cite{LiebWu} famously solved the Hubbard chain at $T=0$ using the Bethe ansatz \cite{Essler,Tsvelik}. Lieb and Wu found that the half-filled Hubbard chain is a Mott insulator for any non-zero $U$. Nevertheless the Bethe ansatz solution is not straightforward to understand and weighty textbooks have been written on the subject \cite{Essler,Tsvelik}.

\subsubsection{Infinite dimensions: dynamical mean-field
theory}\label{DMFT}

As one increases the dimension of a lattice the coordination number
(the number of nearest neighbours for each lattice site) also
increases. In infinite dimensions each lattice site has infinitely
many nearest neighbours. For a classical model mean-field theory
becomes exact in infinite dimensions, as the environment (the
infinite number of nearest neighbours) seen by each site is exactly
the same as the mean-field. However, quantum mechanically things are
complicated by the internal dynamics of the site. In the Hubbard
model each site can contain zero, one or two electrons, and a
dynamic equilibrium between the different charge and spin states is
maintained. However, the environment is still described by a
mean-field, even though the dynamics are not. Therefore, although the
Hartree-Fock theory of the Hubbard model does not become exact in
infinite dimensions, it is possible to construct a theory that
treats the on-site dynamics exactly and the spatial correlations at
the mean-field level; this theory is known as dynamical mean-field
theory (DMFT) \cite{DMFT}.

The importance of DMFT is not in the, somewhat academic, limit of infinite dimensions. Rather, DMFT has become an important approximate theory in the finite numbers of dimensions relevant to real materials \cite{DMFT}. It has been found that DMFT captures a great deal of the physics of  strongly correlated electrons. Typically the most important correlations are on-site and therefore are correctly described by DMFT. These include the correlations that are important in metallic magnetism \cite{Kollar} and many other strongly correlated materials \cite{DMFT,JPCMrev}. Cluster extensions to DMFT, such as cellular dynamical mean-field theory  (CDMFT) and the dynamical cluster approximation (DCA), which capture some of the non-local correlations, have led to further insights into strongly correlated materials \cite{CDMFT}. Considerable success has also been achieved by combining DMFT with density functional theory \cite{Kotliar}.

\subsubsection{The Nagaoka point}

The Nagaoka point in the phase diagram of the Hubbard model is the $U \rightarrow \infty$ limit when we add one hole  to a half filled system. Nagaoka rigourously proved \cite{Nagaoka,Tian} that at this point the state which maximises the total spin of the system (i.e., the state with $\langle S_z\rangle=(N-1)/2$, for an $N$ site lattice) is an extremum in energy, i.e., either the ground state or the highest lying excited state. On most bipartite lattices (cf. Fig. \ref{fig:frust}) one finds that this `Nagaoka state' is indeed the ground state \cite{Tian}. However, on frustrated lattices (cf. Fig. \ref{fig:frust}) the Nagaoka state is typically only the ground state for one sign of $t$ \cite{MPM}. 

It is quite straightforward to understand why the Nagaoka state is often the ground state. As we are considering the $U \rightarrow \infty$ limit there will be strictly no double occupation of any sites. One therefore need only consider the subspace of states with no double occupation. As none of these states contain any potential energy (i.e., terms proportional to $U$), the ground state will be the state that minimises the kinetic energy (the term proportional to $t$). Clearly the ground state is the state that maximises the magnitude of the kinetic energy with a negative sign. In the Nagaoka state all of the electrons align, this means that the holon can hop unimpeded by the Pauli exclusion principle, thus maximising the magnitude of the kinetic energy. It is then a simple matter to check whether this is the ground state or the highest lying excited state as we just compare the energy of the Nagaoka state with that of any other state satisfying the constraint of no double occupation.

Nagaoka's rigourous treatment has not been
extended to doping by more than one hole and it remains an
outstanding problem to further understand this interesting
phenomenon, which shares important features with the magnetism observed in the elemental magnets \cite{Kollar} and many strongly correlated materials \cite{MPM}.

\section{The Heisenberg model}\label{sect:Heisenberg}

Like the Stoner ferromagnetism we discussed above in the context of the Hartree-Fock solution for the Hubbard model (section \ref{sect:Stoner}) and Hund's rules (which we will discuss in section \ref{Hund}), the Heisenberg model is an important paradigm for understanding magnetism. The Heisenberg model does not provide a realistic description of the three elemental ferromagnets (Fe, Co and Ni) as they are metals, whereas the Heisenberg model only describes insulators. However, as we will see in section \ref{Hub->Hei}, the Heisenberg model is a good description of Mott insulators such as La$_{2}$CuO$_4$ (the parent compound of the high temperature superconductors) and $\kappa$-(BEDT-TTF)$_2$Cu[N(CN)$_2$]Cl (the parent compound for the organic superconductors). The Heisenberg model also plays an important role in the valence bond theory of the chemical bond \cite{Shaik}.

In the Heisenberg model one assumes that there is a single (unpaired) electron localised at each site, and that the charge cannot move. Therefore, the only degrees of freedom in the Heisenberg model are the spins of each site (the model can also be generalised to spin $>\frac12$).  The Hamiltonian for the Heisenberg model is
\begin{eqnarray}
\hat{\cal H}_\textrm{Heisenberg}=\sum_{ij}J_{ij}\hat{\bf S}_i\cdot\hat{\bf S}_j,
\end{eqnarray}
where $\hat{\bf S}_i=(\hat S^x_i,\hat S^y_i,\hat S^z_i)=\frac12\sum_{\alpha\beta}\hat c_{i\alpha}^\dagger{\vec \sigma}_{\alpha\beta}\hat c_{i\beta}$ is the spin operator on site $i$, $\vec\sigma=(\sigma_x,\sigma_y,\sigma_z)$ is the vector of Pauli matrices, and $J_{ij}$ is the `exchange energy' between sites $i$ and $j$.

\subsection{Two site model: classical solution}\label{sect:2Heisclass}

In the classical Heisenberg model one replaces the spin operator, $\hat{\bf S}_i$, with a classical spin, i.e., a real vector, ${\bf S}_i$. Thus on two sites, with $J_{12}=J$, the energy of the model is
\begin{eqnarray}
E_\textrm{Heisenberg}^\textrm{(2)}=J{\bf S}_1\cdot{\bf S}_2=J|{\bf S}_1||{\bf S}_2|\cos\phi,
\end{eqnarray}
where $\phi$ is the angle between the two spins (vectors). The classical energy is minimised by $\phi =\pi$ for $J>0$ and $\phi =0$ for $J<0$. Thus for $J>0$ the lowest energy solution is for the two spins to point antiparallel (i.e., in opposite directions to one another); we will refer to this as the antiferromagnetic solution. For $J<0$ the lowest energy solution is for the two spins to point parallel to one another; we will refer to this as the ferromagnetic solution. Note that the difference in energy between the antiferromagnetic solution and the ferromagnetic solution is $2J|{\bf S}_1||{\bf S}_2|$; so for $S=|{\bf S}_1|=|{\bf S}_2|=1/2$ the energy difference is $J/2$.

\subsection{Two site model: exact quantum mechanical solution}\label{sect:2Heisquant}

In order to solve the quantum mechanical version of the two site Heisenberg model it is useful to define the spin raising and lowering operators
\begin{subequations}
\begin{eqnarray}
\hat S_i^+&\equiv& \hat S_i^x+i \hat S_i^y=\hat c_{i\uparrow}^\dagger\hat c_{i\downarrow}\\
\hat S_i^-&\equiv& \hat S_i^x-i \hat S_i^y=\hat c_{i\downarrow}^\dagger\hat c_{i\uparrow}.
\end{eqnarray}
\end{subequations}
Let us denote the state with spin up on site $i$ as $|\uparrow_i\rangle$ and the state with spin down on site $i$ as $|\downarrow_i\rangle$. Therefore,  $\hat S_i^+|\uparrow_i\rangle=0$, $\hat S_i^+|\downarrow_i\rangle=|\uparrow_i\rangle$, $\hat S_i^-|\uparrow_i\rangle=|\downarrow_i\rangle$ and $\hat S_i^-|\downarrow_i\rangle=0$. Further, it is straightforward to confirm that
\begin{eqnarray}
\hat {\bf S}_1\cdot \hat {\bf S}_2=\frac12\left(\hat S_i^+ \hat S_j^- + \hat S_i^-\hat S_j^+\right)+ \hat S_i^z\hat S_j^z.
\end{eqnarray}

We now  note that the Hilbert space of the two site Heisenberg model is spanned by four states (the spin on each site may be up or down; in general for an $N$ site Heisenberg model the Hilbert space is $2^N$-dimensional). Further notice that the total spin of the model ($\hat{\bf S}=\hat{\bf S}_1+\hat{\bf S}_2$) commutes with the Hamiltonian, therefore the eigenstates will also be eigenstates of the total spin. Thus the four eigenstates must be a singlet,
\begin{equation}
|\Psi_s\rangle = \frac{1}{\sqrt{2}}\left(|\uparrow_1\rangle|\downarrow_2\rangle-|\downarrow_1\rangle|\uparrow_2\rangle\right)
\equiv
\frac{1}{\sqrt{2}}\left(|\uparrow\downarrow\rangle-|\downarrow\uparrow\rangle\right),
\end{equation}
and a triplet,
\begin{subequations}
\begin{eqnarray}
|\Psi_t^+\rangle &=& |\uparrow_1\rangle|\uparrow_2\rangle
\equiv
|\uparrow\uparrow\rangle\\
|\Psi_t^0\rangle &=& \frac{1}{\sqrt{2}}\left(|\uparrow_1\rangle|\downarrow_2\rangle+|\downarrow_1\rangle|\uparrow_2\rangle\right)\notag\\
&\equiv&
\frac{1}{\sqrt{2}}\left(|\uparrow\downarrow\rangle+|\downarrow\uparrow\rangle\right)\\
|\Psi_t^-\rangle &=& |\downarrow_1\rangle|\downarrow_2\rangle
\equiv
|\downarrow\downarrow\rangle.
\end{eqnarray}
\end{subequations}

It is now straightforward to calculate the total energy of the model for these states,
\begin{eqnarray}
E_s&=&J\langle\Psi_s|\hat{\bf S}_1\cdot\hat{\bf S}_2|\Psi_s\rangle\notag\\
&=&\frac{J}{2}\Big(\langle\uparrow\downarrow|-\langle \downarrow\uparrow|\Big)\left[ \frac12\left(\hat S_i^+ \hat S_j^- + \hat S_i^-\hat S_j^+\right)+ \hat S_i^z\hat S_j^z \right]  \notag\\&&\hspace{5.3cm} \Big(|\uparrow\downarrow\rangle-| \downarrow\uparrow\rangle\Big)\notag\\
&=&-\frac{3J}{4}
\end{eqnarray}
and
\begin{eqnarray}
E_t&=&J\langle\Psi_t^+|\hat{\bf S}_1\cdot\hat{\bf S}_2|\Psi_t^+\rangle \notag\\
&=&J\langle\Psi_t^0|\hat{\bf S}_1\cdot\hat{\bf S}_2|\Psi_t^0\rangle 
=J\langle\Psi_t^-|\hat{\bf S}_1\cdot\hat{\bf S}_2|\Psi_t^-\rangle\notag\\
&=&J\langle\downarrow \downarrow|\left[ \frac12\left(\hat S_i^+ \hat S_j^- + \hat S_i^-\hat S_j^+\right)+ \hat S_i^z\hat S_j^z \right]|\downarrow\downarrow\rangle\notag\\
&=&+\frac{J}{4}
\end{eqnarray}
Thus we find that the singlet-triplet splitting for the quantum mechanical two site Heisenberg model is $J$.

\subsection{The Heisenberg model as an effective low-energy theory of the Hubbard model}\label{Hub->Hei}

Consider a two site Hubbard model with two electrons and $U\gg t$, which is known as the atomic limit. $U\gg t$ implies $\theta\rightarrow0$ in Eq. \ref{exactgs} and that the ground state is the Heitler-London state, which is a singlet. The two other singlet eigenstates are the charge transfer states, which have energy $\sim U$ and so will not participate in any low-energy processes, i.e., will not be involved in the interesting physics or chemistry. Therefore we can `integrate out' the charge transfer states and derive a simpler model with a smaller Hilbert space. A model derived in this manner is known as an `effective low-energy Hamiltonian' (see section \ref{eh_v_se}). In this case we will use second order perturbation theory to derive our effective effective low-energy Hamiltonian.

We start by writing the two site Hubbard model as
\begin{eqnarray}
\hat{\cal H}_\textrm{Hubbard} = U\left(\hat{\cal H}_0 + \frac{t}{U}\hat{\cal H}_1\right),\label{Hubbard-pert}
\end{eqnarray}
where
%\begin{eqnarray}
$\hat{\cal H}_0 = \sum_i \hat n_{i\uparrow} \hat n_{i\downarrow}$
%\end{eqnarray}
and
%\begin{eqnarray}
$\hat{\cal H}_1 = \sum_{\langle ij\rangle\sigma} \hat c_{i\sigma}^\dagger \hat c_{j\sigma}$.
%\end{eqnarray}
Thus it is clear that the small parameter for our perturbation theory is $t/U$. For $t=0$ the ground state is four-fold degenerate; the four states involved being the Heitler-London state and the triplet states. Formally one should therefore use degenerate perturbation theory. But, the perturbation, $\hat{\cal H}_1$, does not connect any of the four ground states, therefore, to second order (which is all we will consider), non-degenerate perturbation theory will yield the same results. As it simplifies the discussion, we will frame our discussion in terms of non-degenerate perturbation theory. It is left as an exercise to the reader to show that on adding the appropriate projection operators (cf. Ref. \onlinecite{Sakurai}) to perform degenerate perturbation theory the result is unchanged.

\begin{figure}
\includegraphics[width=8cm]{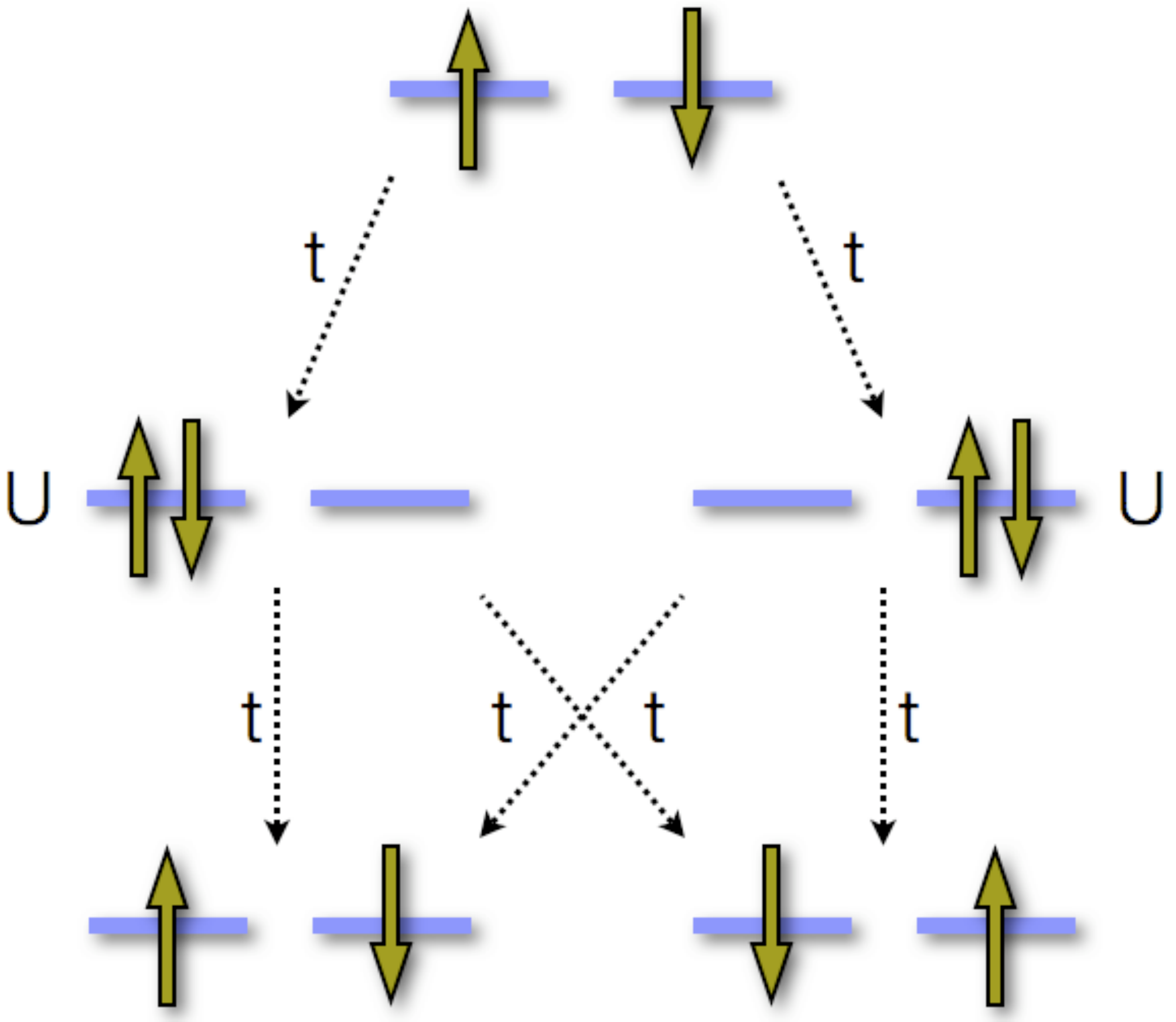}
\caption{Sketch of the superexchange processes that lead to the effective antiferromagnetic Heisenberg coupling between nearest neighbours in the large $U/t$ limit of the half-filled Hubbard model. These processes lower the energy of the singlet state by $4t^2/U$ as their are four paths, the matrix element between the ground state and the intermediate states is $-t$, and the intermediate states are higher in energy by $U$. The energy of the triplet states is unchanged to by perturbations to second order in $t/U$ as the Pauli exclusion principle prevents two electrons with the same spin from occupying the same site.} \label{exchange}
\end{figure}

Consider the related Hamiltonian with $\hat{\cal H}_0\rightarrow\hat{\cal H}_0+\eta|\Psi_\textrm{HL}\rangle\langle\Psi_\textrm{HL}|$ in the limit $\eta\rightarrow0^-$ (i.e., as $\eta$ tends to 0 from below) this has the same properties as Eq. \ref{Hubbard-pert}, except that $|\Psi_\textrm{HL}\rangle$ is the true ground state and we may use non-degenerate perturbation theory. $\langle\Psi_\textrm{HL}|\hat{\cal H}_1|\Psi_\textrm{HL}\rangle=0$ so there is no correction to the ground state energy to first order in $t/U$. The second order change in the ground state energy, $\Delta E^{(2)}$ is given by
\begin{eqnarray}
\Delta E^{(2)}=-\sum_{s\ne\textrm{HL}}\frac{\left|\langle\Psi_\textrm{HL}|\hat{\cal H}_1|\psi_s\rangle\right|^2}{E_s-E_\textrm{HL}},
\end{eqnarray}
where the sum over $s$ runs over all states except the ground state. Note that, as is true in general, the second order contribution to the ground state energy is negative. Evaluating the matrix elements (cf. Fig. \ref{exchange}) one finds that
\begin{eqnarray}
\Delta E^{(2)}=-\frac{4|t|^2}{U}.
\end{eqnarray}
In contrast if we add an infinitesimal term to make one of the triplet states the true ground state, e.g. $0^-|\Psi_1^1\rangle\langle\Psi_1^1|$, we find that the Pauli exclusion principle ensures that $\langle\Psi_1^1|\hat{\cal H}_1|\psi_s\rangle=0$ for all $s$. Thus there is no change in the energy of the triplet state to second order in $t/U$.

Therefore it is clear that for $U/t\rightarrow \infty$ the half-filled Hubbard model reduces to the Heisenberg model, i.e., the eigenstates and energies are the same, if we set $J=4|t|^2/U$. This result is not a special property of the two site model and is true to second order for an arbitrary lattice \cite{Chao} as second order perturbation theory only couples sites $i$ and $j$ if the hopping integral between them is non-zero. For an arbitrary lattice to second order $J_{ij}=4|t_{ij}|^2/U$.

As the Heisenberg model is the large $U/t$ limit of the half-filled Hubbard model electronic correlations are vitally important in the physics of materials described by the Heisenberg model. Therefore weakly correlated theories, such as Hartree-Fock and density functional theory, give qualitatively incorrect results.

\subsection{Frustration: the solution of the three and four site classical Heisenberg models}

\begin{figure}
\includegraphics[width=6cm]{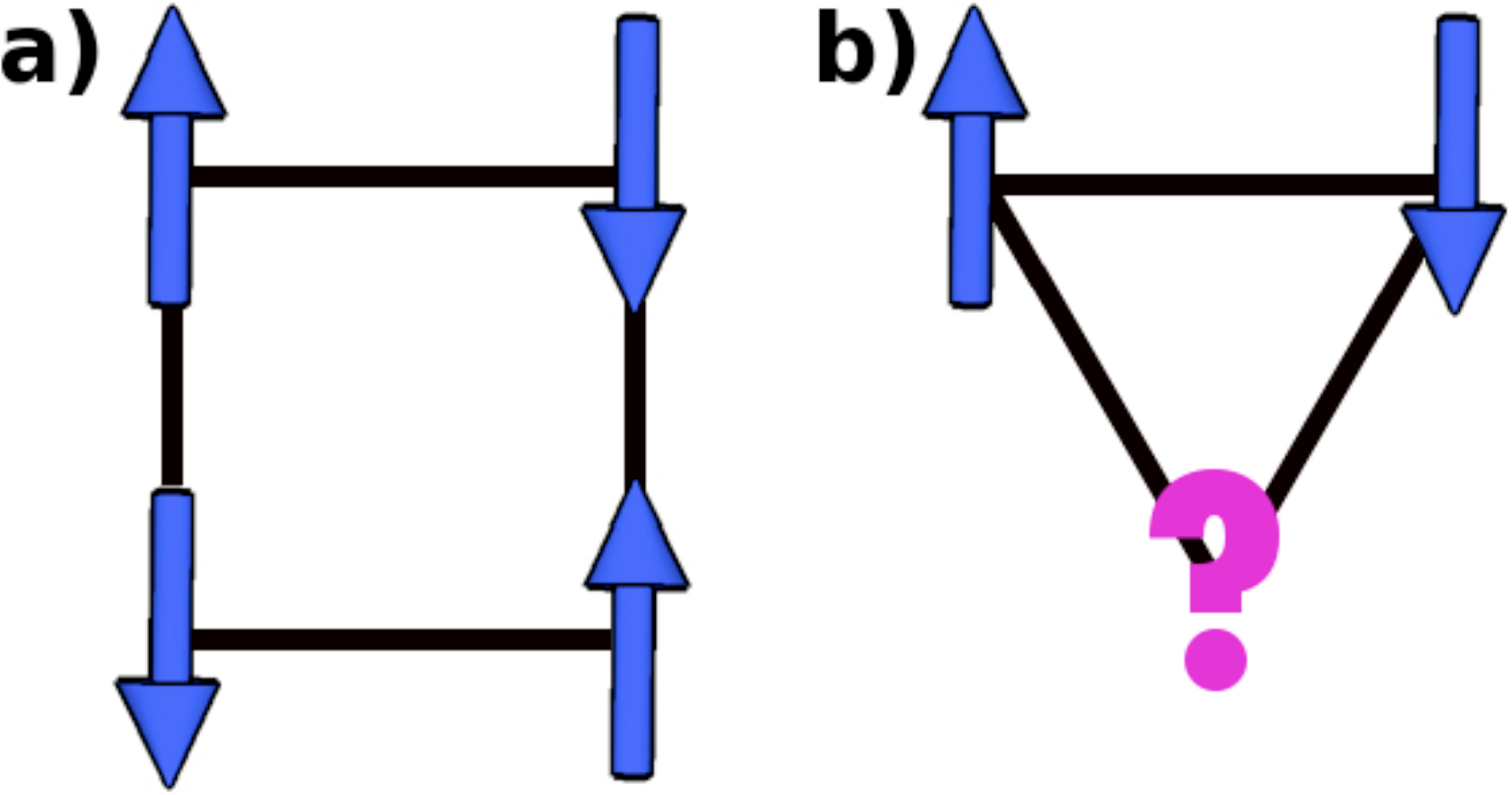}
\caption{Examples of classical spins on (a) a bipartite cluster and (b) a frustrated cluster. On the square cluster (a) one can arrange all of the spins antiferromagnetically, i.e., so that each spin is antiparallel to all of its nearest neighbours. The same is true for the square lattice. This cannot be accomplished on either the triangular cluster (b) or the triangular lattice. In each panel there is an exchange interaction, $J$, between any two spins joined by a line. Modified from Ref. \onlinecite{CiA}.} \label{fig:frust}
\end{figure}

Before considering the classical three site model, let us spend a moment discussing the classical four site model. We assume that the four sites are situated on the vertices of a square and there is an exchange interaction $J$ between nearest neighbours (i.e., along the sides of the square), but no interactions between next nearest neighbours (i.e., along diagonals of the square). The energy of the model is
\begin{eqnarray}
E_\textrm{Heisenberg}^\textrm{(4)}&=&J\sum_{\langle ij\rangle}{\bf S}_i\cdot{\bf S}_j=J\sum_{\langle ij\rangle}|{\bf S}_i||{\bf S}_j|\cos\theta_{ij}\notag\\&=&\frac{J}{4}\sum_{\langle ij\rangle}\cos\theta_{ij},
\end{eqnarray}
where $\theta_{ij}$ is the angle between the spins on sites $i$ and $j$, ${\bf S}_i$ is the spin on the $i^\textrm{th}$ lattice site and in the last equality we have specialised to the case $|{\bf S}_i|=1/2$ for all $i$. Notice that the Hamiltonian is just a sum over the `bonds' (sides of the square).
As for the two site model (section \ref{sect:2Heisclass}), the solution depends on the sign of $J$. For $J<0$ the lowest energy state is ferromagnetic (all of the spins align parallel to one another) and has energy, $E_\textrm{Heisenberg}^\textrm{(4)}=J$. For $J>0$ the lowest energy state is antiferromagnetic (each spin aligns antiparallel to its nearest neighbour; see Fig. \ref{fig:frust}a). Thus the four site cluster is split into two `sublattices' with all of the spins parallel to one another within the same sublattice and antiparallel to spins on the other sublattice. The antiferromagnetic arrangement of spins therefore has energy  $E_\textrm{Heisenberg}^\textrm{(4)}=-J$. Thus we find that the energy difference between the ferromagnetic and antiferromagnetic arrangements is $2J$.

A lattice that can be split, as described above, into two sublattices such that all nearest neighbours are on different sublattices is referred to as bipartite. Both the four site (square) and two site lattices are bipartite. For bipartite lattices the energy difference between the ferromagnetic and antiferromagnetic arrangements is $JNz/4$, where $z$ is the coordination number of  the lattice and $N$ is the number of lattice sites. This is because the energy of each bond can be optimised regardless of what happens to other bonds for either sign of $J$.
It can be seen from Fig. \ref{fig:frust} that the triangular lattice is not bipartite; this leads to significant differences in its physics. 

Before analysing this model mathematically let us consider some of those differences. Clearly for $J<0$ it is straightforward to arrange the spins ferromagnetically. Further, as the energy of each of the three bonds will be optimised in this arrangement we expect the total energy of this state to be $E_\textrm{Heisenberg}^\textrm{(3)}=3J/4$ for $S=1/2$. But, as is shown in Fig. \ref{fig:frust}b, one cannot arrange three spins antiferromagnetically on a triangular lattice. Thus, for $J>0$, we cannot optimise the energy of each bond individually. When this is the case one says that the lattice is `frustrated'. For a frustrated lattice with $S=1/2$ we expect the solution for $J>0$ to have energy $E_\textrm{Heisenberg}^\textrm{(3)}>-3J/4$ and thus one expects  the difference in energy between this state and the ferromagnetic state to be $<JNz/4$.
The concept of frustration can also be generalised to itinerant systems where a similar reduction in the bandwidth of the itinerant electrons is found \cite{MPM}.

Having outlined our expectations, let us now consider the three site Heisenberg model more carefully. The energy is given by
\begin{eqnarray}
E_\textrm{Heisenberg}^\textrm{(3)}=J\sum_{\langle ij\rangle}{\bf S}_i\cdot{\bf S}_j.
\end{eqnarray}
Without loss of generality we can choose ${\bf S}_1=S_1(1,0,0)$; ${\bf S}_2=S_2(\cos\phi_2,\sin\phi_2,0)$; and ${\bf S}_3=S_3(\cos\theta_3\cos\phi_3,\cos\theta_3\sin\phi_3,\sin\theta_3)$. Thus, for $S_1=S_2=S_3=1/2$,
\begin{eqnarray}
E_\textrm{Heisenberg}^\textrm{(3)}&=&\\ \notag&&\hspace{-1cm}\frac{J}4\Big[ \cos\phi_2 + \cos\theta_3\cos(\phi_2-\phi_3) + \cos\theta_3\cos\phi_3 \Big].
\end{eqnarray}
Physically we seek the minimum energy, which yields the conditions
\begin{subequations}\label{constr}
\begin{eqnarray}
\frac{\partial E_\textrm{Heisenberg}^\textrm{(3)}}{\partial \theta_3}=\frac{J}4\sin\theta_3\Big[  \cos(\phi_2-\phi_3) + \cos\phi_3 \Big]=0, \notag\\\\
\frac{\partial E_\textrm{Heisenberg}^\textrm{(3)}}{\partial \phi_3}=\frac{J}4\cos\theta_3\Big[  \sin(\phi_2-\phi_3) - \sin\phi_3 \Big]=0 \notag\\
\end{eqnarray}
and
\begin{eqnarray}
\frac{\partial E_\textrm{Heisenberg}^\textrm{(3)}}{\partial \phi_2}=-\frac{J}4\Big[  \cos\theta_3\sin(\phi_2-\phi_3) + \sin\phi_2 \Big]=0.\notag\\
\end{eqnarray}
\end{subequations}
For $J<0$ the global minimum is, unsurprisingly, $\theta_3=\phi_2=\phi_3=0$, i.e., ferromagnetism. The energy of the ferromagnetic state is $3J/4$. For $J>0$ there are several degenerate minima, which all show the same physics. For simplicity we will just consider the minimum  $\theta_3=0$, $\phi_2={2\pi}/{3}$, and $\phi_3={4\pi}/{3}$. In this solution each of the spins points 120$^\textrm o$ away from each of the other spins, hence this is known as the 120$^\textrm o$ state.
It is left as an exercise to the reader to identify the other solutions of Eqs. \ref{constr}, to show that there are none with lower energy than those discussed above, and to show that all of the degenerate solutions are physically equivalent. The energy of the 120$^\textrm o$ state is $-3J/8$ and hence the energy difference between the ferromagnetic state and 120$^\textrm o$ state is just $9J/8$, i.e., less than we would expect ($JNz/4=3J/2$ for $N=3$, $z=2$) for a bipartite lattice.

\subsection{Three site model: exact quantum mechanical solution}

Group theory, the mathematics of symmetry, allows one to solve the quantum spin $1/2$ three site Heisenberg model straightforwardly. Unfortunately space does not permit an introduction to the relevant group theory. Therefore the reader who is not familiar with the mathematics is advised either to refer to one of the many excellent textbooks on the subject (e.g., Refs. \onlinecite{Tinkham,Lax}) or, failing that, to simply check that the wavefunctions derived by the group theoretic arguments below are indeed eigenstates.

The Hamiltonian is
\begin{eqnarray}
\hat{\cal H}_\textrm{Heisenberg}^\textrm{(3)}&=&J\sum_{\langle ij\rangle}\hat{\bf S}_i\cdot\hat{\bf S}_j \notag\\ &=& J\sum_{\langle ij\rangle}\left[ \frac12\left( \hat S_i^+\hat S_j^- + \hat S_i^-\hat S_j^+\right) + \hat S_i^z\hat S_j^z\right]\hspace{0.5cm}
\end{eqnarray}
We begin by noting that $2\otimes2\otimes2=2\oplus2\oplus4$,\footnote{In this notation the integers are the degeneracy of the state.} i.e., a system formed from three spin $1/2$ particles will have two doublets (with two-fold degenerate spin $1/2$ eigenstates) and one quadruplet (with four-fold degenerate spin $3/2$ eigenstates).

There are only four possible quadruplet states consistent with the $C_{3}$ point group symmetry\footnote{One might, reasonably, take the view that the model has either $D_{3h}$ or $C_{3v}$. In fact the arguments in this section go through almost identically for either of these symmetries (with appropriate changes in notation) due to the homomorphisms from these groups to $C_3$. We will use $C_3$ notation for simplicity.} of the model. Each of these belong to the $A$ irreducible representation of $C_3$. They are
\begin{subequations}
\begin{eqnarray}
\left|\psi_{3/2}^{3/2}\right\rangle &=& \left|\uparrow\uparrow\uparrow\right\rangle \notag\\
\left|\psi_{3/2}^{1/2}\right\rangle &=& \frac1{\sqrt{3}}\left(\left| \downarrow\uparrow\uparrow\right\rangle +\left|\uparrow \downarrow\uparrow\right\rangle +\left|\uparrow\uparrow \downarrow\right\rangle\right) \notag\\
\left|\psi_{3/2}^{-1/2}\right\rangle &=& \frac1{\sqrt{3}}\left(\left| \uparrow \downarrow \downarrow\right\rangle + \left| \downarrow \uparrow \downarrow\right\rangle  + \left| \downarrow \downarrow \uparrow\right\rangle \right) \notag  \\
\left|\psi_{3/2}^{-3/2}\right\rangle &=& \left| \downarrow \downarrow \downarrow\right\rangle,\notag
\end{eqnarray}
\end{subequations}
where $|\alpha\beta\gamma\rangle=|S_1^z,S_2^z,S_3^z\rangle$ and $\alpha$, $\beta$ and $\gamma=\uparrow$ or $\downarrow$. Each of these states have energy $E=3J/4$ and they are the (degenerate) ground states for $J<0$.

We are left with the four doublet states. These belong to the two dimensional $E$ irreducible representation of $C_3$ and, as the Hamiltonian is time reversal symmetric, all four doublet states are degenerate. Explicitly the states are
\begin{subequations}
\begin{eqnarray}
\left|\psi_{1/2}^{1/2}\right\rangle &=& \frac1{\sqrt{3}}\left(\left| \downarrow\uparrow\uparrow\right\rangle +e^{i2\pi/3}\left|\uparrow \downarrow\uparrow\right\rangle +e^{-i2\pi/3}\left|\uparrow\uparrow \downarrow\right\rangle\right) \notag \\
\left|\psi_{1/2}^{-1/2}\right\rangle &=& \frac1{\sqrt{3}}\left(\left| \uparrow \downarrow \downarrow\right\rangle + e^{i2\pi/3}\left| \downarrow \uparrow \downarrow\right\rangle  +e^{-i2\pi/3}\left| \downarrow \downarrow \uparrow\right\rangle \right) \notag \\
\left|\tilde\psi_{1/2}^{1/2}\right\rangle &=& \frac1{\sqrt{3}}\left(\left| \downarrow\uparrow\uparrow\right\rangle +e^{-i2\pi/3}\left|\uparrow \downarrow\uparrow\right\rangle +e^{i2\pi/3}\left|\uparrow\uparrow \downarrow\right\rangle\right) \notag\\
\left| \tilde\psi_{1/2}^{-1/2}\right\rangle &=& \frac1{\sqrt{3}}\left(\left| \uparrow \downarrow \downarrow\right\rangle + e^{-i2\pi/3}\left| \downarrow \uparrow \downarrow\right\rangle  +e^{i2\pi/3}\left| \downarrow \downarrow \uparrow\right\rangle \right) \notag
\end{eqnarray}
\end{subequations}
Each of these states have energy $E=-5J/4$ and they are the (degenerate) ground states for $J>0$. Thus, the energy difference between the highest spin state and the lowest spin state is $2J$. From the solution to the two site model (section \ref{sect:2Heisquant}) we expected each  of the three bonds to yield an energy difference of $J$ between the lowest and highest spin states. Thus the frustration has a similar effect in both the quantum and classical models, i.e. frustration lowers the energy difference between the highest spin and lowest spin states.

\subsection{The Heisenberg model on infinite lattices}

The Heisenberg model can be solved exactly in one dimension, and we will discuss this further below, but not in any other finite dimension. However, in more than one dimension physics of the Heisenberg model is typically very different from that in one dimension, so we will begin by discussing, qualitatively, the semi-classical spin wave approximation for the Heisenberg model, which captures many important aspects of magnetism. A quantitative formulation of this theory can be found in many textbooks, e.g. Refs. \onlinecite{Rossler,A&M}.

\begin{figure*}
\includegraphics[width=12cm]{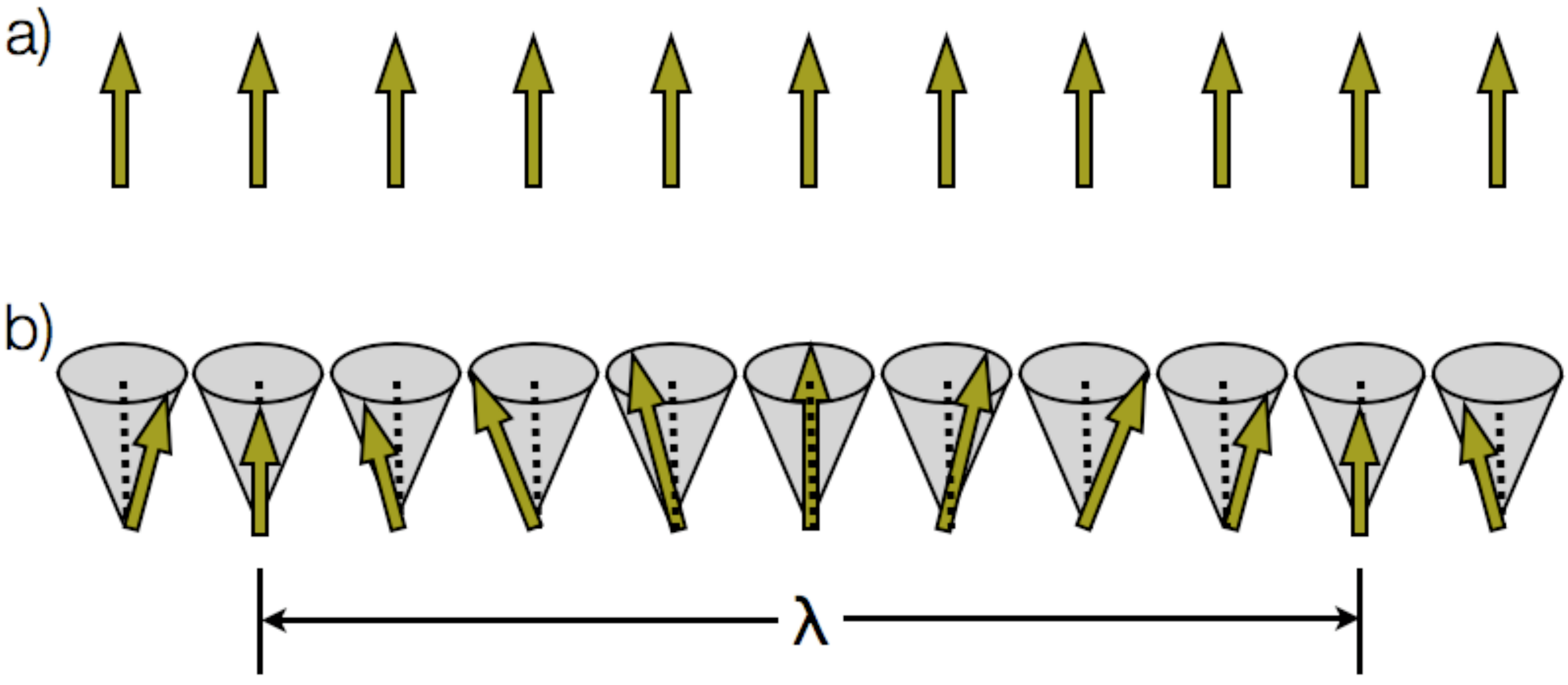}
\caption{Sketches of (a) the classical ground state of a ferromagnetic Heisenberg chain and (b) a spin wave excitation in the same model.} \label{fig:spin-wave}
\end{figure*}

In inelastic neutron scattering experiments  a neutron may have its spin flipped by its interaction with the magnet; this causes a spin 1 excitation in the material. The conceptually simplest spin 1 excitation would be to flip one (spin-$\frac12$) spin; in a one dimensional ferromagnetic Heisenberg model this state has energy $2 |J|$ greater than the ground state. However, a much lower energy excitation is a `spin wave', where each spin is rotated a small amount from its nearest neighbours (cf. Fig. \ref{fig:spin-wave}). In a one dimensional ferromagnetic Heisenberg model spin waves have excitation energies of $\hbar\omega_{k}=2|J|[1-\cos(ka)]$, where $a$ is the lattice constant \cite{Rossler}. Note, in particular, that the excitation energy vanishes for long wavelength (small $k$) spin waves. This spin wave spectrum can indeed be observed directly in neutron scattering experiments from suitable materials \cite{Brookhouse}, and the spectrum is found to be in good agreement with the predictions of the semi-classical theory in many materials. One can also quantise the semi-classical theory by making a `Holstein-Primakoff' transformation \cite{Rossler}. This yields a description of the low-energy physics of the Heisenberg model in terms of non-interacting bosons, which are known as `magnons' and have the same dispersion relation as the classical spin waves. Similar spin wave and magnon descriptions can be straightforwardly constructed for the antiferromagnetic Heisenberg model \cite{Rossler}.

The effective low-energy physics of the one dimensional Heisenberg model is, as noted above, rather different from the semi-classical approximation. To understand this it is helpful to think of the Heisenberg model as a special case of the `XXZ model':
\begin{equation}
H_\textrm{XXZ}=J_{xy}\sum_{i}\left( S_i^{x}S_{i+1}^{x} + S_i^{y}S_{i+1}^{y}\right) + J_{z}\sum_{i} S_i^{z}S_{i+1}^{z}, \end{equation}
which reduces to the Heisenberg model for $J_{xy}=J_{z}=J$. For $J_{z}< J_{xy}<0$ the model displays an exotic quantum phase known as a Luttinger liquid. (At $J_{xy}=J_{z}$ the model undergoes a quantum phase transition from the Luttinger liquid to an ordered phase \cite{Zaliznyak}).

\begin{figure}
\includegraphics[width=8cm]{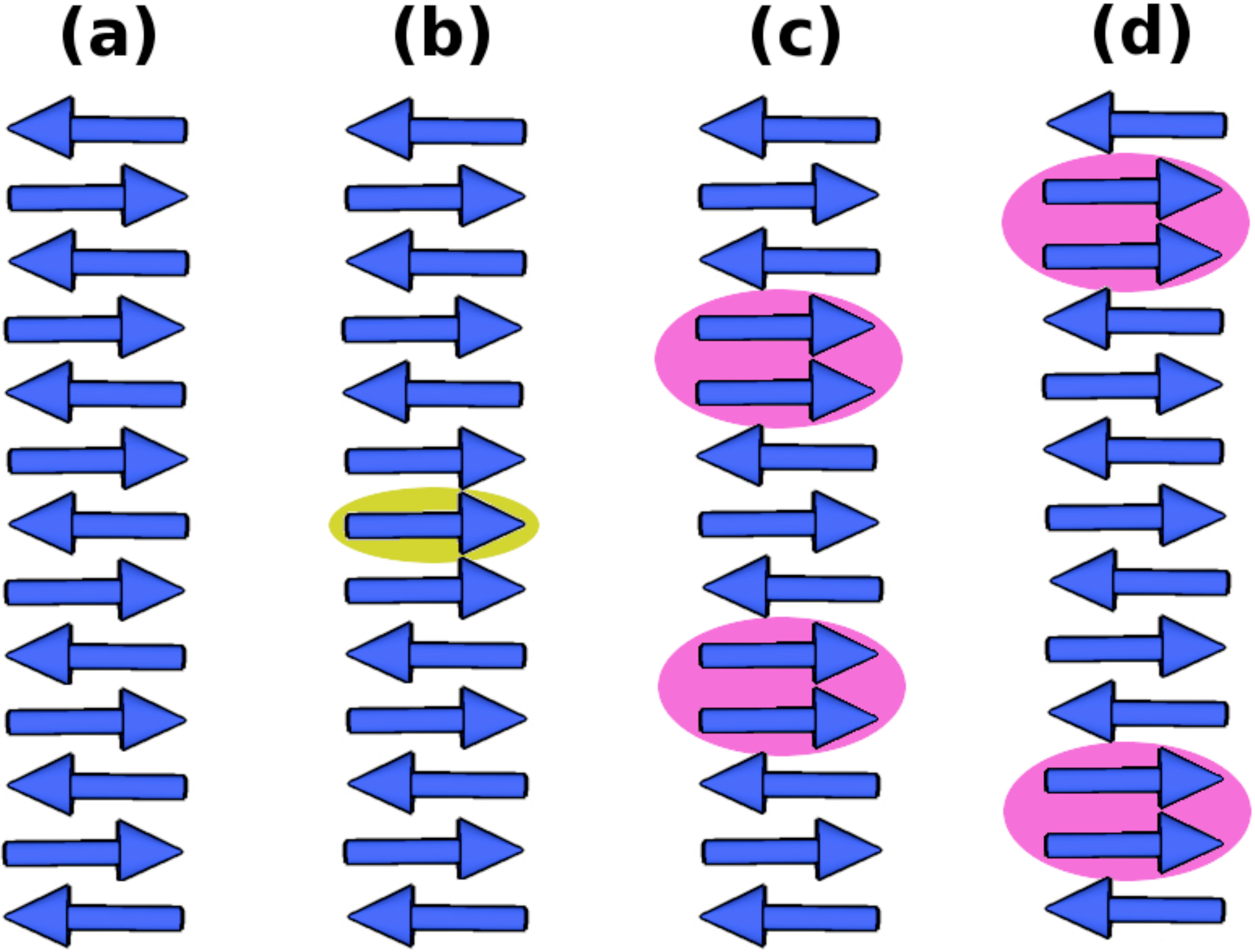}
\caption{Sketch of spinons in a 1D spin chain. (a) Local antiferromagnetic correlations. (b) A neutron scattering off the chain causes one spin (circled) to flip. (c, d) Spontaneous flips of adjacent pairs of spins due to quantum fluctuations allow the spinons (circled) to propagate independently. A key open question is: can this free propagation occur in 2D, or do interactions confine the spinons? Modified from Ref. \onlinecite{CiA}.} \label{fig:spinons}
\end{figure}

On the energy scales relevant to chemistry one does not need to worry about the fact that protons and neutrons are made up of smaller particles (quarks). This is because the quarks are ÔconfinedÕ within the proton/neutron \cite{confinement}. Similarly, in a normal magnet it does not matter that the material is made up of spin-$\frac12$ particles (electrons). As described above, on the energy scales relevant to magnets the spins are confined into spin one particles called magnons. However, magnons can be described in terms of two spin-$\frac12$ ÒspinonsÓ, which are confined inside the magnon. In the Luttinger liquid the  spinons are ÒdeconfinedÓ, i.e., the spinons can move independently of one another (cf. Fig. \ref{fig:spinons}). As the magnon is a composite particle ÔmadeÕ from two spinons this is often referred to as fractionalisation. A key prediction of this theory is that the spinons display a continuum of excitations in neutron scattering experiments (as opposed to the sharp dispersion predicted for magnons). The two spinon continuum has indeed been observed in a number of quasi-one dimensional materials. \cite{Lake.Coldea}

An open research question is: does fractionalisation occur in higher dimensions? Because of the success of spin wave theory (which implies confined spinons) in describing magnetically ordered materials one does not expect fractionalisation in materials with magnetic order. Therefore, one would like to investigate quasi-two or three dimensional materials whose low-energy physics is described by spin Hamiltonians (such as the Heisenberg model), but that do not order magnetically even at the lowest temperatures. Such materials are collectively referred to as spin liquids. There is a long history of theoretical contemplation of spin liquids, which suggests that frustrated magnets and insulating systems near to the Mott transition are strong candidates to display spin liquid physics. However, evidence for real materials with spin liquid ground states has been scarce until very recently, \cite{Lee} but there is now evidence for spin liquids in the triangular lattice compound $\kappa$-(BEDT-TTF)$_{2}$Cu(CN)$_3$ \cite{Shimuzu,JPCMrev}, the kagome lattice (cf. Fig. \ref{lattices}) compound ZnCu$_3$(OH)$_6$Cl$_2$  \cite{Helton} and the hyperkagome lattice compound Na$_4$Ir$_3$O$_8$ \cite{Okamoto}. It remains to be seen whether any of these materials support fractionalised excitations.

\section{Other effective low-energy Hamiltonians for correlated electrons}\label{sect:other}

\subsection{Complete neglect of differential overlap, the Pariser-Parr-Pople model \& extended Hubbard models}

We now consider another model for which the quantum chemistry and condensed matter physics communities have different names. These models belong to class of models known as complete neglect of differential overlap (CNDO). For a pair of orthogonal states, $\phi(x)$ and $\psi(x)$, the integral over all space of the overlap of the two wavefunctions vansishes, i.e., $\int_{-\infty}^\infty\phi(x)\psi(x)dx=0$. If the differential overlap vanishes then the overlap of the two wavefunctions vanishes at every point in space, i.e., $\lim_{\delta\rightarrow0}\int_{x_0}^{x_0+\delta}\phi(x)\psi(x)dx=0$ for all $x_0$. The CNDO approximation is simply to assume that the differential overlap between all basis states is negligible. Thus CNDO implies that $V_{ijkl}=V_{iikk}\delta_{ij}\delta_{kl}$ (cf. section \ref{2lm}). Thus the general CNDO Hamiltonian is
\begin{equation}
\hat{\cal H}_\textrm{CNDO}=-\sum_{ij\sigma}t_{ij}\hat c_{i\sigma}^\dagger \hat c_{j\sigma} + \sum_{ij\sigma\sigma'}V_{ij}\hat n_{i\sigma}\hat n_{j\sigma'},
\end{equation}
where $V_{ij}\equiv V_{iijj}$ and the number operator $\hat n_{i\sigma}\equiv\hat c_{i\sigma}^\dagger\hat c_{i\sigma}$.
The Pariser-Parr-Pople (PPP) model is the CNDO approximation in a basis that only includes the $\pi$ electrons. Often a H\"uckel-like notation is used with $V_{ij}=\gamma_{ij}$, thus
\begin{equation}
\hat{\cal H}_\textrm{PPP}=\sum_{i\sigma}\alpha_{i}\hat c_{i\sigma}^\dagger \hat c_{i\sigma} +\sum_{ij\sigma}\beta_{ij}\hat c_{i\sigma}^\dagger \hat c_{j\sigma} + \sum_{ij\sigma\sigma'}\gamma_{ij}\hat n_{i\sigma}\hat n_{j\sigma'}.
\end{equation}

The extended Hubbard model, as with the plain Hubbard model, is typically studied in a basis with one orbital per site. Further, one often makes the approximation that $V_{ii}=U$, $V_{ij}=V$ if $i$ and $j$ are nearest neighbours and $V_{ij}=0$ otherwise. This yields
\begin{equation}
\hat{\cal H}_\textrm{eH}=-\sum_{\langle ij\rangle\sigma}t_{ij}\hat c_{i\sigma}^\dagger \hat c_{j\sigma} + U\sum_{i}\hat n_{i\uparrow}\hat n_{i\downarrow}+V\sum_{\langle ij\rangle\sigma\sigma'}\hat n_{i\sigma}\hat n_{j\sigma'}.
\end{equation}

%\subsubsection{Beyond CNDO: the most general two site extended Hubbard model}

One can, of course, go beyond CNDO. The most general possible model for two identical sites with a single orbital per site is
\begin{eqnarray}
\hat{\cal H}_\textrm{eH2}&=&-\sum_{\sigma}\left[ t - X\left( \hat n_{1\overline\sigma} + \hat n_{2\overline\sigma} \right) \right] \left(\hat c_{1\sigma}^\dagger \hat c_{2\sigma} + \hat c_{2\sigma}^\dagger \hat c_{1\sigma} \right) \notag\\&&
+ U\sum_{i}\hat n_{i\uparrow}\hat n_{i\downarrow} 
+V\hat n_{1}\hat n_{2}
+ J \hat{\bf S}_1\cdot\hat{\bf S}_2  \notag\\&& + P\left( \hat c_{1\uparrow}^\dagger \hat c_{1\downarrow}^\dagger \hat c_{2\uparrow} \hat c_{2\downarrow} + \hat c_{2\uparrow}^\dagger \hat c_{2\downarrow}^\dagger \hat c_{1\uparrow} \hat c_{1\downarrow} \right),
\end{eqnarray}
where $\hat n_i=\sum_\sigma\hat n_{i\sigma}$, $\hat{\bf S}_i=\sum_{\alpha\beta}\hat c_{i\alpha}^\dagger\vec\sigma_{\alpha\beta}\hat c_{i\beta}$, $\vec\sigma_{\alpha\beta}$ is the vector of Pauli matrices, $J$ is the direct exchange interaction, $X$ is the correlated hopping amplitude, and $P$ is the pair hopping amplitude.

\subsection{Larger basis sets and Hund's rules}\label{Hund}

Thus far we have focused mainly on models with one orbital per site. Often this is not appropriate, for example, if one were interested in chemical bonding or materials containing transition metals. Many of the models discussed in these notes can be straightforwardly extended to include more than one orbital per site. However, while writing down models with more than one orbital per site is not difficult, these models do contain significant additional physics. Some of the most important effects are known as Hund's rules \cite{Fulde}. These rules have important experimental consequences from atomic physics to biology. In order to examine Hund's rules let us consider the atomic limit ($t=0$) of an extended Hubbard model with two electrons in two orbitals per site:
\begin{equation}
\hat{\cal H}_\textrm{eH1s2o}= U\sum_{\mu}\hat n_{\mu\uparrow}\hat n_{\mu\downarrow}
+V_0\hat n_{1}\hat n_{2}
+ J_H \hat{\bf S}_1\cdot\hat{\bf S}_2,
\end{equation}
where $\mu=1$ or 2 labels the orbitals, $\hat n_{\mu\sigma}=\hat c_{\mu\sigma}^\dagger\hat c_{\mu\sigma}$, $\hat n_{\mu}=\sum_\sigma\hat n_{\mu\sigma}$, $\hat{\bf S}_\mu =\sum_{\alpha\beta}\hat c_{\mu\alpha}^\dagger\vec\sigma_{\alpha\beta}\hat c_{\mu\beta}$, $U$ is the Coulomb repulsion between two electrons in the same orbital, $V_0$ is the Coulomb repulsion between two electrons in different orbitals, and $J_H$ is the `Hund's rule coupling' between electrons in different orbitals. Notice that the Hund's rule coupling is an exchange interaction between orbitals. Further, if we compare the Hamiltonian with the definition given in Eq. \ref{eqn:V} we find that
\begin{eqnarray}
-J_{H}&=& \int d^3{\mathbf r}_1  \int d^3{\mathbf r}_2 \, \phi_1^*({\mathbf r}_1) \phi_2({\mathbf r}_1)  \notag\\&&\hspace{2.3cm} \times V({\mathbf r}_1-{\mathbf r}_2)  \phi_2({\mathbf r}_2)^* \phi_1({\mathbf r}_2)  \notag\\
&\sim& \int d^3{\mathbf r}_1  \int d^3{\mathbf r}_2 \, \left|\phi_1({\mathbf r}_1)\right|^2 V({\mathbf r}_1-{\mathbf r}_2) \left| \phi_2({\mathbf r}_2)\right|^2 \notag\\&\geq &0.
\end{eqnarray}
as $V({\mathbf r}_1-{\mathbf r}_2)$ is positive semidefinite. Therefore, typically, $J_H<0$, i.e., the Hund's rule coupling favours the parallel alignment of the spins in a half-filled system.

$U$ is the largest energy scale in the problem, so, for simplicity, let us consider the case $U\rightarrow\infty$. For $J_H=0$ there are four degenerate ground states: a singlet, $\frac1{\sqrt{2}}(|\uparrow\downarrow\rangle-|\downarrow\uparrow\rangle)$ (where the first arrow refers to the spin of the electron in orbital 1 and the second arrow refers to the spin in orbital 2), and a triplet,  $|\uparrow\uparrow\rangle$, $|\downarrow\downarrow\rangle$ and $\frac1{\sqrt{2}}(|\uparrow\downarrow\rangle-|\downarrow\uparrow\rangle)$. But, for $J>0$ the energy of the triplet states is $J_H$ lower than that of the singlet state. Indeed spin symmetry implies that even if we relax the condition $U\rightarrow\infty$ the triplet state remains lower in energy than the singlet state as physically we require $U>J_H$. One can repeat this argument for
 any number of electrons in any number of orbitals and one always finds that the highest spin state has the lowest energy.  However, if one studies models with more than one site and moves away from the atomic limit ($t=0$) one finds that there is a subtle competition between the kinetic (hopping) term and the Hund's rule coupling which means that the high spin state is not always the lowest energy state. Many such interesting effects can be understood on the basis of a two site generalisation of this two orbital model \cite{Oles}.
% This leads to a great deal of interesting physics, chemistry and even biology. For example, the binding of oxygen to haemoglobin effectively changes the ratio $t/J_H$ causing the Fe in haemoglobin to change from a high spin to a low spin state. This causes of the colour difference between oxygenated (red) and deoxygenated (blue) blood.

\subsection{The ionic Hubbard model}\label{sect:ionic}

Thus far we have assumed that all sites are identical. Of course, this is not always true in real materials. In a compound more than one species of atom may contribute to the low-energy physics \cite{Sarma} or different atoms of the same species may be found at crystallographical distinct sites \cite{MPM,MPM2}. A simple model that describes this situation is the ionic Hubbard model:
\begin{eqnarray}
\hat{\cal H}_\textrm{iH}=-t\sum_{\langle ij \rangle\sigma} \hat c_{i\sigma}^\dagger \hat c_{j\sigma} + U\sum_i \hat n_{i\uparrow}\hat n_{i\downarrow} + \sum_{i\sigma}\epsilon_i\hat n_{i\sigma}
\end{eqnarray}
where $\epsilon_i=t_{ii}$ is the site energy, which will be taken to be different on different sites. Note that in the ionic Hubbard model all sites are assumed to have the same $U$.

An important application of the ionic Hubbard model is in describing transition metal oxides \cite{Sarma}. Typically $\epsilon_i$ is larger on the transition metal site than on the oxygen site, therefore the oxygen orbitals are nearly filled. This means that there is a low hole density in the oxygen orbitals and, hence, that electronic correlations are less important for the electrons in the oxygen orbitals than for electrons in transition metal orbitals. If the difference between $\epsilon_i$ on the oxygen sites and $\epsilon_i$ on the transition metal sites  is large enough then the oxygen orbitals are completely filled in all low-energy states and therefore need not feature in the low-energy description of the material. However, just because the oxygen orbitals do not appear explicitly in the effective low-energy description of the material, does not mean that the oxygen does not have a profound effect on the low-energy physics.

\begin{figure}
\includegraphics[width=8cm]{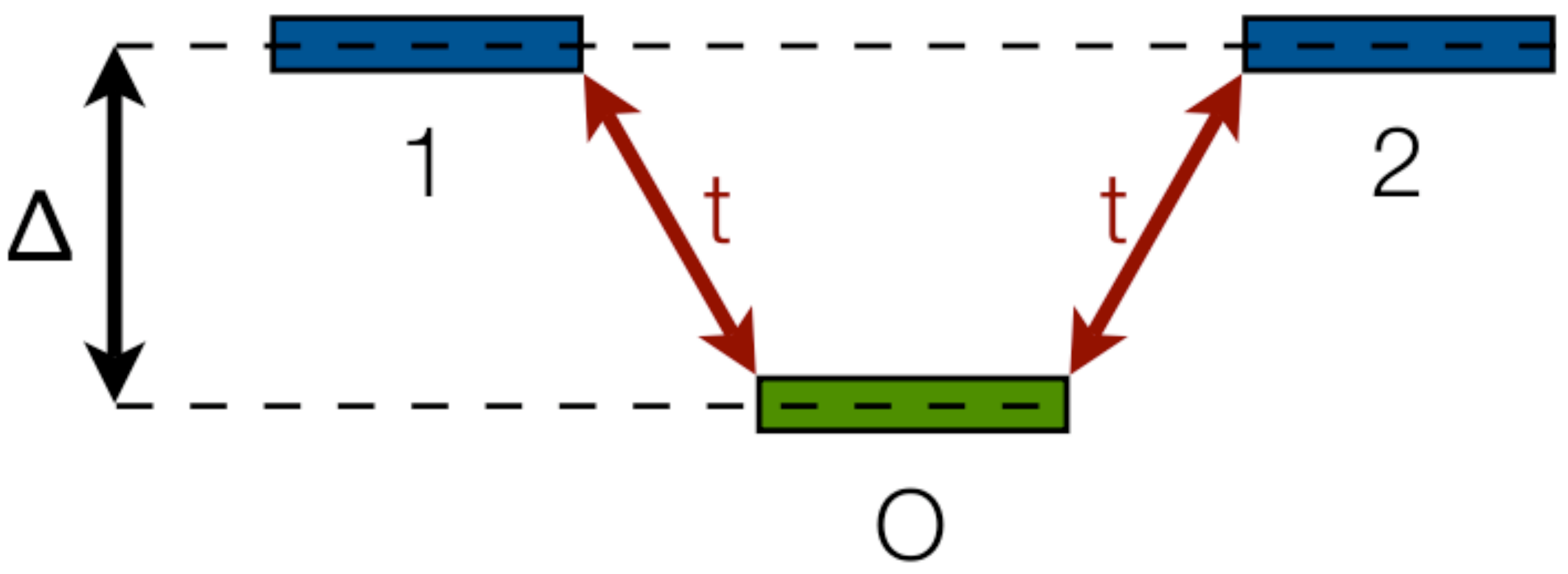}
\caption{Sketch of the toy model for a transition metal oxide, Hamiltonian \ref{eqn:ionicH}, with two transition metal sites (1 and 2) and a single oxygen site (O).} \label{fig:ionicH}
\end{figure}

\begin{figure}
\includegraphics[width=6cm]{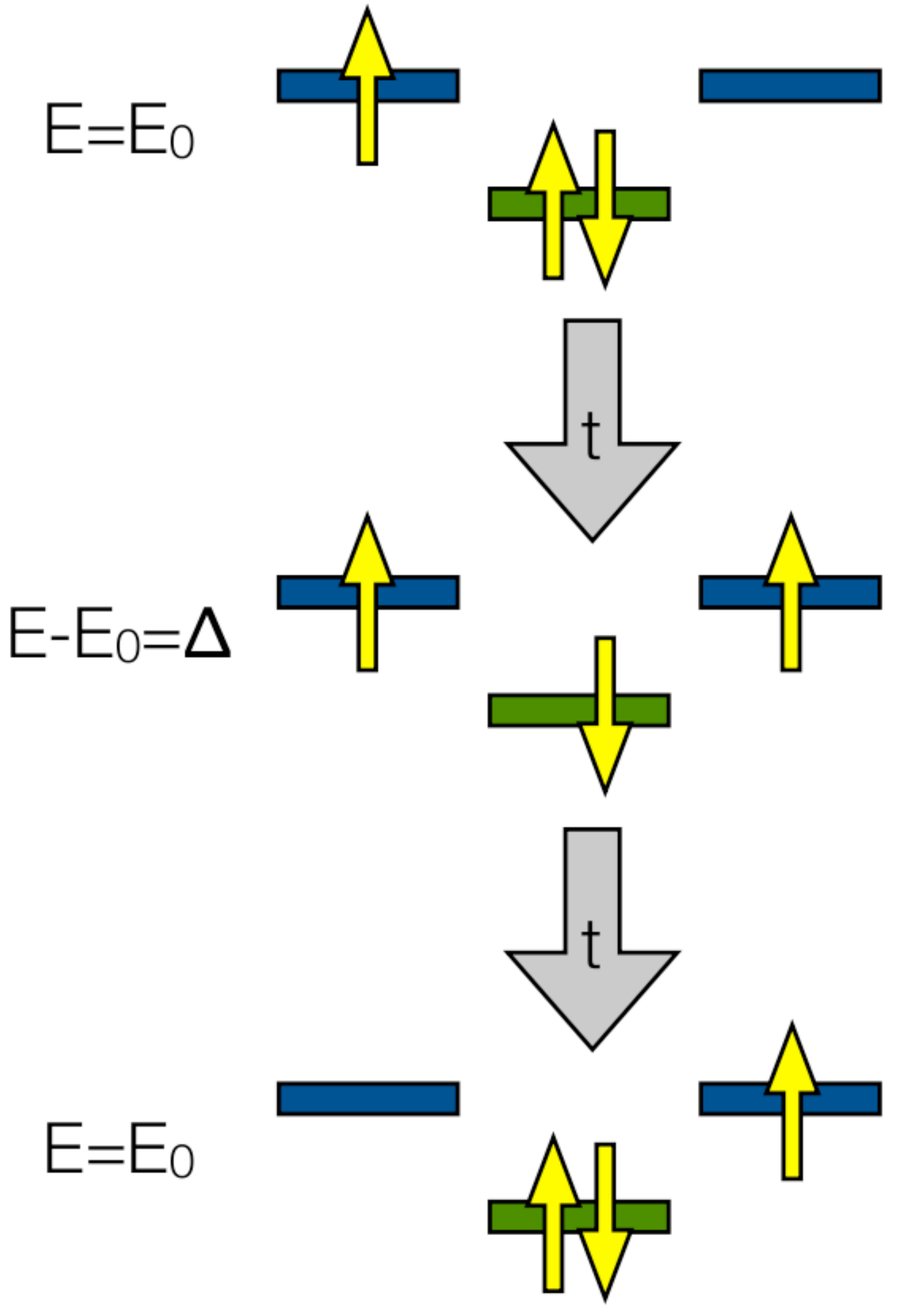}
\caption{Sketch of the processes described by Hamiltonian \ref{eqn:ionicH} that give rise to the effective hopping integral between the two transition metal atom sites.} \label{fig:ionicP}
\end{figure}

To see this consider a toy model with two metal sites (labelled 1 and 2) and one oxygen site (labelled O), whose Hamiltonian is 
\begin{eqnarray}
\hat{\cal H}_\textrm{iH3}&=&-t\sum_{\sigma} \left(\hat c_{1\sigma}^\dagger \hat c_{O\sigma} + \hat c_{O\sigma}^\dagger \hat c_{1\sigma} + \hat c_{2\sigma}^\dagger \hat c_{O\sigma} + \hat c_{O\sigma}^\dagger \hat c_{2\sigma}\right) \notag\\&& + \sum_{i\sigma}\frac{\Delta}{2}\left(\hat n_{1\sigma}+\hat n_{2\sigma}-\hat n_{O\sigma}\right) \label{eqn:ionicH}
\end{eqnarray}
as sketched in Fig. \ref{fig:ionicH}, which is just the ionic Hubbard model with $U=0$ and $\Delta=\epsilon_1-\epsilon_O=\epsilon_2-\epsilon_O>0$. With three electrons in the system and $t=0$ the ground state is four-fold degenerate, the ground states have two electrons on the O atom and the other electron on one of the metal atoms. If we now consider  finite, but small, $t\ll\Delta$ we can construct a perturbation theory in $t/\Delta$. One finds that there is a splitting between the bonding, $\frac{1}{\sqrt 2}(\hat c_{1\sigma}^\dagger+\hat c_{2\sigma}^\dagger)\hat c_{O\uparrow}^\dagger\hat c_{O\downarrow}^\dagger|0\rangle$ and antibonding, $\frac{1}{\sqrt 2}(\hat c_{1\sigma}^\dagger-\hat c_{2\sigma}^\dagger)\hat c_{O\uparrow}^\dagger\hat c_{O\downarrow}^\dagger|0\rangle$, states. The processes that lead to this splitting are sketched in Fig. \ref{fig:ionicP}. Therefore our effective low-energy Hamiltonian is a tight binding model involving just the metal atoms:
\begin{eqnarray}
\hat{\cal H}_\textrm{eff}&=&-t^*\sum_{\sigma} \left(\hat c_{1\sigma}^\dagger \hat c_{2\sigma} + \hat c_{2\sigma}^\dagger \hat c_{1\sigma} \right), 
\end{eqnarray}
where, to second order in $t/\Delta$, the effective metal-to-metal hopping integral is given by
\begin{eqnarray}
t^*=-\frac{t^2}{\Delta}.
\end{eqnarray}
Note that, even though $t$ is positive,  $t^*<0$ (or, equivalently, $\beta^*>0$), in contrast to our na\"ive expectation that hopping integrals are positive ($\beta<0$; cf. section \ref{sect:Huckel}).

\section{The Holstein model}\label{sect:Holstein}

So far we have assumed that the nuclei or ions form a passive background through which the electrons move. However, in many situations this is not the case. Atoms move and these lattice/molecular vibrations interact with the electrons via the electron-phonon/vibronic interaction. One of the simplest models of such effects is the Holstein model, which we discuss below. Electron-vibration interactions play important roles
across science. In physics electron-phonon interactions can give rise to
superconductivity \cite{Ziman}, spin and charge density waves
 \cite{Gruner}, polaron formation \cite{Mott&Alexandrov} and
piezoelectricity \cite{Ziman}. In chemistry vibronic interactions impact
electron-transfer processes \cite{Marcus}, Jahn-Teller effects
 \cite{Bersuker}, spectroscopy \cite{Bersuker}, stereochemistry
 \cite{Bersuker}, activation of chemical reactions \cite{Bersuker}
and catalysis \cite{Bersuker}. In biology the vibronic interactions
play important roles in photoprotection \cite{Olsen},
photosynthesis \cite{Reimers} and vision \cite{rhodopsin}. It is
therefore clear that one of the central tasks for condensed matter
theory and theoretical chemistry is to describe electron-vibration interactions.

In general one may write the Hamiltonian of a system of electrons and nuclei as
\begin{eqnarray}
\hat{\cal H}=\hat{\cal H}_e+\hat{\cal H}_n+\hat{\cal H}_{en},
\end{eqnarray}
where $\hat{\cal H}_e$ contains those terms that only effect the electrons, $\hat{\cal H}_n$ contains those terms that only effect the nuclei and $\hat{\cal H}_{en}$ describes the interactions between the electrons and the nuclei. $\hat{\cal H}_e$ might be any of the Hamiltonians we have discussed above. However, for the Holstein model one assumes a tight-binding form for $\hat{\cal H}_e$. In the normal mode approximation \cite{Bersuker}, which we will make, one treats molecular/lattice vibrations as harmonic oscillators (cf. section \ref{sho}). As the ions carry a charge, any displacement of the ions from their equilibrium positions will change the potential felt by the electrons. The Holstein model assumes that each vibrational mode is localised on a single site. For this to be the case the site must have some internal structure, i.e., the site cannot correspond to a single atom. Therefore the Holstein model is more appropriate for a molecular solids than for simple crystals. For small displacements, $x_{i\mu}$, of the $\mu^\textrm{th}$ mode of the $i^\textrm{th}$ lattice site we can perform a Taylor expansion in the dimensionless normal coordinate of the vibration, $Q_{i\mu}=x_{i\mu}\sqrt{m_{i\mu}\omega_{i\mu}/\hbar}$ where $m_{i\mu}$ and $\omega_{i\mu}$ are, respectively, the mass and the frequency of the $\mu^\textrm{th}$ mode on the $i^\textrm{th}$ site, and we find that
\begin{equation}
\hat{\cal H}_{en}=\sum_{ij\sigma\mu} \frac{\partial t_{ij}}{\partial Q_{i\mu}} Q_{i\mu} \left( \hat c_{i\sigma}^\dagger\hat c_{j\sigma}+\hat c_{j\sigma}^\dagger\hat c_{i\sigma}\right)+\dots.
\end{equation}
In the Holstein model one assumes that the derivative vanishes for $i\ne j$.  We may quantise the vibrations in the usual way (cf. section \ref{sho}) which yields
\begin{equation}
\hat{\cal H}_{en}=\sum_{i\sigma\mu} g_{i\mu} (\hat a_{i\mu}^\dagger +\hat a_{i\mu}) \hat c_{i\sigma}^\dagger\hat c_{i\sigma},
\end{equation}
where $\hat a_{i\mu}^{(\dagger)}$ destroys (creates) a quantised vibration in the $\mu^\textrm{th}$ mode on the $i^\textrm{th}$ site, $g_{i\mu}=2^{-1/2}{\partial t_{ii}}/{\partial Q_{i\mu}}$ and
%\begin{eqnarray}
$\hat{\cal H}_n=\sum_{i\mu}\hbar\omega_{i\mu}\hat a_{i\mu}^\dagger\hat a_{i\mu}$.
%\end{eqnarray}
Thus
\begin{eqnarray}
\hat{\cal H}_\textrm{Holstein}&=&-t\sum_{\langle ij\rangle\sigma}\hat c_{i\sigma}^\dagger\hat c_{j\sigma}+\sum_{i\mu}\hbar\omega_{i\mu}\hat a_{i\mu}^\dagger\hat a_{i\mu}\notag\\&&+\sum_{i\sigma\mu} g_{i\mu} (\hat a_{i\mu}^\dagger +\hat a_{i\mu}) \hat c_{i\sigma}^\dagger\hat c_{i\sigma}
\end{eqnarray}

\subsection{Two site Holstein model}

If we assume that there is only one electron and one mode per site then the Holstein model simplifies to 
\begin{eqnarray}
\hat{\cal H}_\textrm{Holstein}&=&-t\sum_{\sigma}\left(\hat c_{1\sigma}^\dagger\hat c_{2\sigma}+\hat c_{2\sigma}^\dagger\hat c_{1\sigma}\right)+\hbar\omega\sum_{i}\hat a_{i}^\dagger\hat a_{i}\notag\\&&+g \sum_{i} (\hat a_{i}^\dagger +\hat a_{i}) \hat n_{i}
\end{eqnarray}
on two symmetric sites, where $\hat n_i=\sum_\sigma\hat n_{i\sigma}=\sum_\sigma\hat c_{i\sigma}^\dagger\hat c_{i\sigma}$. It is useful to change the basis in which we consider the phonons to that of in phase (symmetric), $\hat s=(\hat a_1 +\hat a_2)/\sqrt{2}$, and out of phase (antisymmetric), $\hat b=(\hat a_1 -\hat a_2)/\sqrt{2}$, vibrations. In this basis one finds that
\begin{subequations}
\begin{eqnarray}
\hat{\cal H}_\textrm{Holstein}=\hat{\cal H}_s+\hat{\cal H}_{be}
\end{eqnarray}
where
\begin{eqnarray}
\hat{\cal H}_{s}=\hbar\omega\hat s^\dagger\hat s+\frac{g}{\sqrt2} (\hat s^\dagger +\hat s) (\hat n_{1} + \hat n_{2})
\end{eqnarray}
and
\begin{eqnarray}
\hat{\cal H}_{be}&=&-t\sum_{\sigma}\left(\hat c_{1\sigma}^\dagger\hat c_{2\sigma}+\hat c_{2\sigma}^\dagger\hat c_{1\sigma}\right)+\hbar\omega\hat b^\dagger\hat b\notag\\&&+\frac{g}{\sqrt2} (\hat b^\dagger +\hat b) (\hat n_{1} - \hat n_{2}).
\end{eqnarray}
\end{subequations}
Note that $\hat n_{1} + \hat n_{2}=N$, the total number of electrons in the problem. As $N$ is a constant of the motion the dynamics of the electrons cannot effect the symmetric vibrations and \emph{vice versa}. Hence all of the interesting effects are contained in $\hat{\cal H}_{be}$ and we need only study this Hamiltonian below.

\subsubsection{Diabatic limit, $\hbar\omega\gg t$}\label{sect:polaron}

In the diabatic limit the vibrational modes are assumed to instantaneously adapt themselves to the particle's position. Thus 
\begin{equation}
\hbar\omega\hat b^\dagger\hat b+\frac{g}{\sqrt2} (\hat b^\dagger +\hat b) (\hat n_{1} - \hat n_{2})=\hbar\omega\hat b^\dagger\hat b\pm\frac{g}{\sqrt2} (\hat b^\dagger +\hat b). 
\end{equation}
The plus sign is relevant when the electron is located on site 1 and the minus sign is relevant when the electron is on site 2. We now introduced the `displaced oscillator transformation',
\begin{eqnarray}
\hat b^\dagger_\pm=\hat b^\dagger \pm \frac1{\sqrt2}\frac{g}{\hbar\omega}. 
\end{eqnarray}
Therefore we find that
\begin{eqnarray}
\hat{\cal H}_{be}&=&-t\sum_{\sigma}\left(\hat c_{1\sigma}^\dagger\hat c_{2\sigma}+\hat c_{2\sigma}^\dagger\hat c_{1\sigma}\right)\notag\\&&+\hbar\omega\left(\hat b_+^\dagger\hat b_++\hat b_-^\dagger\hat b_-\right)-\frac{g^2}{\hbar^2\omega^2}.
\end{eqnarray}

It is important to note that the operators $\hat b_+$ and $\hat b_-$ satisfy the same commutation relations as the $\hat b$ operator, therefore they describe bosonic excitations. We define the ground states of the displaced oscillators by $\hat b_-|0_-\rangle=0$ and $\hat b_+|0_+\rangle=0$. Therefore
\begin{eqnarray}
\hat b |0_+\rangle=-\frac1{\sqrt2}\frac{g}{\hbar\omega} |0_+\rangle 
\end{eqnarray}
and hence
\begin{subequations}\label{eqn:coh-both}
\begin{eqnarray}
\hat b_- |0_+\rangle=-\frac{{\sqrt2}g}{\hbar\omega} |0_+\rangle; \label{eqn:coh+}
\end{eqnarray}
similarly
\begin{eqnarray}
\hat b_+ |0_-\rangle=\frac{{\sqrt2}g}{\hbar\omega} |0_-\rangle,\label{eqn:coh-}
\end{eqnarray}
\end{subequations}
i.e., $|0_\pm\rangle$ is an eigenstate of $\hat b_\mp$ with eigenvalue $\mp\sqrt2g/\hbar\omega$. The eigenstates of bosonic annihilation operators are known as coherent states \cite{Walls}. Eqs.  \ref{eqn:coh-both} therefore show that the ground state of one of the $\hat b_\pm$ operators may be written as a coherent state of the other operator \cite{Weiss}, i.e., 
\begin{eqnarray}
 |0_\pm\rangle=\exp\left[ -\frac{\sqrt2g}{\hbar\omega} \left(\frac12 \pm \hat b_\mp^\dagger \right)\right]  |0_\mp\rangle.
 \end{eqnarray}
Therefore
\begin{eqnarray}
\langle 0_+ |0_-\rangle=\exp\left[ -\frac{g^2}{\hbar^2\omega^2} \right],
\end{eqnarray}
which is known as the Franck-Condon factor.

The Franck-Condon factor describes the fact that, in the diabatic limit, the bosons cause a `drag' on the electronic hopping. That is, we can describe the solution of the diabatic limit in terms of an effective two site tight binding model if we replace $t$ by
\begin{eqnarray}
t^*=t\langle 0_+ |0_-\rangle=t\exp\left[ -\frac{g^2}{\hbar^2\omega^2} \right].
\end{eqnarray}
Thus the hopping integral is `renormalised' by the interactions of the electron with the vibrational modes (cf. section \ref{eh_v_se}). This renormalisation is also found in the solution of an electron moving on a lattice in the diabatic limit. In this context the exponential factor is known as the polaronic band narrowing \cite{Mott&Alexandrov}. The exponential factor results from the small overlap of the two displaced operators, and may be thought of as an increase in the effective mass of the electron.

\subsubsection{Adiabatic limit, $\hbar\omega\ll t$}

We begin by noting that, as there is only one electron the spin of the electron only leads to a trivial two-fold degeneracy, and therefore can be neglected without loss of generality. A useful notational change is to introduce a pseudospin notation where
we define $\hat\sigma_z=\hat c_{1\sigma}^\dagger\hat c_{1\sigma}-\hat c_{2\sigma}^\dagger\hat c_{2\sigma}$ and $\hat\sigma_x=\hat c_{1\sigma}^\dagger\hat c_{2\sigma}+\hat c_{2\sigma}^\dagger\hat c_{1\sigma}$. Therefore the one electron, two site Holstein model Hamiltonian becomes
\begin{eqnarray}
\hat{\cal H}_{sb}=-t \hat\sigma_x+\hbar\omega\hat b^\dagger\hat b+\frac{g}{\sqrt2} (\hat b^\dagger +\hat b) \hat\sigma_z.
\end{eqnarray}
which is often referred to as the spin-boson model.

Let us now replace the bosonic operators by position and momentum operators for the harmonic oscillator defined as
\begin{subequations}
\begin{eqnarray}
\hat x=\sqrt{\frac{\hbar}{2m\omega}}\left(\hat b^\dagger + \hat b\right)
\end{eqnarray}
and 
\begin{eqnarray}
\hat p=i\sqrt{\frac{m\hbar\omega}{2}}\left(\hat b^\dagger - \hat b\right).
\end{eqnarray}
\end{subequations}
Therefore
\begin{eqnarray}
\hat{\cal H}_{sb}=-t \hat\sigma_x+\frac{\hat p^2}{2m}+\frac12m\omega\hat x^2+g\sqrt{\frac{m\omega}{\hbar}} \hat x \hat\sigma_z.
\end{eqnarray}

\begin{figure}
\includegraphics[width=8cm]{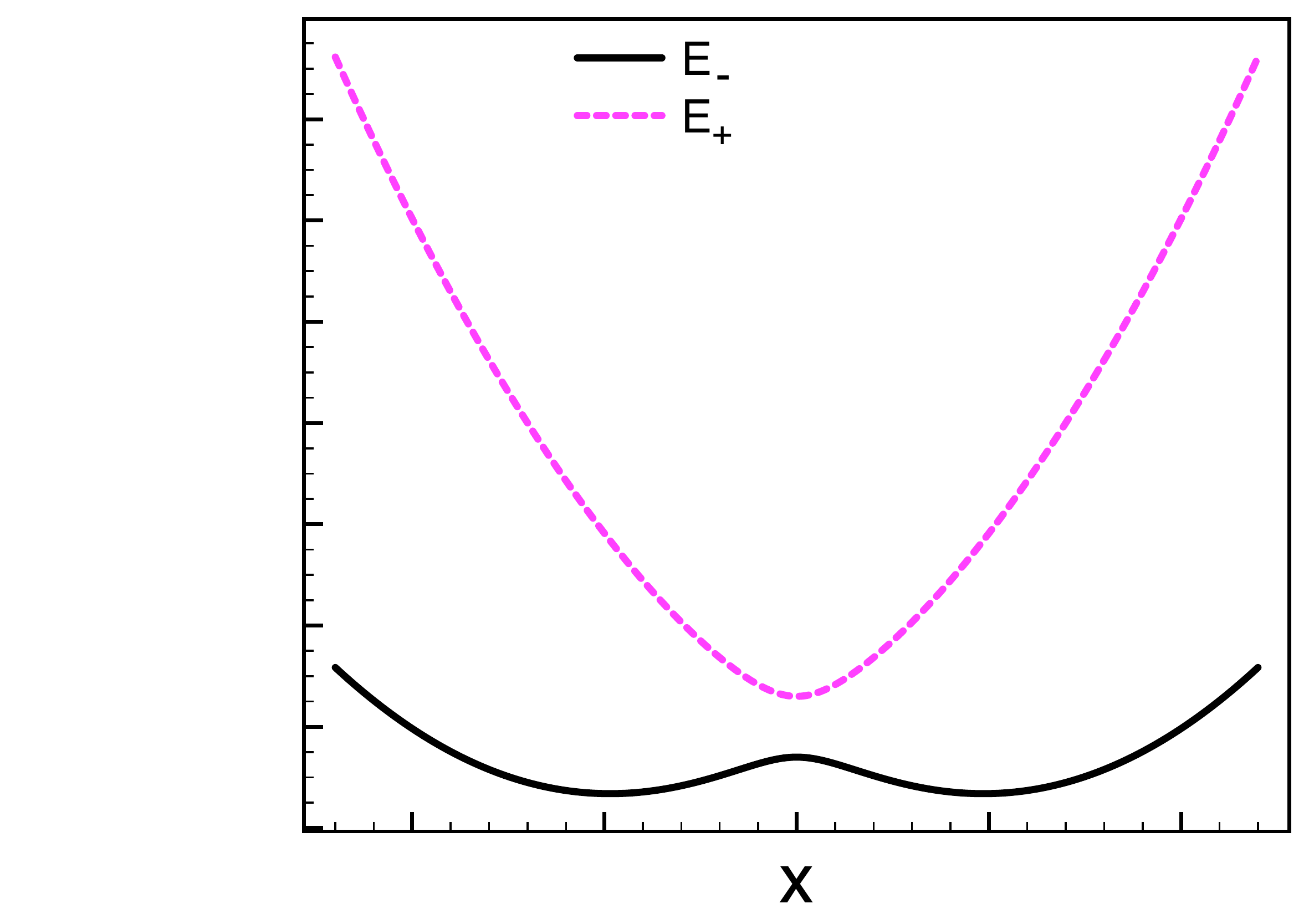}
\caption{The energies of the ground and excited states for a single electron in the two site Holstein model in the adiabatic weak coupling limit ($t\gg g\gg\hbar\omega$). Calculated from Eq. \ref{adia}. $x$ is the position of the harmonic oscillator describing out of phase vibrations.} \label{fig:holstein}
\end{figure}

The adiabatic limit is characterised by a sluggish bosonic bath that responds only very slowly to the motion of the electron, i.e.,  ${\hat p^2}/{2m}\rightarrow0$, which it is often helpful to think of as the $m\rightarrow\infty$ limit. Further, in the adiabatic limit the Born-Oppenheimer approximation \cite{Weiss,SchatzRatner} holds, which implies that the total wavefunction of the system, $|\Psi\rangle$, is a product of a electronic (pseudospin) wavefunction, $|\phi_e\rangle$, and a vibrational (bosonic) wavefunction, $|\psi_v\rangle$, i.e., $|\Psi\rangle=|\phi_e\rangle\otimes|\psi_v\rangle$. Therefore, the harmonic oscillator will be in a position eigenstate and we may replace the position operator, $\hat x$, by  a classical position $x$, yielding
\begin{subequations}
\begin{eqnarray}
\hat{\cal H}_{sb}&=&-t \hat\sigma_x+g\sqrt{\frac{m\omega}{\hbar}} x \hat\sigma_z+\frac12m\omega x^2 \\
&=& 
\left(
\begin{array}{cc}
g\sqrt{\frac{m\omega}{\hbar}} x  &   -t   \\
-t  &    -g\sqrt{\frac{m\omega}{\hbar}} x  
\end{array}
\right)
   +\frac12m\omega x^2,
\end{eqnarray}
\end{subequations}
where in the second line we have simply switched to the matrix representation of the Pauli matrices. This is easily solved and one finds that the eigenvalues are 
\begin{subequations}
\begin{eqnarray}
E_\pm&=&\frac12m\omega x^2\pm\sqrt{t^2+\frac{m\omega}{\hbar}g^2x^2}\label{adia}\\
&\approx&\frac12m\omega x^2\pm\frac{m\omega g^2x^2}{2 \hbar t}\pm t,\label{lim} 
\end{eqnarray}
\end{subequations}
where Eq. \ref{lim} holds in the weak coupling limit, $gx\ll t$. We plot the variation of these eigenvalues with $x$ in this limit in Fig. \ref{fig:holstein}. Notice that for the electronic ground state, $E_-$, the lowest energy states have $x\ne0$. This is an example of spontaneous symmetry breaking \cite{Anderson}, as ground state of a system has a lower symmetry than the Hamiltonian of the system. Thus the system must ``choose'' either the left well or the right well (but not both) in order to minimise its energy.

\section{Effective Hamiltonian or semi-empirical model?}\label{eh_v_se}

The models discussed in these notes are generally known as semi-empirical models in a chemical context and as effective Hamiltonians in the physics community. Here the difference is not just  nomenclature, but is also indicative of an important difference in the epistemological status awarded to these models by the two communities. In this section I will describe two different attitudes towards semi-empirical models/effective Hamiltonians and discuss the epistemological views embodied in the work of two of the greatest physicists of the twentieth century.

\subsection{The Diracian worldview}

Paul Dirac famously wrote \cite{Dirac} that ``the fundamental laws necessary for the mathematical treatment of a large part of physics and the whole of chemistry are thus completely known, and the difficulty lies only in the fact that application of these laws leads to equations that are too complex to be solved.'' There is clearly a great deal of truth in the statement. In solid state physics and chemistry we know that the Schr\"odinger equation provides an extraordinarily accurate description of the observed phenomena. Gravity, the weak and strong nuclear forces and relativistic corrections are typically unimportant, thus all of the interactions boil down to non-relativistic electromagnetic effects.

Dirac's worldview is realised in the  \emph{ab initio} approach to electronic structure. Wherein one starts from the Hartree-Fock solution to the full Schr\"odinger equation in some small basis set. One then adds in correlations via increasingly complex approximation schemes and increases the size of the basis set, in the hope that with a sufficiently large computer one will find an answer that is ``sufficiently close'' to the exact solution (full CI in an infinite complete basis set).

In the last few decades rapid progress has been made in  \emph{ab initio} methods due to an exponential improvement in computing technology, methodological progress, and the widespread availability of  implementations of these methods \cite{Pople}. However, this progress is unsustainable: the complexity recognised by Dirac eventually limits the accuracy possible from \emph{ab initio} calculations. Indeed, solving the Hamiltonian given in Eq. \ref{eqn:latticemodelH} is known to be computationally hard. Feynman proposed building a computer that uses the full power of quantum mechanics to carry out quantum simulations \cite{Feynman}. Indeed, the simplest of all quantum chemical problems, the H$_2$ molecule in a minimal basis set, has been solved on a prototype quantum computer \cite{Lanyon}. But, while  even a rather small scale quantum computer [containing just a few hundred qubits \cite{Lanyon}] would provide a speed-up over classical computation, it is believed that the solution of Hamiltonian \ref{eqn:latticemodelH} remains hard even on a quantum computer (i.e., it is believed that even a quantum computer could not solve Hamiltonian \ref{eqn:latticemodelH} in a time that grows only polynomially with the size of the system \cite{Schuch}). Further, simple extensions of these arguments provide strong reasons to believe that there is no efficiently computable approximation to the exact functional in density functional theory \cite{Schuch}.  Therefore it appears that the equations will always remain ``too complex to be solved'' directly. This suggests that semi-empirical models will always be required for large systems.

\subsection{The Wilsonian project}

Typically one is only interested in a few low-energy states of a system, perhaps the ground state and the first few excited states. Therefore, so long as our model gives the correct energies for these low-energy states we should regard it as successful. This, apparently simple, realisation, particularly as embodied by Wilson's renormalisation group \cite{Goldenfeld}, has had profound implications throughout modern physics from high energy particle physics to condensed matter physics.

The basic idea of renormalisation in remarkably simple. Imagine starting with some system that has a large number of degrees of freedom. As we have noted, for practical purposes we only care about about the lowest energy states. Therefore one might be tempted to simplify the description of the system by discarding the highest energy states. However, simply discarding such states will cause a shift in the low-energy spectrum. Therefore, one must remove the high energy states that complicate the description and render the problem computationally intractable in such as way as to preserve the low-energy spectrum. This is often referred to as `integrating out' the high energy degrees of freedom (because of the way this process is carried out in the path integral formulation of quantum mechanics \cite{Wen}). 
Typically integrating out the high energy degrees of freedom causes the parameters of the Hamiltonian to `flow' or `run', i.e., change their values. When this happens one says that the parameters are renormalised. 

A simple example is the Coulomb interaction between the two electrons in a neutral Helium atom. For simplicity lets imagine trying to calculate just the ground state energy. We begin by analysing the problem in the absence of a Coulomb interaction between the two electrons. In the ground state both electrons occupy the $1s$ orbital. We would like to work in as small a basis set as possible. The simplest approach is just to work in the minimal basis set, which, in this case, is just the two $1s$ spin-orbitals, $\phi_{1s\sigma}({\bf r})$. The total energy of a He atom neglecting the inter-electron Coulomb interaction is -108.8 eV (relative to the completely ionised state). Now we restore the Coulomb repulsion between electrons. A simple question is: how much does this change the total energy of the He atom? In the minimal basis set the solution seems straightforward:
\begin{eqnarray}
\langle 1s^2|V|1s^2\rangle&=&\int_{-\infty}^\infty d^3{\bf r}_1\int_{-\infty}^\infty d^3{\bf r}_2 \frac{e^2|\phi_{1s\uparrow}|^2 |\phi_{1s\downarrow}|^2}{4\pi\epsilon_0|{\bf r}_1-{\bf r}_2|}\notag\\&\simeq &34.0\textrm{ eV}.
\end{eqnarray}
Therefore it is tempting to conclude that we can model the He atom by a one site Hubbard model with $U=\langle 1s^2|V|1s^2\rangle$. However, this yields a total energy for the He atom of -74.8 eV, which is not particularly close to the experimental value of -78.975 eV \cite{Gasiorowicz}.

Let us then continue to consider the problem in the basis set of the hydrogenic atom, which is complete due to the spherical symmetry of the Hamiltonian. One can now straightforwardly carry out a perturbation theory around the non-interacting electron solution where we take
\begin{subequations}
\begin{eqnarray}
H_0=\sum_{i=1}^2\left(-\frac{\hbar^2\nabla_i^2}{2m}-\frac{e^2}{\pi\epsilon_0|{\bf r}_i|}\right),
\end{eqnarray}
and
\begin{eqnarray}
H_1=\frac{e^2}{4\pi\epsilon_0|{\bf r}_1-{\bf r}_2|}.
\end{eqnarray}
\end{subequations}
A detailed description of this perturbation theory is given in chapter 18 of Ref. \onlinecite{Gasiorowicz}. However, for our discussion, the key point is that, in this perturbation theory, the term $\langle 1s^2|V|1s^2\rangle$ is simply the first order correction to the ground state energy. It is therefore clear why the minimal basis set gives such a poor result, it ignores all the higher order corrections to the total energy.

The failure of the simple minimal basis set calculation does not, however, mean that the effective Hamiltonian approach also fails, despite the fact that the effective Hamiltonian is also in an extremely small basis set. Rather, one must realise that, as well as the first order contributions, $U$ also contains contributions from higher orders in perturbation theory. It is therefore possible, although extremely computationally demanding, to calculate the parameters for effective Hamiltonians from this kind of perturbation theory \cite{Freed}.

A more promising approach, which has been applied to a number of molecular crystals \cite{notEdan,Edan}, is to use atomistic calculations to parameterise an effective Hamiltonian. For example, density functional theory gives quite reasonable values for the total energy of the ground state of many molecules. Therefore one approach to calculating the Hubbard $U$ is to calculate the ionisation energy, $I=E_0(N-1)-E_0(N)$, and the electron affinity, $A=E_0(N)-E_0(N+1)$, of the molecule, where $E_0(n)$ is the ground state energy of the molecule when it contains $n$ electrons and $N$ is the filling corresponding to half-filled band. One finds that $U=I-A=E_0(N+1)+E_0(N-1)-2E_0(N)$. A simple way to see this is that if we assume the molecule is neutral when it contains $N$ electrons then $U$ corresponds to the energy difference in the charge disproportionation reaction $2M\rightleftharpoons M^++M^-$ for two well separated molecules, $M$. A more extensive discussion of this approach is given in Ref. \onlinecite{Edan}.

It is worth noting that we have actually carried out this program of parametrising effective Hamiltonians three times in the discussion above. In section \ref{Hub->Hei} we showed that the Heisenberg model is an effective low-energy model for the half-filled Hubbard model in the limit $t/U\rightarrow0$.  In section \ref{sect:ionic} we derived an effective tight binding model that involved only the metal sites from an ionic Hubbard model of a transition metal oxide.  Finally, in section \ref{sect:polaron} we showed that vibronic interactions lead to an effective tight binding model describing the low-energy physics of the Holstein model in the diabatic limit, and that in this model the quasiparticles (electron-like excitations) are polarons, a bound state of electrons and vibrational excitations with a mass enhanced over that of the bare electron.

However, to date, the most important method for parametrising effective Hamiltonians has been to fit the parameters to a range of experimental data, whence the name `semi-empirical'. Of course experimental data contains all corrections to all orders therefore this is indeed an extremely sensible thing to do.  But, it is important to understand that empiricism is not a dirty word. Indeed empiricism is what distinguishes science from other belief systems. Further, this empirical approach is exactly the approach that the mathematics tells one to take. It is also important to know that no quantum chemical or solid state calculation is truly \emph{ab initio} -  the nuclear and electronic masses and the charge on the electron are all measured rather than calculated. Indeed the modern view of the `standard model' of particle physics is that it too is an effective low-energy model \cite{confinement}. For example, in quantum electrodynamics (QED), the quantum field theory of light and matter, the bare charge on the electron is, for all practical purposes, infinite. But, the charge is renormalised to the value seen experimentally in a manner analogous to the renormalisation of the Hubbard $U$ of He discussed above. Therefore, as we do not, at the time of writing, know the correct mathematical description of processes at higher energies, all of theoretical science should, perhaps, be viewed as the study of semi-empirical effective low-energy Hamiltonians \cite{Laughlin}.

Finally, the most important point about effective Hamiltonians is that they promote understanding. Ultimately the point of science is to understand the phenomena we observe in the world around us. While the ability to perform accurate numerically calculations is important, we should not allow this to become our main goal. The models discussed above provide important insights into the chemical bond, magnetism, polarons, the Mott transition, electronic correlations, the failure of mean field theories,  etc. All of these effects are much more difficult to understand simply on the basis of atomistic calculations. Further, many important effects seen in crystals, such as the Mott insulator phase, are not found methods such as density functional theory or Hartree-Fock theory, while post Hartree-Fock methods are not practical in infinite systems. Thus effective Hamiltonians have a vital role to play in developing the new concepts that are required to understand the emergent phenomena found in molecules and solids \cite{moreisdifferent}. 

\acknowledgements

I would like to thank Balazs Gy\"orffy, who taught me that ``you can't not know'' many of things discussed above. I also thank James Annett, Greg Freebairn, Noel Hush, Anthony Jacko, Bernie Mostert, Seth Olsen, Jeff Reimers, Edan Scriven, Mike Smith, Eddy Yusuf, and particularly Ross McKenzie, for many enlightening conversations about the topics discussed above and for showing me that chemistry is a beautiful and rich subject with many simplifying principles. I would also like to thank Bernd Braunecker, Karl Chan, Anthony Jacko, Sergio Di Matteo, Ross McKenzie, Seth Olsen, Eddie Ross and Kristian Weegink for their insightful comments on an early draft of these notes. I am supported by a Queen Elizabeth II fellowship from the Australian Research Council (project DP0878523).


\begin{thebibliography}{999}

\bibitem{Fulde}
Fulde, P. \emph{Electron correlations in molecules and solids;} Springer: Berlin, 1995.

\bibitem{Jeff}
Reimers, J. R. \emph{Computational Methods for Large systems: Electronic Structure Approaches for Biotechnology and Nanotechnology;} Wiley: Hoboken, in press.

\bibitem{SchatzRatner}
Schatz, G. C.; Ratner, M. A. \emph{Quantum mechanics in chemistry;} Prentice Hall: Englewoods Cliffs, 1993.

\bibitem{Mahan}
Mahan, G. D.; \emph{Many-particle physics;} Kluwer Academic: New York, 2000.


\bibitem{Goldstein}
Goldstein, H.; Poole, C.; Safko, J. \emph{Classical mechanics;} Addison Wesley: San Francisco, 2002.

\bibitem{Atkins}
Atkins, P.; de Paula, J. \emph{Atkin's physical chemistry;} Oxford University Press: Oxford, 2006.

\bibitem{Rae}
See, for example, Rae,  A. I. M. \emph{Quantum mechanics;} Institute of Physics Publishing: Bristol, 1996.

\bibitem{Gasiorowicz}
See, for example, Gasiorowicz, S. \emph{Quantum physics;} Wiley: Hoboken, 2003.

\bibitem{JordanWigner}
Jordan P.; Wigner,  E. \emph{Z. Phys.} {\bf1928,} {\it47}, 631-651.

\bibitem{Lowe}
Lowe J. P.; Peterson, K. A. \emph{Quantum chemistry;} Elsevier: Amsterdam, 2006.

\bibitem{A&M}
Ashcroft N.W.; Mermin, N.D. \emph{Solid state physics;} Holt, Rinehart and Winston: New York, 1976.

\bibitem{Tinkham}
Tinkham, M. \emph{Group theory and quantum mechanics;} McGraw-Hill: New York, 1964.

\bibitem{Lax}
Lax, M. \emph{Symmetry principles in solid state and molecular physics;} Wiley: New York, 1974.

\bibitem{Coulson}
McWeeny, R. \emph{Coulson's valence;} Oxford University Press: Oxford, 1979.

\bibitem{Brogli}
Brogli, F.; Heilbronner, E. {\it Theor. Chim. Acta} {\bf1972,} {\it26}, 289-299.

\bibitem{Arfkin}
See, e.g., Arfken, G. \emph{Mathematical methods for physicists}, 3$^\textrm{rd}$ ed.; Academic Press: Orlando, 1985.

\bibitem{Mandl}
Mandl, F. \emph{Statistical physics;} Wiley: Chichester, 1998.

\bibitem{Arfkin799}
See pp 799-800 of Ref. \cite{Arfkin}.

\bibitem{Castro}
(a) Castro Neto, A. H.; Guinea, F.; Peres, N. M. R.; Novoselov, K. S.; Geim, A. K. \emph{Rev. Mod. Phys.} {\bf2009,} {\it81}, 109-162. (b) Castro Neto, A. H.; Guinea, F.; Peres, N. M. R. \emph{Phys. World} {\bf2006,} {\it19}, 33-37.

\bibitem{Geim}
(a) Novoselov, K. S.; Geim, A. K.; Morozov, S. V.; Jiang, D.; Zhang, Y.; Dubonos, S. V.; Gregorieva, I. V.; Firsov, A. A. {\it Science} {\bf2004,} {\it306}, 666-669. (b) Choucair, M.; Thordarson P.; Stride, J. A. {\it Nature Nanotech.} {\bf2009,} {\it4}, 30-33.

\bibitem{Schrodinger}
Schr\"odinger, E.  {\it Ann. Physik} {\bf1926,} {\it79}, 361-428.

\bibitem{HeitlerLondon} 
Heitler, W.; London,   F. {\it Z. Phys.} {\bf1927,} {\it44}, 455-472.

\bibitem{Pauling}
Pauling, L. \emph{The nature of the chemical bond and the structure of molecules and crystals}; Cornell Univ. Press: Ithaca, 1960.

\bibitem{Mott}
Mott, N.F. {\it Proc.  Roy. Soc. A} {\bf1949,} {\it62}, 416-422.

\bibitem{JPCMrev}
Powell, B. J.; McKenzie,  R. H. {\it J. Phys.: Condens. Matter} {\bf2006,} {\it18}, R827-R865.

\bibitem{Anderson87}
(a) Anderson, P. W. {\it Science} {\bf 1987,} {\it235}, 1196-1198. (b) Zhang, F. C.;
Gross, C.; Rice T. M.; Shiba, H. {\it Supercond. Sci. Technol.} {\bf 1988,} {\it1},
36-46.

\bibitem{Anderson-who-or-what}
Anderson, P. W. {\it Phys. Today} {\bf 2008,} {\it61} (4), 8-9.

\bibitem{RVBorganics}
Powell B. J.; McKenzie, R. H. {\it Phys. Rev. Lett.} {\bf2005,} {\it94}, 047004; Gan,  J. Y.; Chen, Y.; Su, Z. B.; Zhang, F. C. \ibid {\bf2005,} {\it94}, 067005; Liu, J.; Schmalian, J.; Trivedi, N. \ibid {\bf2005,} {\it94}, 127003.



\bibitem{Yang}
Cohen, A. J.; Mori-Sanchez, P.; Yang, W. T.   {\it Science}  {\bf2008,} {\it321}, 792-794. 

\bibitem{Rossler}
R\"ossler, U.  \emph{Solid state theory;} Springer: Berlin, 2004.

\bibitem{Mohn}
Mohn P.; Wohlfarth, E. P. {\it J. Mag. Mag. Mat.} {\bf1987,} {\it68}, L283-L285.

\bibitem{Jacko}
Jacko, A. C.; Fj\ae restad, J. O.; Powell, B. J. {\it Nature Phys.} {\bf2009,} {\it5}, 422-425.

\bibitem{Gutzwiller}
Gutzwiller, M. C. {\it Phys. Rev. Lett.} {\bf 1963,} {\it10}, 159-162.

\bibitem{BR}
Brinkmann W. F.; Rice, T. M. {\it Phys. Rev. B} {\bf 1970,} {\it2}, 4302-4304.

\bibitem{LiebWu}
Lieb, E. H.; Wu,  F. Y. {\it Phys. Rev. Lett.} {\bf1968,} {\it20}, 1445-1448.

\bibitem{Essler}
 Essler, F. H. L.; Frahm, H.; G\"ohmann, F.; Kl\"umper,  A.; Korepin, V. E. \emph{The one-dimensional Hubbard model;} Cambridge University Press: Cambridge, 2005.

\bibitem{Tsvelik}
Tsvelik, A. M. \emph{Quantum field theory in condensed matter physics;} Cambridge University Press: Cambridge, 1996.

\bibitem{DMFT}
Kotliar G.; Vollhardt,  D. {\it Phys. Today} {\bf 2004,} {\it57} (3), 53-59.

\bibitem{Kollar} 
Kollar, M.; Strack, R.; Vollhardt, D. {\it Phys. Rev. B}  {\bf1996,} {\it53}, 9225-9231.

\bibitem{CDMFT}
Maier, T.; Jarrell, M.; Pruschke, T.; Hettler, M. H. {\it Rev. Mod. Phys.} {\bf 2005,} {\it77}, 1027-1080.

\bibitem{Kotliar}
Kotliar, G.; Savrasov, S. Y.; Haule, K.; Oudovenko, V. S.; Parcollet, O.; Marianetti, C. A. {\it Rev. Mod. Phys.} {\bf2006,} {\it78}, 865-951.

\bibitem{Nagaoka}
Nagaoka, Y. {\it Phys. Rev.} {\bf 1966,} {\it145}, 392-405.

\bibitem{Tian} 
Tian, G. {\it J. Phys. A} {\bf 1990,} {\it23}, 2231-2236.

\bibitem{MPM}
Merino, J.;  Powell, B. J.; McKenzie, R. H. {\it Phys. Rev. B} {\bf2006,} {\it73},
235107.

\bibitem{Shaik}
Shaik, S.; Hiberty, P. C. {Valence bond theory, its history, fundamentals, and applications: a primer}. In \emph{Reviews in computational chemistry;} Lipkowitz, K. B., Larter, R., Cundari,  T. R., Eds.;  Wiley-VCH: Hoboken, NJ, 2004; pp. 1-100.

\bibitem{Sakurai}
Sakurai, J. J. \emph{Modern quantum mechanics;} Addison-Wesley: New York, 1994.

\bibitem{Chao}
Chao, K. A.; Spa\l ek, J.; Ole\'s, A. M. {\it J. Phys. C} {\bf 1977,} {\it10}, L271-L276.

\bibitem{Brookhouse}
Brockhouse, B. N. {Slow neutron spectroscopy and the grand atlas of 
the physical world}. In \emph{Nobel lectures in physics, 
1991$-$1995;} Ekspong,  G., Ed.; World Scientific: Singapore, 1997. Also available from \url{http://nobelprize.org/nobel_prizes/physics/laureates/1994/brockhouse-lecture.html}. 

\bibitem{Zaliznyak}
Zaliznyak, I. A. {\it Nature Mat.} {\bf2005,} {\it4}, 273-275.

\bibitem{confinement}
Griffiths, D. \emph{Introduction to elementary particles;} Wiley-VCH: Weinheim, 2008.

\bibitem{Lake.Coldea}
(a) Coldea, R.; Tennant, D. A.; Tylczynski, Z. {\it Phys. Rev. B} {\bf2003,} {\it68}, 134424. (b) Lake, B.; Tennant, D. A.; Frost, C. D.; Nagler, S. E. {\it Nature Mat.} {\bf2005,} {\it4}, 329-334.

\bibitem{Lee}
Lee, P. A.  {\it Science} {\bf2008,} {\it321}, 1306-1307.

\bibitem{Shimuzu}
Shimizu Y.; \etal {\it Phys. Rev. Lett.} {\bf 2003,} {\it91}, 107001.

\bibitem{Helton}
Helton J.; \etal {\it Phys. Rev. Lett.} {\bf 2007,} {\it98}, 107204.

\bibitem{Okamoto}
Okamoton Y.; \etal {\it Phys. Rev. Lett.} {\bf 2007,} {\it99}, 137207.

\bibitem{Oles}
Raczkowski, M.; Fr\'esard, R.; Oles, A. M. {\it J. Phys.: Condens. Matter} {\bf 2006,} {\it18}, 7449-7469.

\bibitem{Sarma} 
Sarma, D. D. {\it J. Sol. State Chem.} {\bf 1990,} {\it88}, 45-52.

\bibitem{MPM2}
(a) Merino, J.; Powell, B. J.; McKenzie, R. H.
{\it Phys. Rev. B} {\bf2009,} {\it79}, 161103(R); (b) Merino, J.; McKenzie, R. H.; Powell, B. J. {\it Phys. Rev. B} {\bf2009,} {\it80}, 045116. 
(c) Powell, B. J.; Merino, J.; McKenzie, R. H. {\it Phys. Rev. B} {\bf2009,} {\it80}, 085113.

\bibitem{Ziman}
See for example, Ziman, J. M. {\it Electrons and phonons;} Oxford
University Press: Oxford, 1960.

\bibitem{Gruner}
For a review see Gr{\"u}ner, G. {\it Density waves in
solids;} Perseus Publishing: Cambridge, 1994.

\bibitem{Mott&Alexandrov}
See, for example, Alexandrov, A. S.; Mott, N. F. {\it Polarons and
biploarons;} World Scientific: Singapore, 1995.

\bibitem{Marcus}
For a review see Marcus, R. A. {\it Rev. Mod. Phys.} {\bf1993,} {\it65}, 599-610.

\bibitem{Bersuker}
See, for example, Bersuker, I. B. {\it The Jahn-Teller effect and
vibronic interactions in modern chemistry;} Plenum: New York,
1984.

\bibitem{Olsen}
(a) Olsen, S.; Riesz,  J.; Mahadevan, I.; Coutts, A.; Bothma, J. P.; Powell, B. J.; McKenzie, R. H.; Smith, S. C.; Meredith, P. %Convergent proton-transfer photocycles violate mirror-image symmetry in a 
%key melanin monomer
 {\it J. Am. Chem. Soc.}  {\bf2007,} {\it129}, 6672-6673. (b) Meredith, P.;  Powell, B. J.; Riesz, J.; Nighswander-Rempel, S.; Pederson,  M. R.; Moore, E.;
{\it Soft Matter} {\bf2006,} {\it 2}, 37-44.


\bibitem{Reimers}
Reimers, J. R.; Hush,  N. S. {\it J. Am. Chem. Soc.} {\bf2004,} {\it126}, 4132Ð4144.

\bibitem{rhodopsin} 
Hahn, S.; Stock, G. {\it J. Phys. Chem. B} {\bf2000,} {\it104}, 1146-1149.

\bibitem{Walls}
Walls D. F.; Milburn, G. J. \emph{Quantum optics;} Springer: Berlin, 2006.

\bibitem{Weiss}
Weiss, U. \emph{Quantum dissipative systems;} World Scientific: Singapore, 2008.

\bibitem{Anderson}
For an introductory discussion of broken symmetry see, for example, Blundell, S. J. {\it Magnetism in condensed matter} Oxford University Press, Oxford, 2001. For a more advanced discussion see, for example, Anderson, P. W. \emph{Basic notions of condensed matter physics} Benjamin-Cummings: Menlo Park, 1984.

\bibitem{Dirac}
Dirac, P. {\it Proc. Roy. Soc. A} {\bf1929,} {\it123}, 714-733.

\bibitem{Pople}
(a) Pople, J. A. {\it Rev. Mod. Phys.} {\bf1999,} {\it71}, 1267-1274. (b) Truhlar, D. G. {\it J. Am. Chem. Soc.,} {\bf2008,} {\it130}, 16824-16827.

\bibitem{Feynman}
Feynman, R. P. {\it Int. J. Theor. Phys.} {\bf1982,} {\it21}, 467-488.

\bibitem{Lanyon}
Lanyon, B. P.; Whitfield, J. D.; Gillet, G. G.; Goggin, M. E.; Almeida, M. P.;  
Kassal, I.; Biamonte, J. D.; Mohseni, M.; Powell,  B. J.;
Barbieri, M.; Aspuru-Guzik, A.; White, A. G. {\it Nature Chem.} {\bf2010,} {\it2,} 106-111.

\bibitem{Schuch}
Schuch N.; Verstraete, F. {\it Nature Phys.} published online: 23 August 2009, doi:10.1038/nphys1370.

\bibitem{Goldenfeld}
Goldenfeld, N. D.  {\it Lectures on Phase Transitions and the Renormalisation Group;} Addison-Wesley, 1992.

\bibitem{Wen}
See, for example, Wen, X.-G. \emph{Quantum field theory of many-body systems;} Oxford Univ. Press: Oxford, 2004.

\bibitem{Freed} 
(a) Freed, K. F. {\it Acc. Chem. Res.} \textbf{1983,} {\it16}, 137-144.
(b) Gunnarsson, O. {\it Phys. Rev. B} {\bf 1990,} {\it41}, 514-518.
(c) Iwata S.; Freed,  K. F. {\it J. Chem. Phys.} {\bf1976,} {\it65}, 1071-1088. 
(d) Graham R. L.; Freed, K. F. {\it J. Chem. Phys.} {\bf1992,} {\it96}, 1304-1316.
(e) Martin C. M.; Freed, K. F. {\it J. Chem. Phys.} {\bf1994,} {\it100}, 7454-7470.
(f) Stevens, J. E.; Freed, K. F.; Arendt, F.; Graham, R. L. {\it J. Chem. Phys.} {\bf1994,} {\it101}, 4832-4841.
(g) Finley J. P.; Freed,  K. F. {\it J. Chem. Phys.} {\bf1995,} {\it102}, 1306-1333.
(h) Stevens, J. E.; Chaudhuri, R. K.; Freed, K. F. {\it J. Chem. Phys.} {\bf1996,} {\it105}, 8754-8768.
(i) Chaudhuri, R. K.; Freed, K. F. {\it J. Chem. Phys.} {\bf2003,} {\it119}, 5995-6002. 
(j) Chaudhuri, R. K.; Freed, K. F. {\it J. Chem. Phys.} {\bf2005,} {\it122}, 204111.

\bibitem{Edan}
(a) Scriven, E.; Powell, B. J.
{\it J. Chem. Phys.} {\bf2009,} {\it130}, 104508; (b) {\it Phys. Rev. B} {\bf 2009,} {\it80,} 205107.

\bibitem{notEdan}
(a) Martin, R. L.; Ritchie, J. P. {\it Phys. Rev. B} {\bf 1993,} {\it48}, 4845-4849.
(b) Antropov, V. P.; Gunnarsson, O.; Jepsen, O. {\it Phys. Rev. B} {\bf ,1992} {\it46}, 13647-13650.
(c) Pederson, M. R.; Quong, A. A. {\it Phys. Rev. B} {\bf 1992,} {\it46}, 13584-13591.
(d) Brocks,  G.; van den Brink, J.; Morpurgo, A. F. {\it Phys. Rev. Lett.} {\bf2004,} {\it93}, 146405.
(e) Cano-Cort\'es, L.; Dolfen, A.; Merino, J.; Behler, J.; Delley, B.; Reuter, K.; Koch, E. {\it Eur. Phys. J. B} {\bf2007,} {\it56}, 173-176.


\bibitem{Laughlin}
For an accessible, and highly outspoken, discussion of these ideas see Laughlin, R. B.; Pines, D. {\it Proc. Natl. Acad. Sci.} {\bf2000,} {\it97}, 28-31 and Laughlin, R. B.  \emph{A different universe} Basic Books: New York, 2005.

\bibitem{moreisdifferent}
Anderson, P. W. {\it Science} {\bf 1972,} {\it177}, 393-396.

%only in figs

\bibitem{CiA}
Powell,  B. J. {\it Chem. Aust.} {\bf2009,} {\it76}, 18-21.

\end{thebibliography}
\end{document}